\pdfoutput=1

\documentclass[final,3p,times,twocolumn]{elsarticle}
\usepackage{amsmath}
\journal{Nucl. Instr. Meth. Phys. Res.}
\begin{document}

\begin{frontmatter}


\author[ut]{R.~Abbasi}
\author[ut]{M.~Abou~Bakr~Othman}
\author[ks]{C.~Allen}
\author[pu]{L.~Beard}
\author[ut]{J.~Belz}
\author[ks,mep]{D.~Besson}
\author[ut]{M.~Byrne}
\author[ut]{B.~Farhang-Boroujeny}
\author[ut]{A.~Gardner}
\author[bg]{W.H.~Gillman}
\author[ut]{W.~Hanlon}
\author[ks]{J.~Hanson}
\author[ut]{C.~Jayanthmurthy}
\author[ks]{S.~Kunwar}
\author[us]{S.L.~Larson}
\author[ut]{I.~Myers\corref{cor1}}
\cortext[cor1]{Corresponding Author. Tel.: +01 801 5879986. Addr.: 115 S 1400 E \#201 JFB}
\ead{isaac@cosmic.utah.edu}
\author[ks]{S.~Prohira}
\author[ks]{K.~Ratzlaff}
\author[ut]{P.~Sokolsky}
\author[bh]{H.~Takai}
\author[ut]{G.B.~Thomson}
\author[ut]{D.~Von~Maluski}
\address[ut]{University of Utah, Salt Lake City, UT 84112 U.S.A.}
\address[ks]{University of Kansas, Lawrence, KS 66045 U.S.A.}
\address[pu]{Purdue University, West Lafayette, IN 47907 U.S.A.}
\address[bg]{Gillman \& Associates, Salt Lake City, UT 84106 U.S.A.}
\address[us]{Utah State University, Logan, Utah 84322 U.S.A.}
\address[bh]{Brookhaven National Laboratory, Upton, NY 11973 U.S.A.}
\address[mep]{Moscow Engineering and Physics Institute, 31 Kashirskaya Shosse, Moscow 115409 Russia}

\title{Telescope Array Radar (TARA) Observatory for Ultra-High Energy Cosmic Rays}

\begin{abstract}
%% Define abbrev. exphlicitly here !!!!!
Construction was completed during summer 2013 on the Telescope Array RAdar (TARA) bi-static radar observatory for Ultra-High-Energy Cosmic Rays (UHECR). TARA is co-located with the Telescope Array, the largest ``conventional'' cosmic ray detector in the Northern Hemisphere, in radio-quiet Western Utah. TARA employs an 8~MW Effective Radiated Power (ERP) VHF transmitter and smart receiver system based on a 250~MS/s data acquisition system in an effort to detect the scatter of sounding radiation by UHECR-induced atmospheric ionization. TARA seeks to demonstrate bi-static radar as a useful new remote sensing technique for UHECRs. In this report, we describe the design and performance of the TARA transmitter and receiver systems.

\end{abstract}

\begin{keyword}
cosmic ray \sep FPGA \sep radar \sep digital signal processing \sep chirp

\end{keyword}

\end{frontmatter}

%\linenumbers

%%%%%%%%%%%%%%%%%%%%%%%%%%%%%%%%%%%  MAIN TEXT  %%%%%%%%%%%%%%%%%%%%%%%%%%%%%%%%%%%%%%%%%
%%
\section{Introduction}
\label{sec:intro}
% ~\footnote{Ultra-High Energy Cosmic Rays (UHECR), Telescope Array RAdar (TARA), Effective Radiated Power (ERP), Extensive Air Shower (EAS), Field Programmable Gate Array (FPGA), Surface Detector (SD), Fluorescence Detector (FD), Continuous Wave (CW), Radio Frequency (RF), Signal to Noise Ratio (SNR), Analog to Digital Converter (ADC)}
%% Isaac: In combining EAS Intro. with this intro, which reviewer said was the same and should be merged, I lost
%% a good place to use the watson and KampUng citations, which deal with GZK and composition, respectively. Not sure 
%% why they were here in the first place?  
Cosmic rays with energies per nucleon in excess of $\approx 10^{14}$ eV~\cite{auger_1939} create cascades of particles with electromagnetic and hadronic components in the atmosphere, known as Extensive Air Showers (EAS).  Conventional cosmic ray experiments detect events through coincident shower front particles in an array of surface detectors~\cite{tasd,pao_sd} or through fluorescence photons that radiate from the shower core~\cite{ta_fd,pao_fd,Tunka} which permit fluorescence telescopes to study shower longitudinal development.  Another technique takes advantage of two naturally-emitted radio signals: the Askaryan effect~\cite{Askaryan} and geomagnetic radiation from interactions with the Earth's magnetic field~\cite{kahn}. 

With ground arrays, air shower particles are observed directly. The land required to instrument ground arrays is large, {\em cf.} Telescope Array's 700~km$^2$ surface detector covers roughly the same land area as New York City. The costs of the equipment required to instrument such a large area are substantial, and the available land can only be found in fairly remote areas.

A partial solution to the difficulties and expense involved in ground arrays is found in the fluorescence technique. Here, the atmosphere itself is part of the detection system, and air shower properties may be determined at distances as remote as 40~km. Unfortunately fluorescence observatories are typically limited to a ten percent duty cycle by the sun, moon and weather. 

The possibility of radar observation of cosmic rays dates to the 1940's, when Blackett and Lovell~\cite{blovell} proposed cosmic rays as an explanation of anomalies observed in atmospheric radar data. At that time, a radar facility was built at Jodrell Bank to detect cosmic rays, but no results were ever reported. Recent experimental efforts utilizing atmospheric radar systems were conducted at Jicamarca~\cite{wahl2008} and at the MU-Radar~\cite{terasawa2009}. Both observed a few signals of short duration indicating a relativistic target. However in neither case were the measurements made synchronously with a conventional cosmic ray detector. 

\begin{figure}[h]
\centerline{\includegraphics[trim=0.0cm 0.0cm 0.0cm 0.0cm,clip=true,width=.47\textwidth,natwidth=5764,natheight=4598]{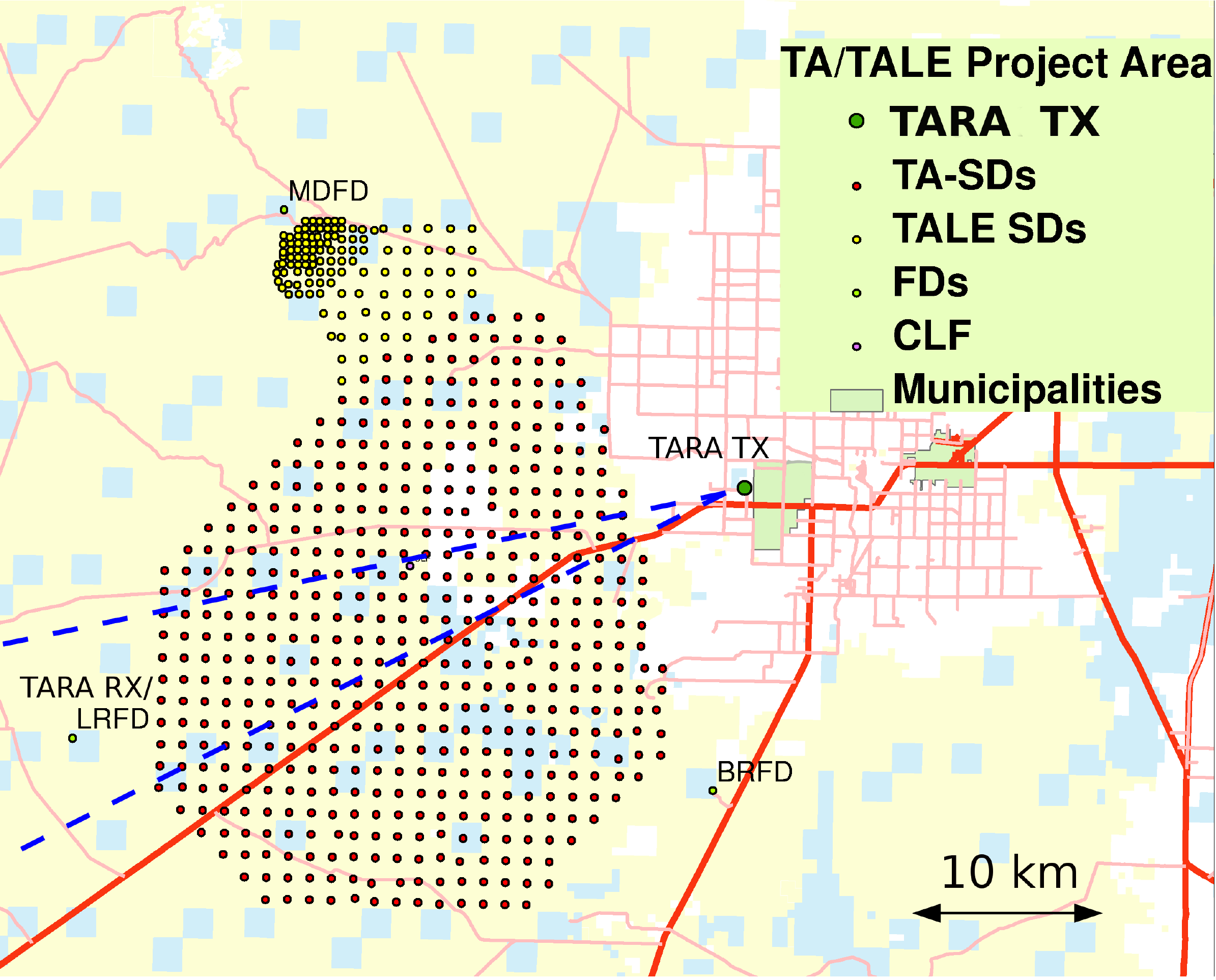}}
{\caption{Map of TARA Observatory sites (transmitter and receiver) along with the Telescope Array (TA) detector facilities. The transmitter broadcasts as station WF2XZZ near Hinckley, Utah, towards a receiver site located at the TA Long Ridge Fluorescence Detector (FD). The sounding radiation illuminates the air over the central portion of the TA Surface Detector array, shown with dashed blue lines that indicate the beamwidth 3~dB below the peak gain.}
\label{fig:tamap}}
\end{figure}

A new approach, first attempted by the MARIACHI~\cite{mariachi1,mariachi2} project, is to utilize {\em bi-static} or two-station radar in conjunction with a conventional set of cosmic ray detectors. Air shower particles move very close to the speed of light, so the Doppler shift is large compared with airplanes or meteors. The bi-static configuration in which the sounding (interrogating) wave Poynting vector is generally perpendicular to shower velocity (as shown in Figure~\ref{fig:geometry}) minimizes the large Doppler shift in frequency expected of the reflected signal (see~\cite{underwood,bakunov}, and Section~\ref{sec:eas_echoes} below.) This scenario is unlike that explored in~\cite{bakunov} in which the two vectors are roughly anti-parallel. In the latter case, the relativistic frequency shift is maximized. Also, depending on the size of the radar cross section relative to the square of the sounding wavelength, scattering in the forward direction might be enhanced relative to back scatter~\cite{willis}, thus providing an advantage in detecting the faintest echoes in comparison to mono-static radar (ranging radar). 

Co-location with a conventional detector allows for definitive coincidence studies to be performed. If coincidences are detected, the conventional detector's information on the shower geometry will allow direct comparison of echo signals with the predictions of air shower Radio Frequency (RF) scattering models. 

The Telescope Array Radar (TARA) project is the next logical step in the development of the bi-static radar technique. Whereas MARIACHI made parasitic use of commercial television carriers as a source of sounding radiation (now impossible due to the transition to digital broadcasts), TARA employs a single transmitter in a vacant VHF band which is under the experimentalists' control. The TARA receiver consists of broadband log-periodic antennas, which are read out using a 250~MS/s digitizer. TARA is co-located with the Telescope Array, a state-of-the-art ``conventional'' cosmic ray detector, which happens to be located in a low-noise environment. The layout of the TA and TARA detection facilities are shown in Figure~\ref{fig:tamap}. 

This work begins with a brief description of the nature of air shower echoes expected for the TARA configuration. Next, we describe the transmitter and receiver system in some detail, including tests of system performance. Finally we describe upgrades to the system which are currently in progress. 
 % John Belz
% Make footnote with all non-standard abberviations in the paper
%%

%%
\section{Extensive Air Showers, Radar Echoes}
\label{sec:eas_echoes}
\begin{figure}[h]
	\begin{center}
		\includegraphics[trim=0.2cm 0.2cm 0cm 0.4cm,clip=true,scale=0.41]{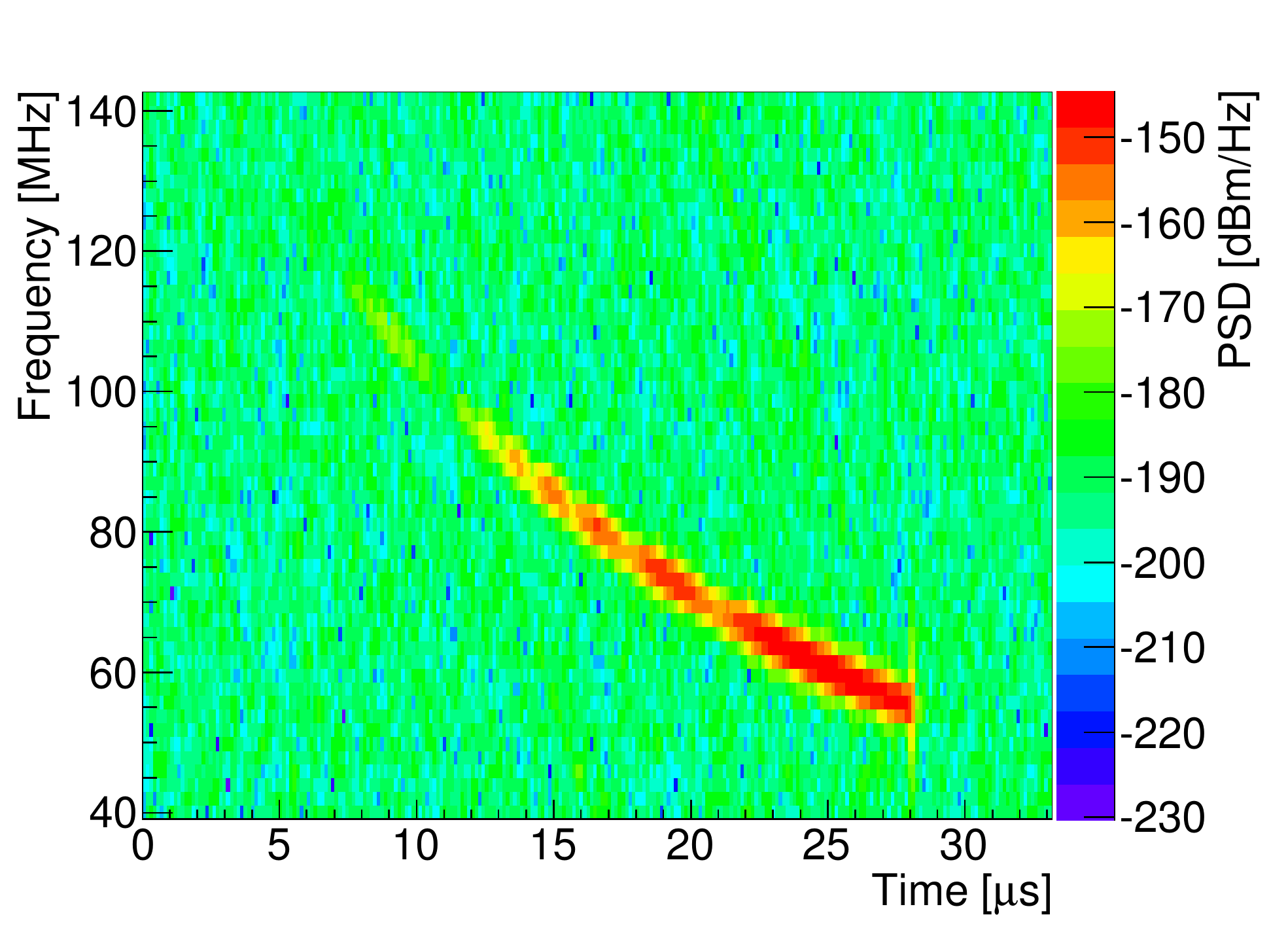}
	\end{center}
\caption{Spectrogram of a chirp signal produced by the radar echo simulation for an EAS located midway between the transmitter and receiver with a zenith angle of $30^{\circ}$ out of the transmitter-receiver plane. A weighted fit to the power of this signal gives a -2.3~MHz/$\mu$s chirp rate. Color scale is Power Spectral Density (PSD) given as dBm/Hz.}	
\label{fig:chirp}
\end{figure}

As the EAS core ionizes the atmosphere, liberated charges form a plasma with plasma frequency $\nu_p = (2\pi)^{-1} \sqrt{n_e e^2/m_e \epsilon_0}$, where $n_e$ is the electron number density, $e$ is the charge of the electron, and $m_e$ is the electron mass.  A shower is denoted $\textit{under-dense}$ or $\textit{over-dense}$ (See Figure~3 in ~\cite{Gor}) relative to the sounding frequency $\nu$ depending on whether $n_e$ corresponds to $\nu_p>\nu$ or $\nu_p<\nu$. The radar cross-section of the underdense region is expected to be greatly attenuated due to collisional damping~\cite{vidmar,itikawa71,itikawa73}. Therefore, we expect the dominant contribution to EAS radar cross-section $\sigma_{EAS}$ to be the over-dense region, which is modeled as a thin-wire conductor~\cite{crispin}. Figure~\ref{fig:chirp} displays a ``typical'' EAS echo from simulation, where standard shower models of particle production and energy transport have been assumed~\cite{GaisHill}.

\begin{figure}[h]
	\begin{center}
		\includegraphics[trim=0.0cm 1.0cm 0cm 0.0cm,clip=true,width=0.48\textwidth]{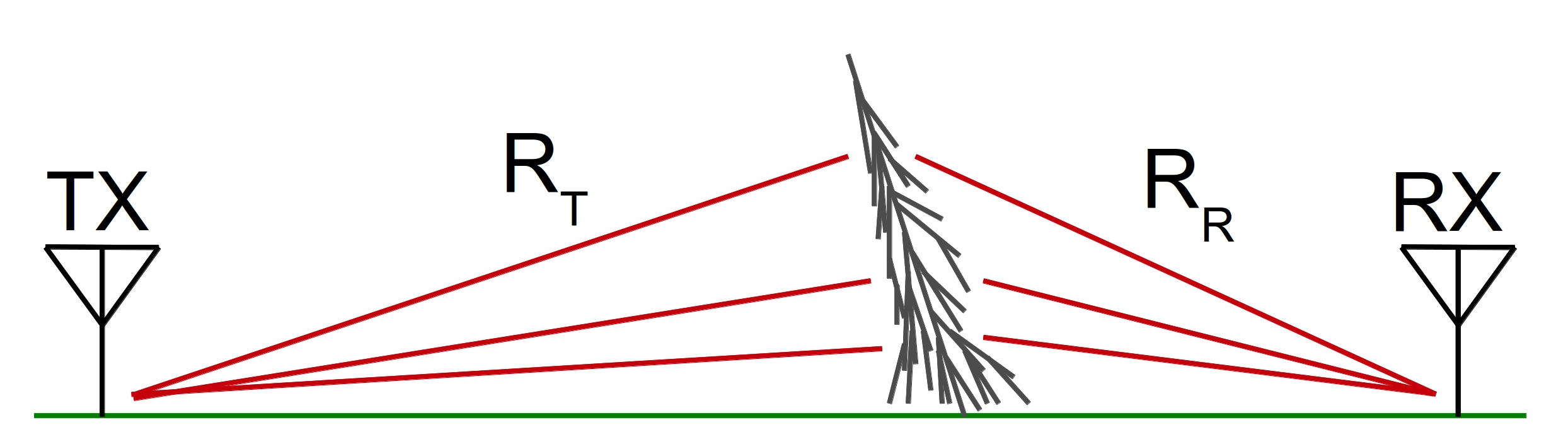}
	\end{center}
\caption{Bi-static geometry of a radar sounding wave interrogating an EAS to scale in the TARA geometry. $R_T$ and $R_R$ are the distances from transmitter (TX) to shower and shower to receiver (RX), respectively. The TX/RX antenna symbols represent location only. Actual antenna sizes are smaller than a pixel if represented to scale.}	
\label{fig:geometry}
\end{figure}

The mechanism of radar echo detection of EAS differs from other radio applications because the target is small (i.e., small RCS) and moving near the speed of light.  However, letting $R_T$ and $R_R$ represent the transmitter/shower and receiver/shower distances, respectively, the bi-static geometry (Figure~\ref{fig:geometry}) minimizes the phase shift because the total path length $L = R_R + R_T$ evolves slowly with time.  The time-dependence of the path length causes the phase of the echo to evolve, while the transmission maintains a constant frequency.  The result is an echo that has a time-dependent frequency -- a $\emph{chirp}$ signal~\cite{underwood} (Figure~\ref{fig:chirp}).

%Radar echo detection of cosmic rays is starkly contrasted with traditional radar applications. The moving body of shower particles traveling through the atmosphere presents as a very small radar target moving at the speed of light. At time $t_{0}$, the distance between the shower front and transmitter $R_{T,0}$ and the distance between the shower core and receiver $R_{R,0}$, which sum to the total path length $L_{0}$ is different than the path length $L_{1}$ at time $t_{1}$. Differing path lengths result in received signals with the same frequency, but different phase. The succession of scattered waves from an evolving path length $L$ and changing phase results in a signal whose frequency varies with time -- \emph{chirp} signals~\cite{underwood}. 

\begin{figure}[!h]
	\begin{center}
		\includegraphics[trim=0cm 0.5cm 1.5cm 0.9cm,clip=true,width=0.48\textwidth]{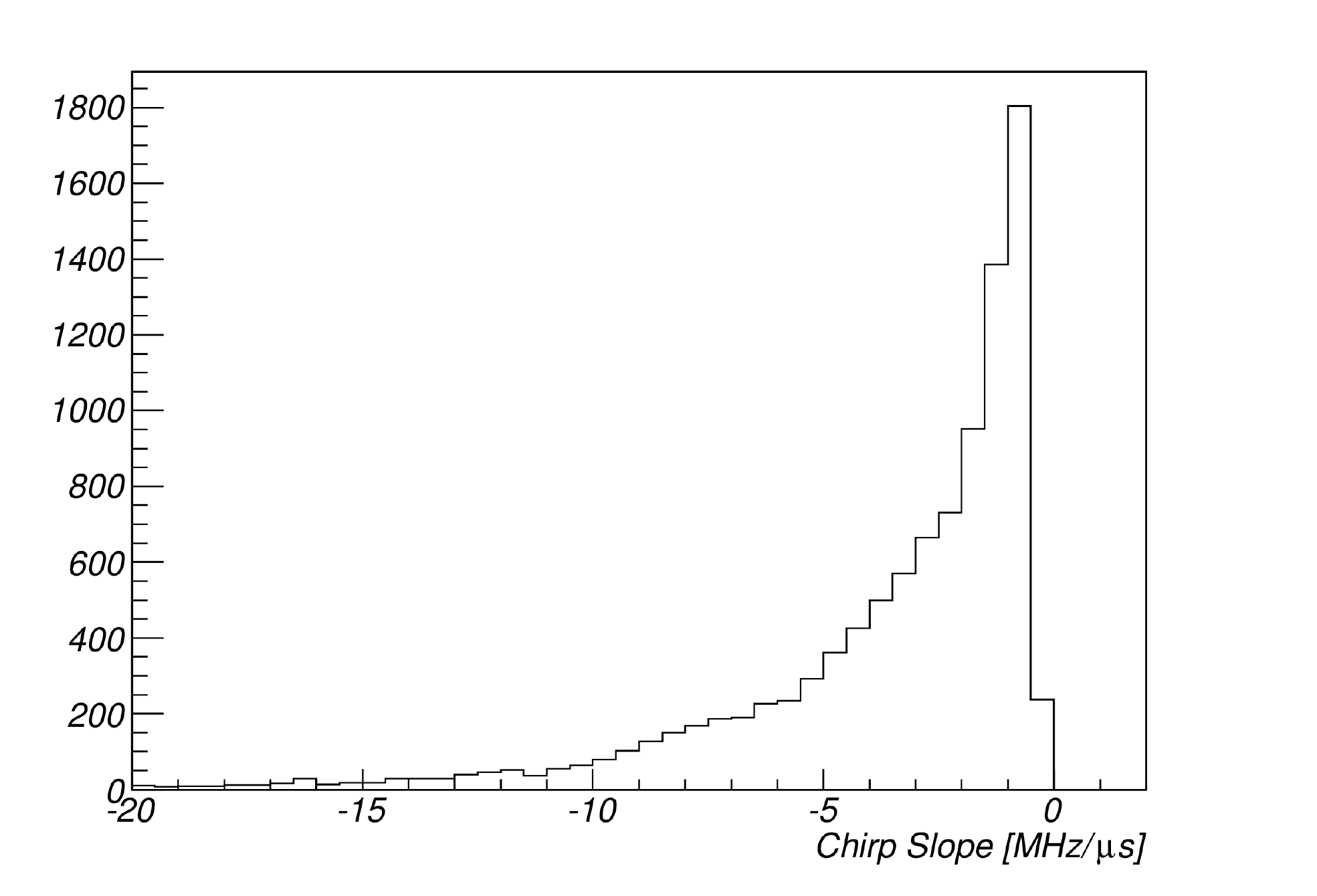}
	\end{center}
\caption{Simulated chirp rate distribution from a set of 10000 TA cosmic ray events. The rate is calculated from a weighted fit (by power) to the spectrogram of the simulated signal.}	
\label{fig:chirpslope}
\end{figure}

\begin{figure}[!h]
	\begin{center}
		\includegraphics[trim=0cm 0.5cm 1.5cm 0.9cm,clip=true,width=0.48\textwidth]{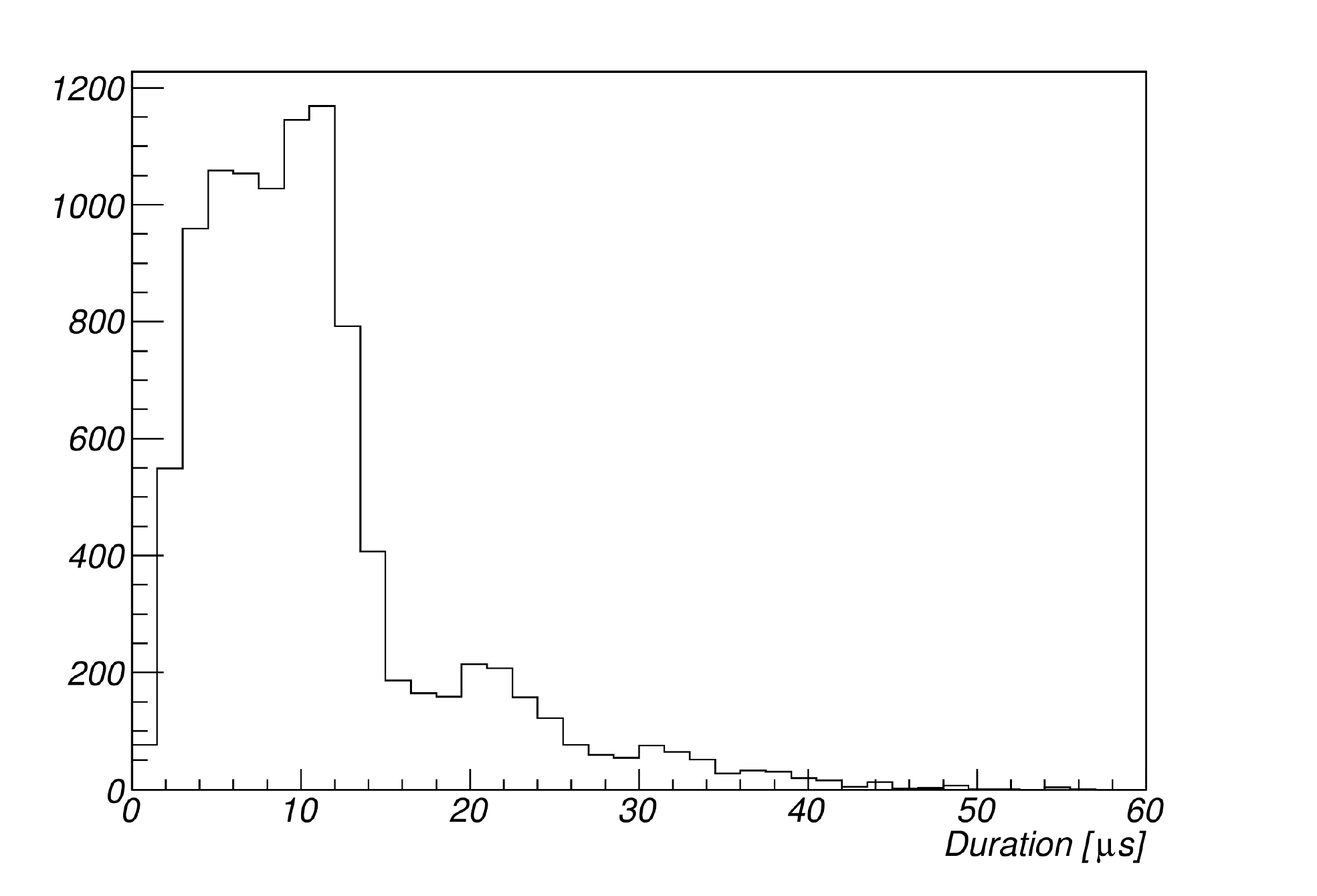}
	\end{center}
\caption{Chirp duration distribution from 10000 simulated radar echoes from TA cosmic ray events. Duration is defined as $d = t_1 - t_0$, where $t_0$ is the time where the maximum power is received and $t_1$ is the \emph{later} time when the received power drops by 20~dB below the maximum, which approximates the end of the shower.}	
\label{fig:chirpduration}
\end{figure}

Chirp signals are ubiquitous in nature, although CR radar echos have very unique signatures. A simulation~\cite{takai2011} has been designed that requires as inputs the CR energy, geometry and transmitter and receiver details, and which evolves an EAS according to standard particle production and energy transport models~\cite{GaisHill} while tracking the phase and amplitude of the radar echo. Shower parameters are functions of the primary particle energy~\cite{zech}. The simulation indicates (see, for example, a ``typical'' TARA geometry simulation spectrogram in Figure~\ref{fig:chirp}) that CR radar echoes are short in duration (comparable to the shower life-time, $\approx 10$~$\mu$s), have  chirp rates of a few times $1$~MHz/$\mu$s and span a bandwidth on order of the sounding frequency (see Figures~\ref{fig:chirpslope} and \ref{fig:chirpduration}). 

The energy and geometry of a distribution of 10000 cosmic rays detected at the TA surface detector array have been simulated. Figures~\ref{fig:chirpslope} and \ref{fig:chirpduration} show distributions of the chirp rate and duration for these events. Data obtained from the simulation have been used to guide the design of the DAQ, transmitter system, and the receiver antennas.  A 54.1~MHz radar sounding frequency (the TARA licensed frequency) implies the need to resolve a bandwidth of roughly 100~MHz and therefore implement a DAQ with at least 200~MS/s ADC. An FPGA based design is necessary to implement real-time digital filters that will trigger the DAQ on signals that resemble chirp radar echos.

Air showers are uniquely defined by their radar echo signatures with the exception of a lateral symmetry with respect to a plane connecting the transmitter and receiver and also a rotational symmetry about a line connecting the transmitter and receiver. Stereo detection is necessary (at minimum) to break this symmetry. Section~\ref{sec:remoterx} discusses the remote station prototype that will supplement our primary receiver for stereo detection.

\emph{The actual radar cross section $\sigma_{EAS}$ is currently unknown}. The bi-static radar equation gives the received power $P_R$ as a function of transmitter power $P_T$.  Given the transmitter wavelength $\lambda$ and receiver and transmitter antenna gains $G_R$ and $G_T$, the bi-static radar equation is written as 

\begin{equation} \label{eq:bistatic_radar_equation_1}
	\frac{P_R}{P_T} = \left( \frac{G_T}{4\pi R_T^2} \right) \sigma_{EAS} \left( \frac{G_R}{4\pi R_R^2} \right) \frac{\lambda^2}{4\pi}\,.
\end{equation}

Detection possibility depends on the signal-to-noise ratio (SNR), defined as 
\begin{equation}
\label{eq:snr}
\text{SNR}=\frac{P_{\text{c}}}{\sigma_\nu^2}\,,
\end{equation}
where $P_{\text{c}}$ is the chirp signal power and $\sigma_{\nu}$ is the standard deviation of the background noise. A second definition is necessary for signals with time-varying amplitude like those predicted by the EAS radar echo simulation. For such signals we use the amplitude signal-to-noise ratio (ASNR)
\begin{equation}
\label{eq:asnr}
\text{ASNR} = \frac{\nu_{\text{max, c}}^2}{\sigma_{\nu}^2}\,.
\end{equation}
$\nu_{\text{max, c}}$ is the maximum chirp amplitude. The TARA DAQ can trigger on realistic chirp signals as low as 7~dB below the noise (-7~dB~ASNR, see Section~\ref{subsub:simshower}). A simple calculation will show that, if the thin wire approximation $\sigma_{tw}$ is assumed to correctly model the actual radar cross section (RCS) $\sigma_{EAS}$, TARA expects radar echoes with positive SNR (in dB).

TARA detector parameters are given in Table~\ref{tab1}. Consider a 60~MHz Doppler shifted tone, scattered from an EAS located midway between the transmitter and receiver, which have a 39.5~km separation distance. Received power can be calculated from Equation~\ref{eq:bistatic_radar_equation_1} if $\sigma_{EAS} \simeq \sigma_{tw}$ is known. Some basic assumptions allow a quick calculation of $\sigma_{tw}$: Shower $X_{\text{max}}$ occurs roughly 2~km from the ground; the antennas' polarization vector and shower axis are in the same plane; the length $L$ of the scattering region of the shower is the speed of light multiplied by the electron attachment/recombination lifetime $\tau = 10$~ns~\cite{vidmar}, which implies $L = 3$~m; the over-dense region radius near $X_{\text{max}}$ is the thin wire radius~\cite{Gor} $a = 0.01$~m. With these assumptions the thin-wire cross section~\cite{crispin} is $\sigma_{tw} \sim 1$~m$^2$ and the received power is -79~dBm. 
\begin{table}[ht]
\begin{center}
\begin{tabular}{| c | c |}
	\hline
	Parameter & Value \\ \hline \hline
	UHECR energy & 10$^{19}$~eV \\ \hline
	$P_{T}$ & 40~kW \\ \hline
	$G_{T}$ & 22.6~dBi (Section~\ref{sub:txantenna_measured_performance}) \\ \hline
	$G_{R}$ & 12.6~dBi (Section~\ref{sec:rxantenna}) \\ \hline
	$R_{T} = R_{R}$ & 19.75~km \\
	\hline
\end{tabular}
\caption{The list of the parameters assumed for calculating received power from a 60~MHz Doppler shifted radar echo scattered from an EAS.}
\label{tab1}
\end{center}
\end{table}

Section~\ref{sec:rxantenna}, Figure~\ref{fig:LPDA_noise_floor} shows a plot of receiver system background noise superimposed with galactic noise. Receiver sample rate and Fourier transform window size used in the calculation were 250~MS/s and 32768, respectively. The power spectral density (PSD) of a -79~dBm tone detected by this system is $-79 + 10\,\text{log}_{10}\,(32768/250\cdot10^{6}) = -117$~dBm/Hz. The reader should note that antenna patterns are both assumed to be at their maximum, which rarely occurs in practice. Further, polarization angle differences ($\phi$) between the shower axis and antennas can yield another reduction in power $\propto \text{cos}^4\,(\phi)$. 

The receiver background noise plot demonstrates that, in the TARA frequency band of interest, backgrounds are dominated by galactic noise. At 60~MHz, the background noise PSD is -160~dBm/Hz, much lower than that of a narrow-band Doppler shifted radar echo at -117~dBm/Hz scattered from an ideal thin wire. Under reasonable assumptions for signal parameters, combined with our measured irreducible backgrounds and system response, the thin wire approximation for the radar cross-section $\sigma_{EAS}$ implies high values of signal-to-noise.

\section{Transmitter}
\label{sec:transmitter}
\subsection{Hardware}
\label{sub:hardware}
TARA operates a high power, Continuous Wave (CW), low frequency radar transmitter built from re-purposed analog TV transmitter equipment with FCC call sign WF2XZZ, an experimental license. The transmitter site ($39^{\circ}~20'~19.82400''$~N, $112^{\circ}~42'~3.24000''$~W) is just outside Hinckley, UT city limits where human exposure to RF fields is of little concern. A high gain Yagi array (Section~\ref{sec:txantenna}) focuses the radar wave toward the receiver station (Section~\ref{sec:rxantenna}) located 40~km away. Figure~\ref{fig:tamap} shows the transmitter location near Hinckley and relative to the TA SD array~\cite{tasd}. The geometry was chosen to maximize the possibility of coincident SD and radar echo events. 

\begin{figure}[!h]
\centerline{\mbox{\includegraphics[trim=0mm 0mm 0mm 0mm, clip, width=0.49\textwidth]{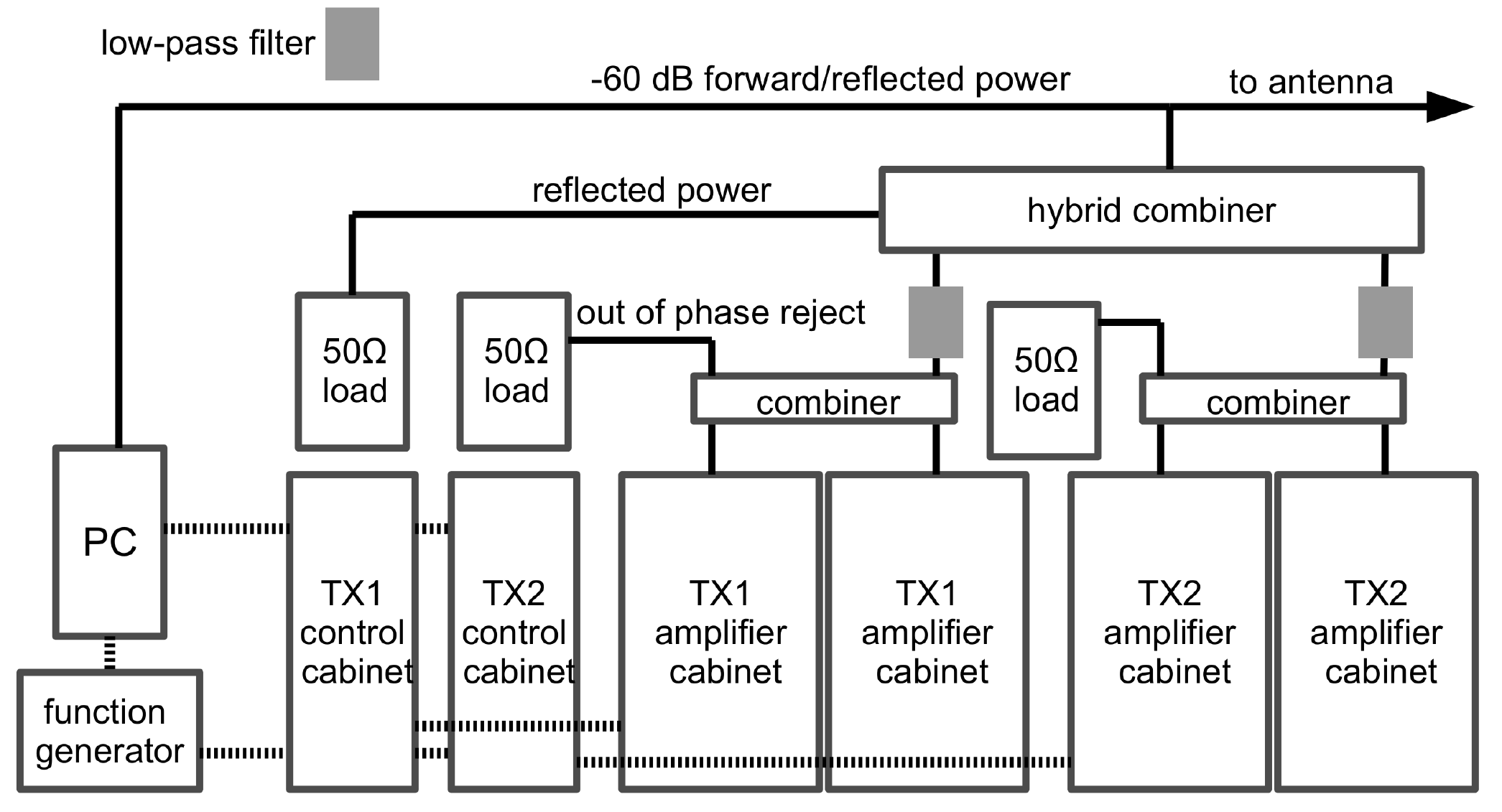}}}
\caption{Schematic of the transmitter hardware configuration. A computer connected to RF sensor equipment, an arbitrary function generator and transmitter control electronics orchestrates the two distinct transmitters and provides remote control and logging. RF power from each transmitter's two amplifier cabinets is combined with out of phase power rejected into a 50~$\Omega$ load. A hybrid combiner sums the combined output of each transmitter and sends that power to the antenna. Power reflected back into the hybrid combiner is directed to a third RF load.
\label{fig:schematic}}
\end{figure}

Figure~\ref{fig:schematic} shows a schematic of the transmitter hardware configuration. A Tektronix arbitrary function generator (AFG 3101; Tektronix, Inc.) provides the primary sine wave, which is amplified over nine orders of magnitude before reaching the antenna. 54.1~MHz was chosen as the sounding frequency because of the lack of interference in the vacated analog channel two TV band and the 100~kHz buffer between it and the amateur radio band which ends at 54.0~MHz. 

Two 20~kW analog channel 2 TV transmitters have a combined 40~kW power output. The primary signal from the function generator is split to feed both transmitters (Harris Platinum HT20LS, p/n 994-9236-001; Harris Broadcast) with the same level of gain. Each transmitter includes a control cabinet and two cabinets of power amplifier modules. RF power from each cabinet is combined in a passive RF combiner (620-2620-002; Myat, Inc.) that routes any out-of-phase signal to a 50~$\Omega$ load. The combined output of each transmitter is sent to a $90^{\circ}$ hybrid combiner (RCHC-332-6LVF; Jampro, Inc.) that sums the total output of each transmitter.  Between the final combined input and each transmitters' combined output there is an inline analog channel 2 low pass filter (visual low-pass filter, 3 1/8''; Myat, Inc.) to minimize harmonics. RF power leaves the building through 53~m of semi-flexible 3 1/8" circular air-dielectric wave guide (HJ8-50B; Andrew, Inc.).

%\begin{figure}[h]
%\centerline{\mbox{\includegraphics[width=0.45\textwidth]{transmitter.jpg}}}
%\caption{Two 20~kW analog channel two TV transmitters combined for the Keck Radar Observatory radar wave. Modifications %have been made to remove frequency modulation and allow amplification of a pure tone 54.1~MHz output. 
%\label{fig:transmitter}}
%\end{figure}

Modifications were made to the transmitters to bypass interlocks that detect the presence of aural and visual inputs and video sync pulses necessary for standard TV transmission. Control cabinet electronics were calibrated to measure the correct forward and reflected power of the 54.1 MHz tone instead of the RF envelope during the sync pulse. Currently, total power output is limited to 25~kW because of limitations that arise from amplifying a single tone versus the full 6~MHz TV band. 

Air conditioning and ventilation are critical to high power transmitter performance. Currently, transmitter efficiency is slightly better than 30\%, which implies that nearly 75~kW of heat must be removed from the building. The environment at the site is very dry and dusty, so all of the air brought into the building is filtered and positive gauge pressure is maintained. A single 25~ton AC unit filters and pumps cool air into the building. An economizer will shut down the compressor if the outside air temperature drops below $15.6^{\circ}$~C ($60^{\circ}$~F). However, if the room is not cooling quickly with low outside ambient temperature, the compressor will be turned back on. Hot air near the ceiling is vented as necessary to maintain a slight positive pressure. 

Future improvements to the transmitter will include biasing the power amplifiers for class B operation, in which amplification is applied to only half the 54.1~MHz cycle. Resonance in the transmitter and antenna allow the second half of the wave to complete the cycle. Efficiency will nearly double compared with the current configuration.  

\subsection{Remote Monitoring and Control}
\label{sub:monitoring}
Remote monitoring and control of the transmitter is important for two reasons. First, Federal Communications Commission (FCC) regulations require that non-staffed transmitter facilities be remotely controlled and several key parameters monitored. Second, forward power and other parameters must be logged for receiver data analysis. 

A computer interfaces with digital I/O and analog input devices that, in turn, are connected to the transmitters' built in digital I/O and analog output interface. RF power sensors (PWR-4GHS; Mini-Circuits) measure the final forward and reflected power via strongly attenuating sample ports on the wave guide near the building exit port. The sum of the two control cabinets' forward and reflected power measurements can be compared with the separate RF final forward and reflected power measurements. 

The host computer monitors transmitter digital status, analog outputs and RF power sensors and controls the function generator. Logs are updated every five minutes with forward and reflected power for each transmitter, \emph{final} (re: antenna) forward and reflected power, room temperature and various transmitter status and error states. Warning and error thresholds can trigger emails to the operators and initiate automatic shutdown. The program also provides a simple interface that allows the operator to remotely turn the transmitter on and off, increase or decrease forward power, and add a text log entry. 

\subsection{Performance}
\label{sub:performance}
\begin{figure}[!h]
\centerline{\mbox{\includegraphics[trim = 0mm 0.0mm 0.0mm 0.0mm, clip, width=0.5\textwidth]{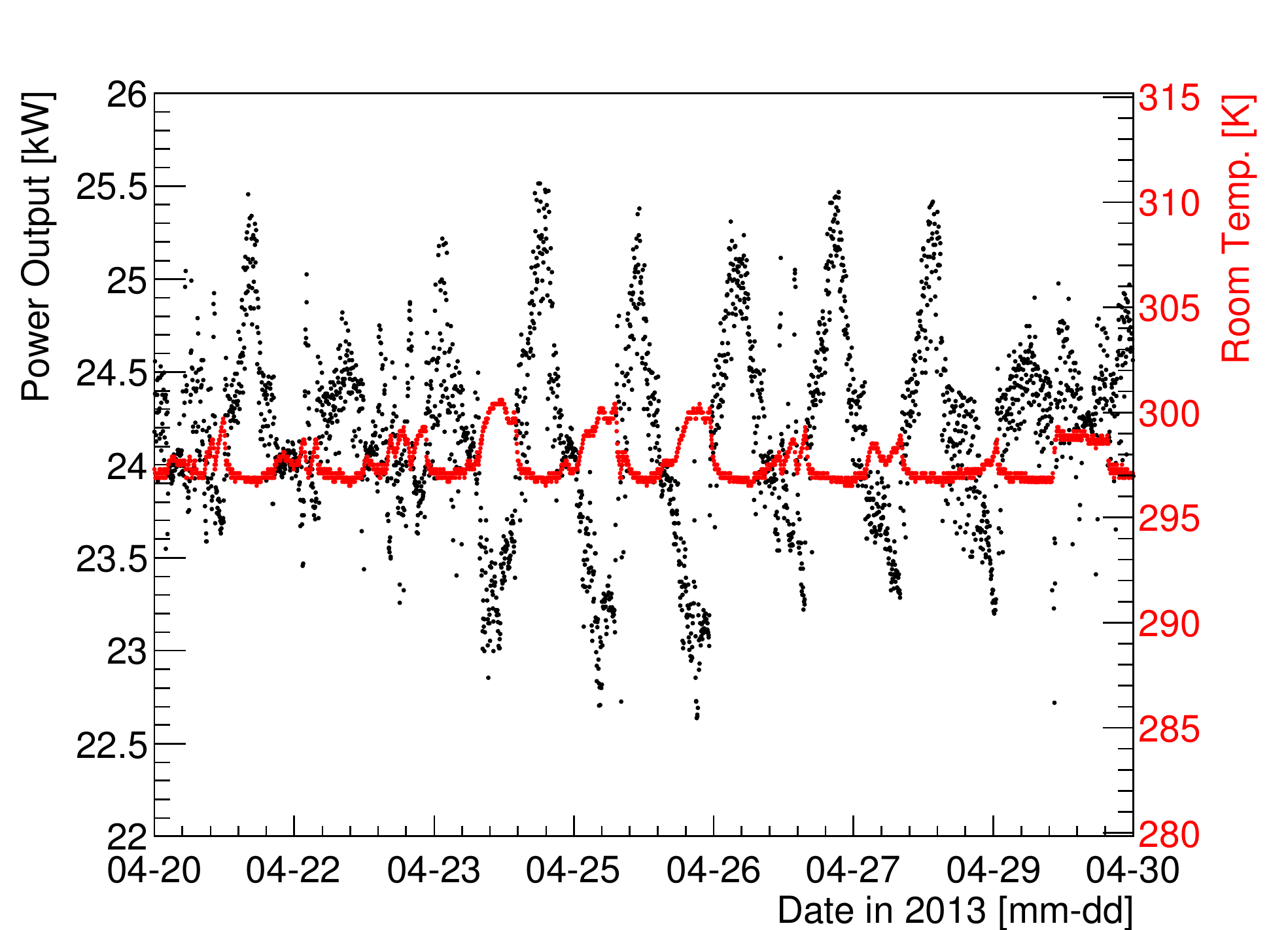}}}
\caption{Transmitter forward power (black) and room temperature (red) during April 2013. Poor air conditioning calibration resulted in daily temperature fluctuations which caused large output power modulation.   
\label{fig:tx_temp}}
\end{figure}

TV transmitters are designed for 100\% duty cycle operation. Similarly, the TARA transmitter is intended for continuous operation to maximize the probability of detection of UHECRs. With fixed gain and input signal, power is strongly correlated with transmitter room ambient temperature. Large temperature fluctuations in April 2013 resulted in a $\sim3$~kW spread in output power (Figure~\ref{fig:tx_temp}). 

\begin{figure}[!h]
\centerline{\mbox{\includegraphics[trim = 0mm 0.0mm 0.0mm 0.0mm, clip, width=0.5\textwidth]{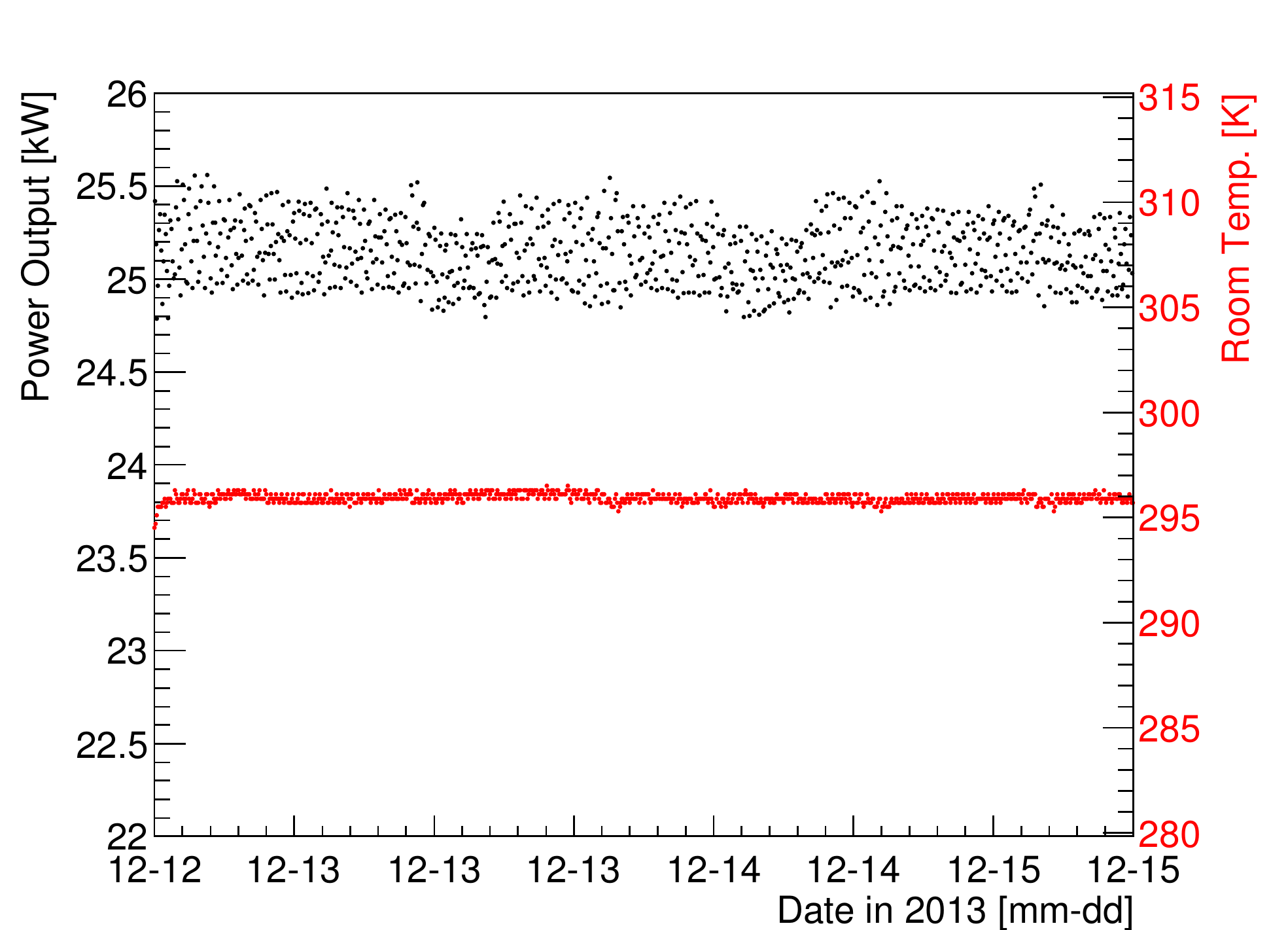}}}
\caption{Transmitter forward power (black) and room temperature (red) during December 2013. A well-calibrated air conditioning system keeps room temperature stable and increased automatic gain control minimizes forward power fluctuations. 
\label{fig:tx_temp_stable}}
\end{figure}

Transmitter forward power is more stable if room temperature is kept lower than 300~K ($80^{\circ}$~F).  Figure~\ref{fig:tx_temp_stable} shows forward power fluctuations in August 2013 are much smaller than April. Built-in automatic gain control was increased during this period as well. The average power in December is higher than the average power in April because a slightly higher power input signal was used in later months. Reflected power is typically $\sim100$~W, which is very low for such a high power system. This can be attributed to very good impedance matching with the extremely narrow-band Yagi antenna array. 

\begin{figure}[!h]
\centerline{\mbox{\includegraphics[trim = 0mm 0mm 0mm 0mm, clip, width=0.5\textwidth]{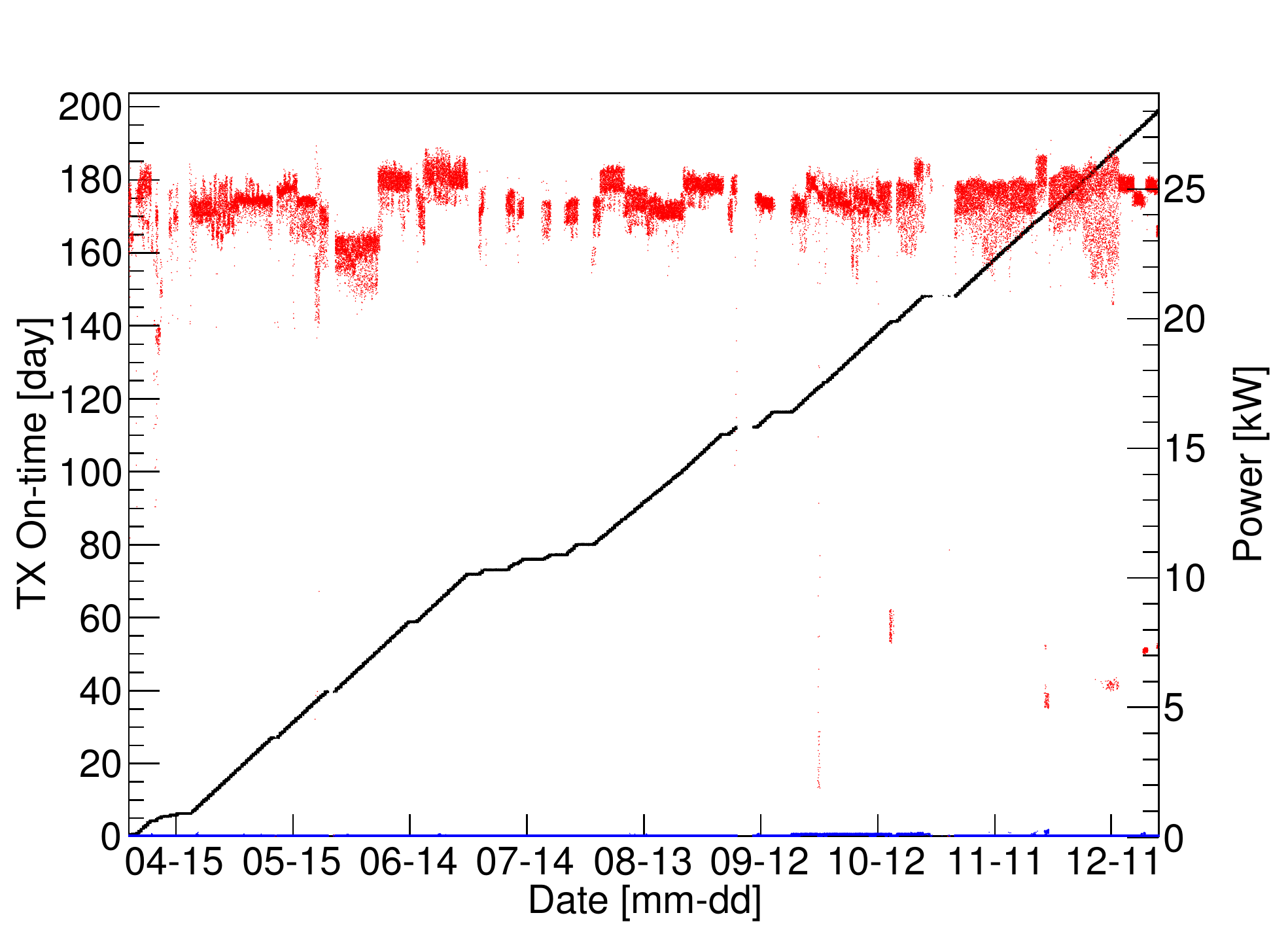}}}
\caption{Transmitter on-time in days (black, left vertical axis) and forward and reflected power in units of kW (red and blue, right vertical axis) during 2013. Total duty cycle during this period is 83\%.
\label{fig:on_time}}
\end{figure}

Figure~\ref{fig:on_time} shows the total forward and reflected power in red and blue, respectively, referenced to the right vertical axis and the integrated on-time in black, referenced to the left vertical axis, since its commissioning in late March, 2013. The transmitter has been turned off several times for maintenance and testing and during periods when our receiver equipment was removed from the field for upgrades. Although forward power is not continuous and fluctuations were large in the past, we consider 200 days of operation in the first year to bode well for future data collection.

Harmonics have been measured to confirm compliance with FCC regulations and to avoid interfering with other stations. With total forward output at 25~kW, the fundamental and several harmonic frequencies were measured from a low power RF sample port. The first five harmonics are about 60~dB below the fundamental (see Table~\ref{tab:harmonics}).  Harmonics will be further attenuated by about 30~dB by the intrinsic bandpass of the antenna. 

\begin{table}[!h]
\begin{center}
\begin{tabular}{|l|c|}
\hline
Frequency (MHz) & Power (dBm) \\ \hline \hline
54.1 & 8.5 \\ \hline
108.2 & -66.0 \\ \hline
162.3 & -68.3 \\ \hline
216.4 & -84.4 \\ \hline
270.5 & -89* \\ \hline
324.6 & -77* \\ \hline
378.7 & -94* \\ \hline
432.8 & -87* \\ \hline
486.9 & -98* \\ \hline
541.0 & -91* \\ \hline 
\end{tabular}
\end{center}
\caption{Power of fundamental frequency and first ten harmonics for the 54.1~MHz radar sounding wave. These measurements were taken from a highly attenuated final forward power RF sample port. Total transmitted power was approximately 25~kW. FM and TV stations are required by the FCC to limit the first ten harmonics to at least 60~dB below their approved total transmitted power. Experimental station WF2XZZ is exempt from this requirement although it readily meets it. (*fluctuating value, $\pm 5$~dB) \label{tab:harmonics}}
\end{table}
 % Isaac
%%

%%
\section{Transmitting Antenna}\label{sec:txantenna}
\subsection{Physical Design}\label{sub:txantenna_physical_design}

As the bi-static radar equation
(Equation~\ref{eq:bistatic_radar_equation_1}) shows, the received
power is the product of the scattering cross section, transmitted
power, transmitter antenna gain, receiver antenna gain and receiver aperture. Because the
physics of the radar scattering cross section is not well understood,
an antenna with high gain and
directivity was chosen to maximize received power. 

The TARA transmitting antenna is composed of 8 narrow band Yagi antennas
designed and manufactured by M2 Antenna Systems, Inc. Each Yagi is
constructed of aluminum and capable of handling 10~kW of continuous RF
power. The specifications for each Yagi are a frequency range of 53.9 - 54.3~MHz, 12~dBi free space
gain, front to back ratio of 18~dB, and beamwidths (defined as the angle in the plane under consideration over which the radiated power is within three dB of the maximum) of $27^\circ$ and
$23^\circ$ in the vertical and horizontal planes respectively. 

Each Yagi antenna is composed of five elements: a reflector, driven element,
and three directors, and are mounted on a 21.6~ft long, $2\,''$ diameter
boom. A balanced t-match is fed from a 4:1 coaxial balun
which transforms the unbalanced 50~$\Omega$ input to the balanced 200~$\Omega$
used to drive the antenna. A 50~$\Omega$ $7/8\,''$ coaxial waveguide connects the
balun to the four port power
dividers. Table~\ref{tab:txantenna_dimensions} describes the lengths
and positions of the antenna elements on the boom. All elements are
constructed of aluminum tubing of $3/4\,''$ outer diameter. Each element,
except for the driven element is constructed of two equal sections that
are joined at the boom via $7/8\,''$ outer diameter sleeve
elements. The weight is 35~lbs when completely assembled.

\begin{table}[h]
\centering
\begin{tabular}{|l|r|r|}\hline
Element         &Length (in) &Position (in) \\ \hline\hline
Reflector       &107.625     &-44.375 \\ \hline
Driven Element  &100.500     &  0.000 \\ \hline
Director 1      & 99.500     & 51.125 \\ \hline
Director 2      & 97.250     &131.625 \\ \hline
Director 3      & 97.000     &193.625 \\ \hline
\end{tabular}
\caption{Length and relative boom position of antenna elements of the
  TARA Yagi antennas. All elements have a diameter of
  $0.75\,''$.\label{tab:txantenna_dimensions}}
\end{table}

Transmitter output power is delivered to the antenna array
via approximately 100~feet of CommScope HJ8-50B 3~$1/8\,''$ Heliax air
dielectric coaxial wave guide. The Heliax then connects to a two port power
divider located at the base of the antenna array. Each output port of
the power divider feeds equal length 1~$5/8\,''$ coaxial cables, which
in turn feed a four port power divider. Each four port power divider then
delivers power to the individual Yagi antennas via equal length $7/8\,''$ coaxial
cables. All components in the transmission line chain are impedance
matched to 50~$\Omega$.

%{\it Note: statement about attenuation losses between
%transmitter and array would be nice here. Do we have that data?
% IJM - No, but it can be computed pretty straightforward from tables that show the attenuation of Heliax per 100 ft.}
\begin{figure}[!h]
\centering
\includegraphics[width=0.46\textwidth]{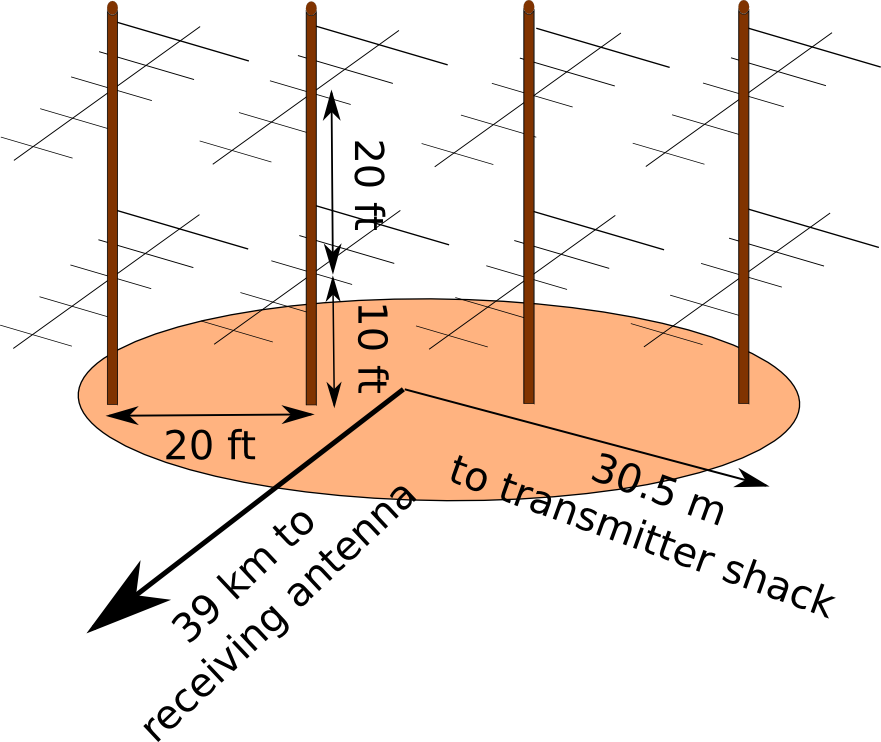}
\caption{Configuration of the eight Yagi antennas and mounting poles which
  comprise the TARA transmitting antenna array.\label{fig:txantenna_configuration}}
\end{figure}

The antennas are mounted on four wooden telephone poles, two stacked vertically on
each pole. The bottom and top antennas on each pole are located 10 ft
and 30 ft above the ground respectively. Currently, the antennas are
mounted in a configuration that provides a horizontally polarized signal. Wooden poles were used to allow a change of polarization. The poles, separated by 20~ft, are aligned in a plane perpendicular to
the line pointing toward the receiver site located at
the Long Ridge fluorescence detector 39 km to the
southwest. Figure~\ref{fig:txantenna_configuration} shows the antenna array configuration.

\subsection{Theoretical Performance}\label{sub:txantenna_theoretical_performance}
The eight Yagi antennas are operated as a phased array to take advantage
of pattern multiplication to improve gain and directivity relative to
the individual antennas. The design philosophy of the antenna array
is to deliver a large amount of power in the forward direction in a very narrow
beam to maximize the power density over the TA surface detector. High
power density is equivalent to a large $P_{T} G_{T}$ factor in the bi-static radar equation, which is needed to increase the chance of
detection of a cosmic ray air shower via radar echo given
the uncertainty in the radar scattering cross section $\sigma_{EAS}$. Before construction, modeling of the array was performed using version two of NEC~\cite{nec}, an antenna modeling and optimization
software package.

\begin{figure}[h!]
\centerline{
\includegraphics[trim=2.8cm 0.0cm 2.8cm 0.0cm,clip=true,width=0.48\textwidth]{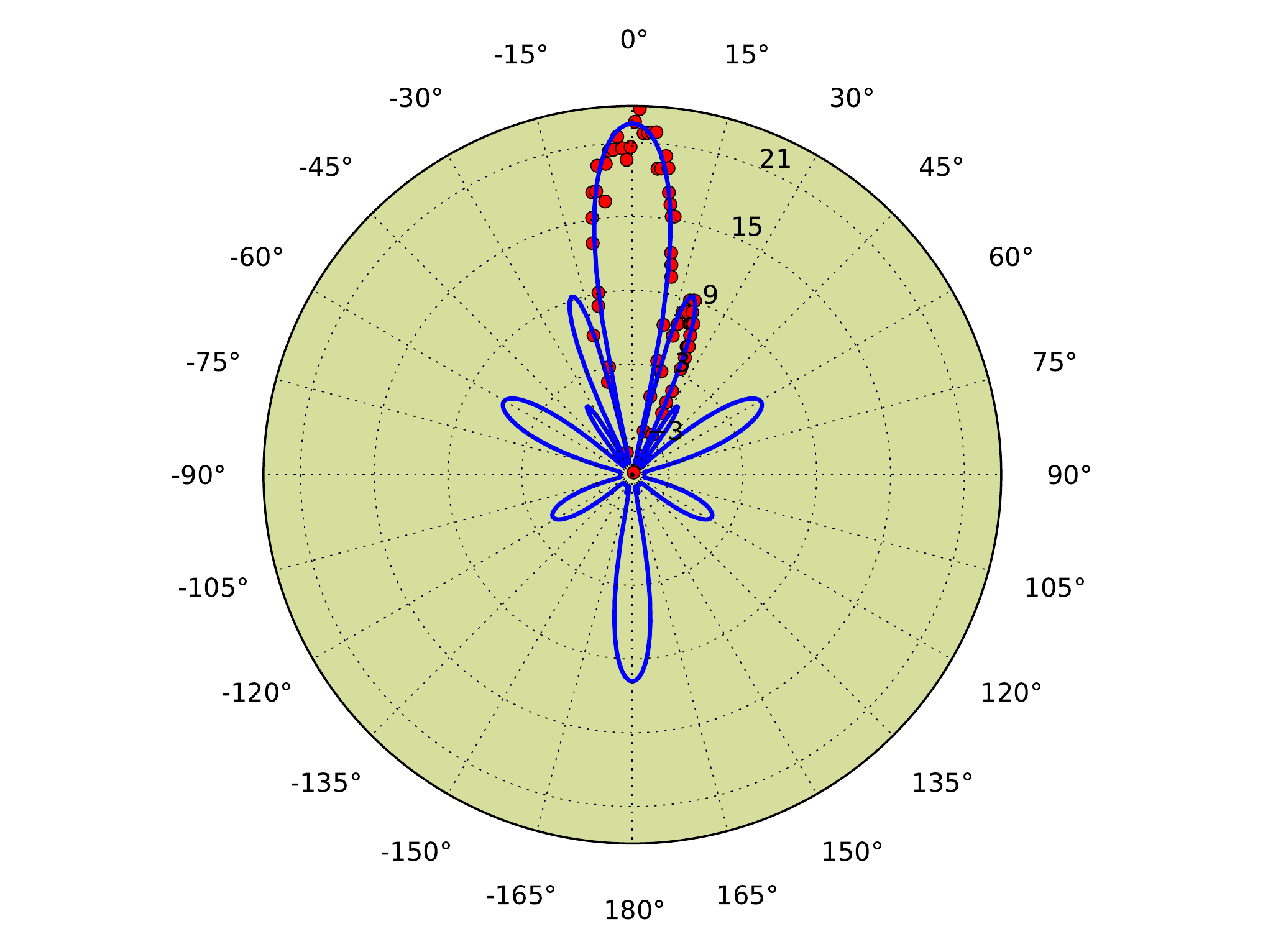}
}
\centerline{
\includegraphics[trim=2.8cm 0.0cm 2.8cm 0.0cm,clip=true,width=0.48\textwidth]{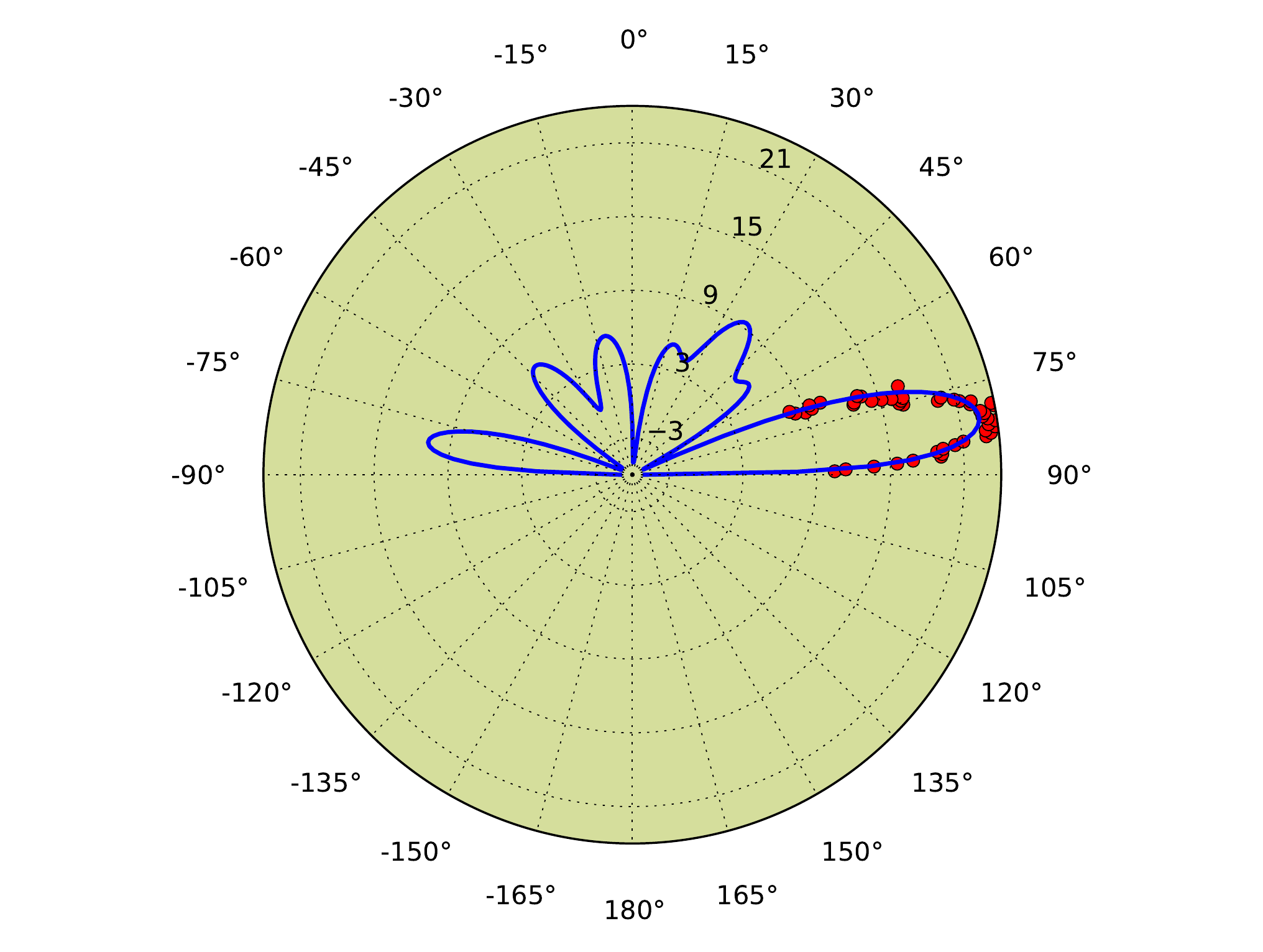}
}
\caption{
Simulated horizontal (top) and vertical (bottom) radiation patterns of the eight Yagi TARA antenna array shown in blue. Red points are measured data that have been uniformly scaled to best fit the model. Forward gain is 22.6 dBi, beam width is 12$^{\circ}$ horizontal, 10$^{\circ}$ vertical, and the F/B ratio is 11.8 dB.
}
\label{fig:txantenna_radiation_patterns}
\end{figure}

Figure~\ref{fig:txantenna_radiation_patterns} shows the radiation
pattern of the full eight Yagi array when configured as shown in
Figure~\ref{fig:txantenna_configuration}. Forward gain is 22.6~dBi, horizontal beam width is 12$^{\circ}$, vertical beam width is
10$^{\circ}$, the front-to-back (F/B) ratio is 11.8~dB and the
elevation angle of the main lobe is 9$^{\circ}$. 

%\begin{figure}[h!]
%\centering
%\begin{subfigure}{0.5\textwidth}
%\includegraphics[trim=2.5cm 0.0cm 2.5cm 0.0cm,clip=true,width=0.95\linewidth]{TXant_pattern_wfit_horizontal.eps}
%\end{subfigure}

%\begin{subfigure}{0.5\textwidth}
%\includegraphics[trim=2.5cm 0.0cm 2.5cm 0.0cm,clip=true,width=0.95\linewidth]{TXant_pattern_wfit_vertical.eps}
%\end{subfigure}
%\caption{Theoretical horizontal (top) and vertical (bottom) radiation patterns of the eight Yagi TARA antenna
%  array shown in blue. Red points are measured data that have been uniformly scaled to best fit the model. Forward gain is 22.6 dBi, beam width is 12$^{\circ}$
%  horizontal, 10$^{\circ}$ vertical, and the F/B ratio is 11.8 dB.
%\label{fig:txantenna_radiation_patterns}}
%\end{figure}

Simulations were performed to find the best spacing between the
mounting poles, vertical separation of antennas and height above ground to shape and direct the main lobe in a preferred direction. Antenna pole spacing influences the main lobe beam width. A narrower beam width can be obtained at the expense of transferring power to the side lobes
which do not direct RF energy over the TA surface detector. Elevation angle is manipulated by antenna height above ground. Changing this parameter does little else to the main lobe. Elevation angle and beam width were selected to increase the
probability that air shower $X_{\mathrm{max}}$ would fall in the path of the main
lobe where the charged particle
density is the greatest. The $9^{\circ}$ main lobe elevation angle is chosen such that the sounding wave illuminates the mean $X_{\mathrm{max}}$ midway between transmitter and receiver for a distribution of showers (varying zenith angle) of order $10^{19}$~EeV~\cite{abbasi}.

\subsection{Measured Performance}\label{sub:txantenna_measured_performance}
\begin{figure}
\centering
\includegraphics[trim=1.0cm 0.2cm 1.0cm 0.5cm,clip=true,width=0.49\textwidth]{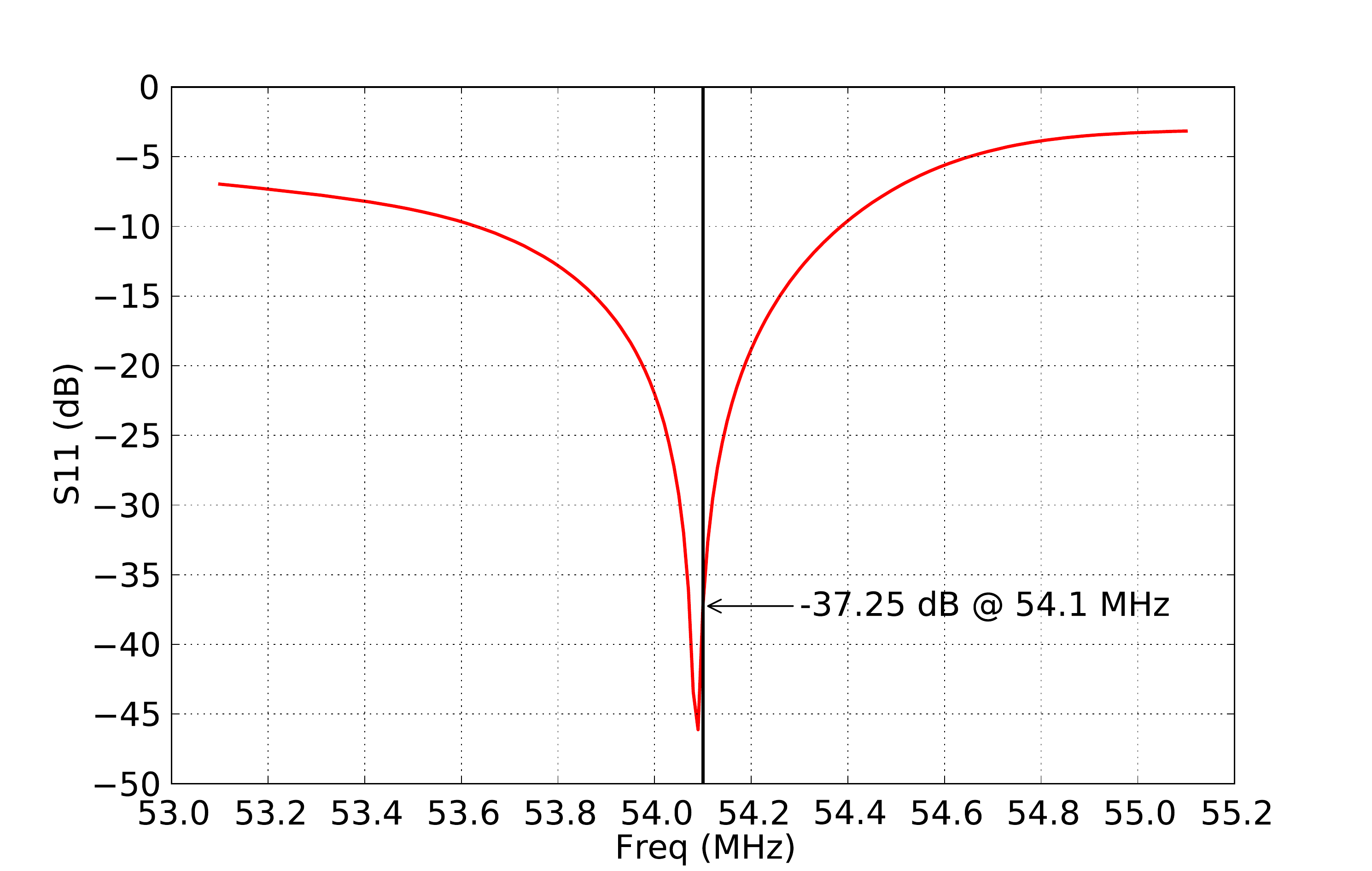}
\caption{Reflection coefficient ($S_{11}$) for the eight Yagi array.\label{fig:txant_s11}}
\end{figure}

The ability of an antenna to transmit energy is best characterized by the reflection coefficient $S_{11}$ (also called return loss when expressed in dB). It is a measure of the ratio of the voltage reflected from a transmission line relative to input. Large reflection coefficient implies significant energy is reflected back into the transmitter building which can interfere with other electronics, elevate ambient temperature and even damage the transmitter. Figure~\ref{fig:txant_s11} shows the reflection coefficient for the Yagi array. It shows a return loss of -37.25~dB at the sounding frequency, which is excellent. $S_{11}$ of -20~dB or less is considered good. 

To verify that the transmitting antenna is operating as designed, an RF
power meter or similar device can be used to measure the power as a
function of position relative to the antenna. This measurement is challenging because it must be performed in the far field of the antenna
(typically $r \gg\lambda$). To fully probe the radiation pattern of
the TARA transmitting antenna, power measurements must be made high above the
ground since the main lobe is inclined $9^\circ$ relative to horizontal. 

Vertical radiation pattern measurements were taken by using antenna transmitting/receiving symmetry. A tethered weather balloon was floated with a custom battery powered 54.1~MHz signal generator that fed a dipole antenna. Over a range of discrete heights, received power was recorded at the output (normally the input) of the Yagi array. 
%Several methods have been proposed to do this measurement
%including flying a weather balloon with a power meter and measuring the antenna output, flying a balloon with a
%small transmitter and measuring the received signal at the transmitting antenna, or flying a small
%UAV (Unmanned Aerial Vehicle) carrying similar payloads. In addition to a power
%measuring device (or transmitter), the location of the payload must also be recorded either by a GPS receiver or
%calculated from multiple grounded observers using theodolites.

The horizontal (azimuthal) radiation pattern was measured using a spectrum analyzer on the
ground to determine the pointing direction and
shape of the main lobe. Measurements of transmitted RF power were taken at
distances between 650 and 1000~m radially from the center of the array. Power was measured along a road that does not run perpendicular to the pointing direction of
the transmitter so a $1/r^2$ correction was made.
Figure~\ref{fig:txantenna_radiation_patterns} shows the measured points for the horizontal and vertical patterns overlayed on the models.
% (IJM - Is this line necessary? Not sure.)The measured pattern in dBm
%was then scaled to the modeled pattern by first smoothing it, then
%adding the difference between the peak of the modeled pattern and the
%smoothed measured pattern. 
These measurements are all relative, not absolute, so a uniform scale factor was determined by minimizing $\chi^2$ between the model and data. The measured pattern agrees very well with
the model in pointing direction and shape.

%\begin{figure}
%\centering
%\includegraphics[width=0.46\textwidth]{measured_horizontal_pattern.pdf}
%\caption{Modeled horizontal radiation pattern compared to a measurement performed
%  along the ground. The ground measurement was performed using a spectrum
%  analyzer while the transmitter broadcast its sounding wave. Red points show the measurements and the red line
%  is the smoothed representation of those
%  points.\label{fig:measured_horizontal_pattern}}
%\end{figure}

 % Bill
%%

%%
\section{Receiver Antenna}
\label{sec:rxantenna}

\begin{figure}[h]
\centerline{\mbox{\includegraphics[width=0.45\textwidth]{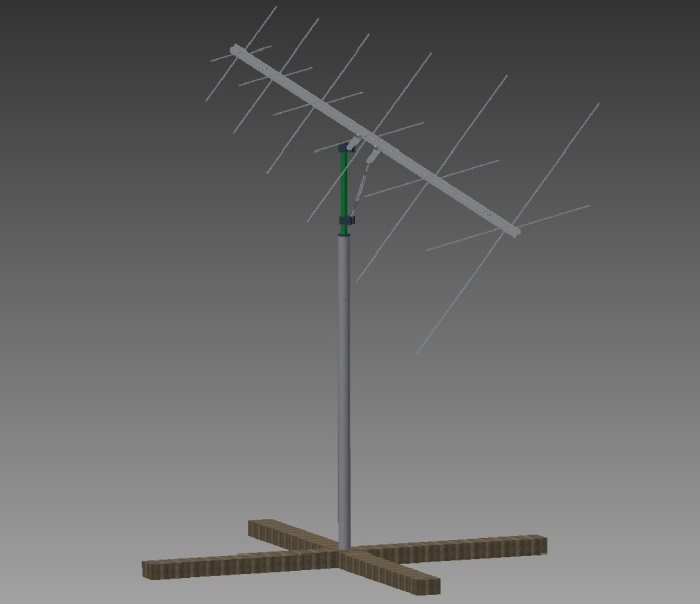}}}
\caption{Dual polarized TARA Log Periodic Dipole Antenna (LPDA).
\label{fig:LPDA}}
\end{figure}

\begin{table}[h]
\centering
\begin{tabular}{|l|r|r|}\hline
Element         &Length (in) &Position (in) \\ \hline\hline
1         &21.875     &3.625 \\ \hline
2  &26.625     & 18.0625 \\ \hline
3      & 32.5     &35.625 \\ \hline
4      &39.625     &57.0 \\ \hline
5     & 48.3125     &83.125 \\ \hline
6     & 58.3125     &115.0 \\ \hline
\end{tabular}
\caption{Length and relative boom position of antenna elements of the
  TARA Log Periodic Dipole Antennas. All elements have a diameter of
  $0.25\,''$.\label{tab:rxantenna_dimensions}}
\end{table}

The TARA receiver antenna site is located at the Telescope Array Long Ridge Fluorescence Detector ($39^{\circ}~12'~27.75420''$~N, $113^{\circ}~7'~15.56760''$~W). Receiver antennas are dual-polarized log periodic dipole antennas (LPDA) designed to match the expected $< 100$~MHz signal frequency characteristics. Due to noise below 30~MHz and the FM band above 88~MHz, the effective band is reduced to 40 to 80~MHz. Each antenna channel is comprised of a series of six $\lambda/2$ dipoles.  The ratio of successive dipole lengths is equal to the horizontal spacing between two dipoles (the defining characteristic of LPDA units), with the longest elements farthest from the feed-point to mitigate large group delay across the passband. Table~\ref{tab:rxantenna_dimensions} gives the lengths and positions of the antenna elements on the boom from the front edge to the back. All elements are constructed of aluminum tubing of $1/4\,''$ outer diameter. Figure ~\ref{fig:LPDA} shows a schematic of the receiver LPDA.

\begin{figure}[!h]
\centerline{\mbox{\includegraphics [width=0.45\textwidth]{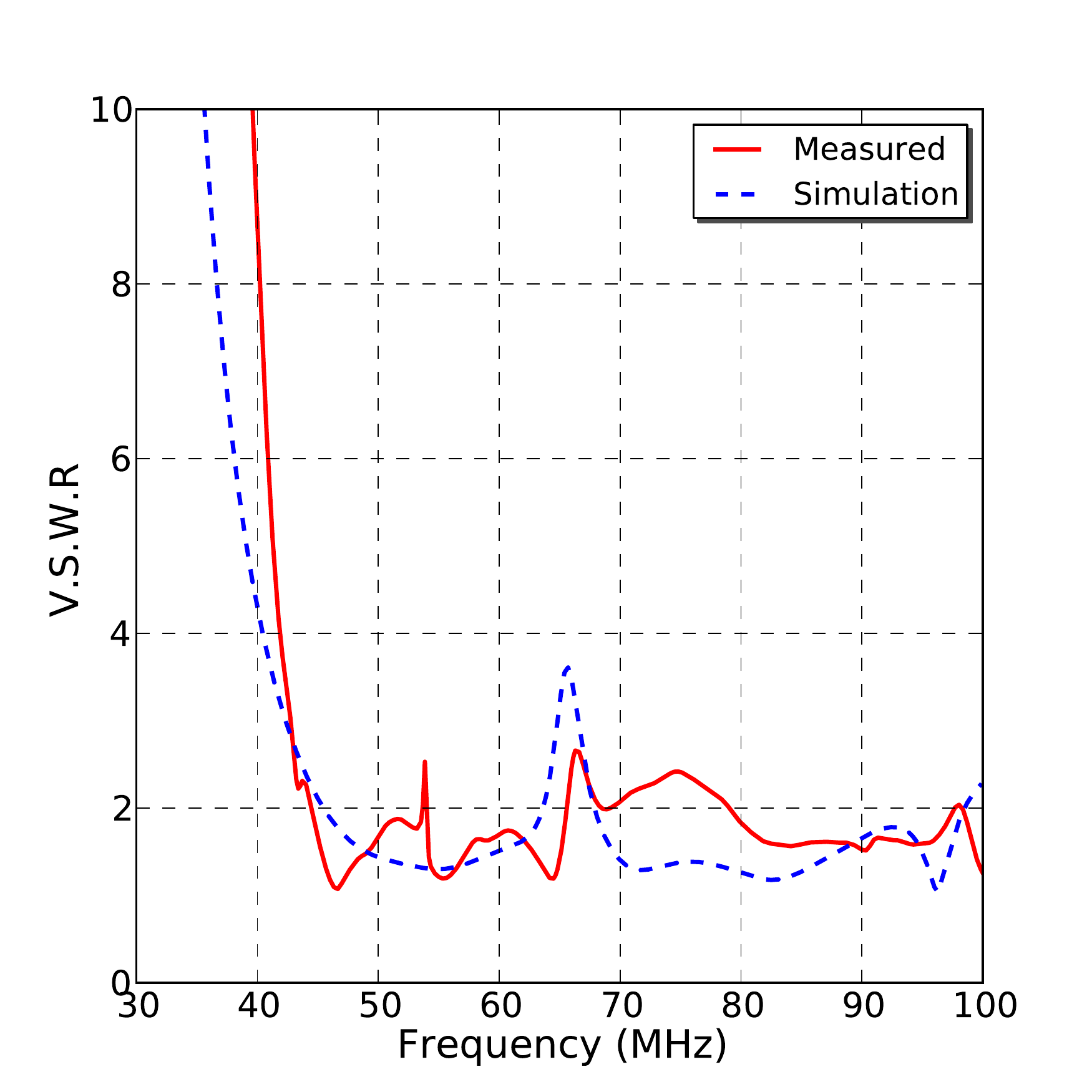}}}
\caption{SWR of a horizontally polarized TARA LPDA as measured in an anechoic chamber.
\label{fig:LPDA_SWR}}
\end{figure}

\begin{figure}[!h]
\centerline{\mbox{\includegraphics[trim=0.0cm 0.0cm 0.0cm 0.0cm,clip=false,angle=270,width=0.45\textwidth]{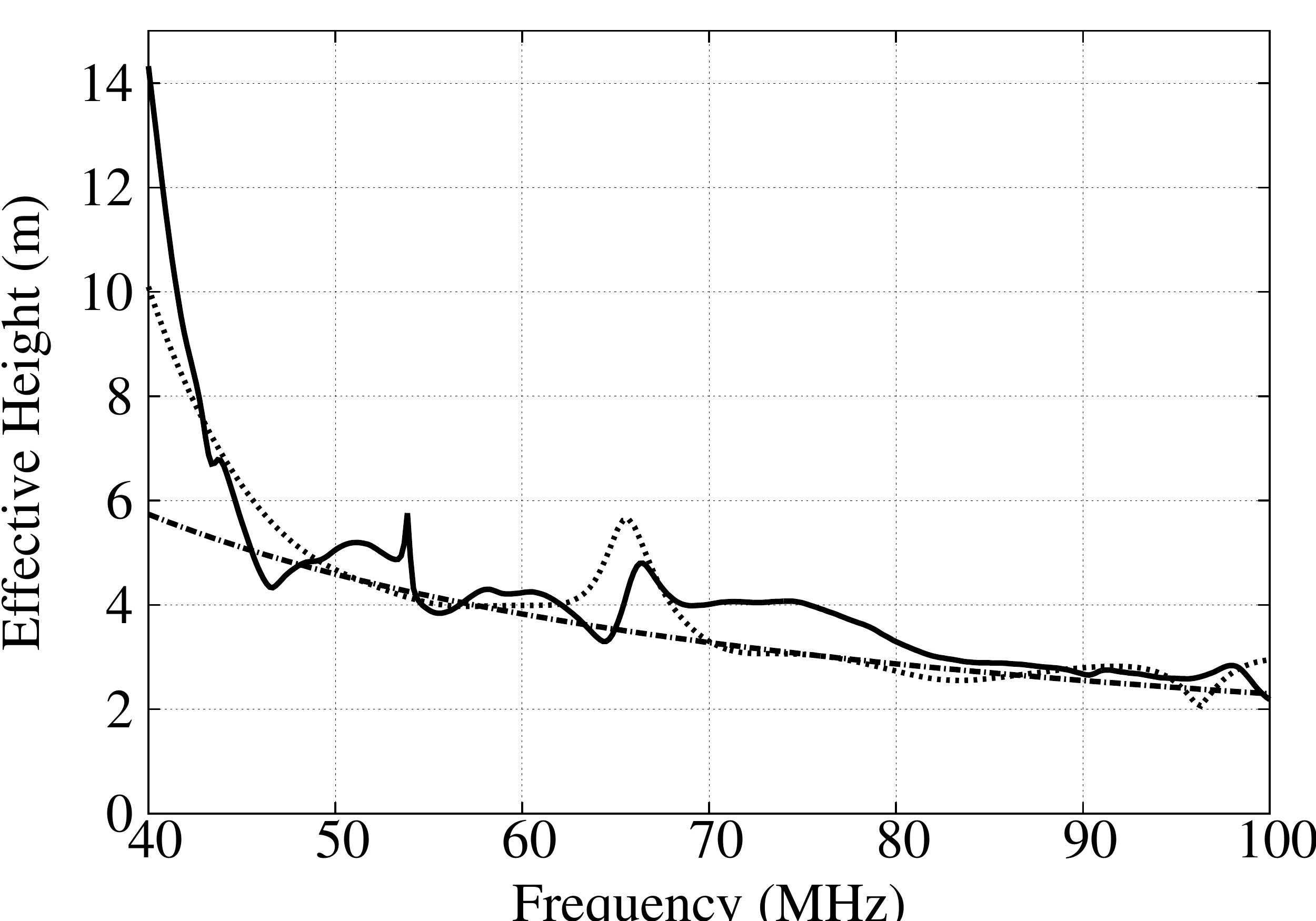}}}
\caption{Effective height in meters vs. frequency in MHz of the TARA receiver LPDA.  The $S_{11}$ parameter and gain of the receiver antenna are inserted into Equation~\ref{eq:hkraus} and plotted vs. frequency using the anechoic chamber data (solid line), simulated data from NEC (fine dashed), and simulated data with the 54.1~MHz values of $S_{11}$ and gain held constant (dot-dashed line).
\label{fig:heff}}
\end{figure}

The impedance of the antenna against a 50~$\Omega$ transmission line was measured in an anechoic chamber at the University of Kansas. The standing wave ratio (SWR), the magnitude of the complex reflection coefficient ($S_{11}$), is shown as a function of frequency in Figure~\ref{fig:LPDA_SWR}. An SWR of 3.0 implies greater than 75\% signal power is transmitted from the antenna to the receiver at a given frequency.

The complex $S_{11}$ measurement also quantifies the $\textit{effective height}$ of the LPDA.  The effective height translates the incident electric field strength in V/m to a voltage at the antenna terminals.  It is given as ${\bf E_{inc}} \cdot {\bf h_{eff}} = |E_{inc}| |h_{eff}| \cos(\theta) = V$, where $\theta$ is the polarization angle and the antenna is assumed to be horizontally polarized.  The boresight effective height can be expressed~\cite{Antennas} as % (ref. Kraus ch 2)

\begin{figure}[h!]
\centerline{
\includegraphics[trim=11.5cm 0.0cm 10.5cm 0.0cm,clip=true,width=0.48\textwidth]{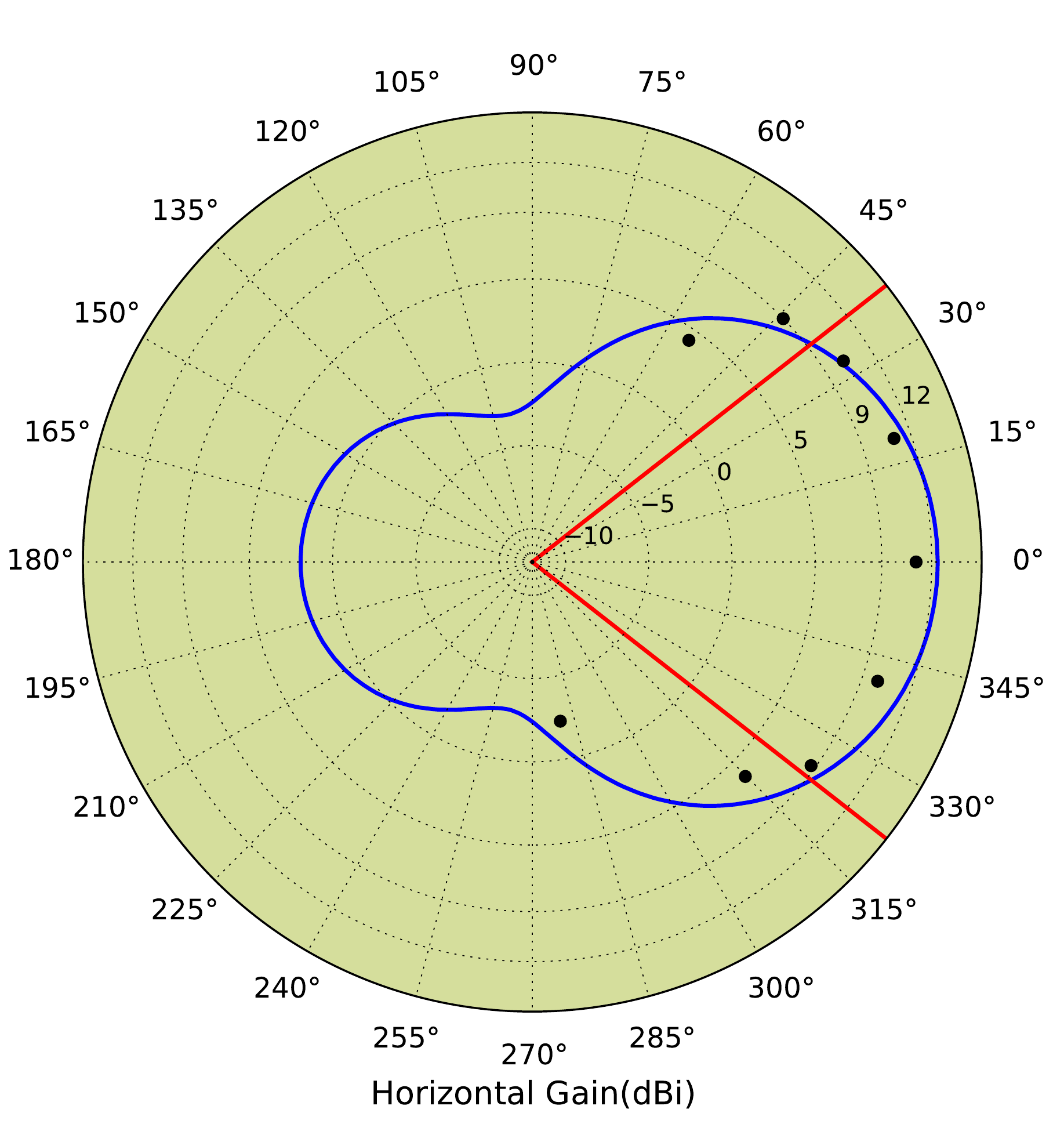}
}
\centerline{
\includegraphics[trim=11.5cm 0.0cm 10.5cm 0.0cm,clip=true,width=0.48\textwidth]{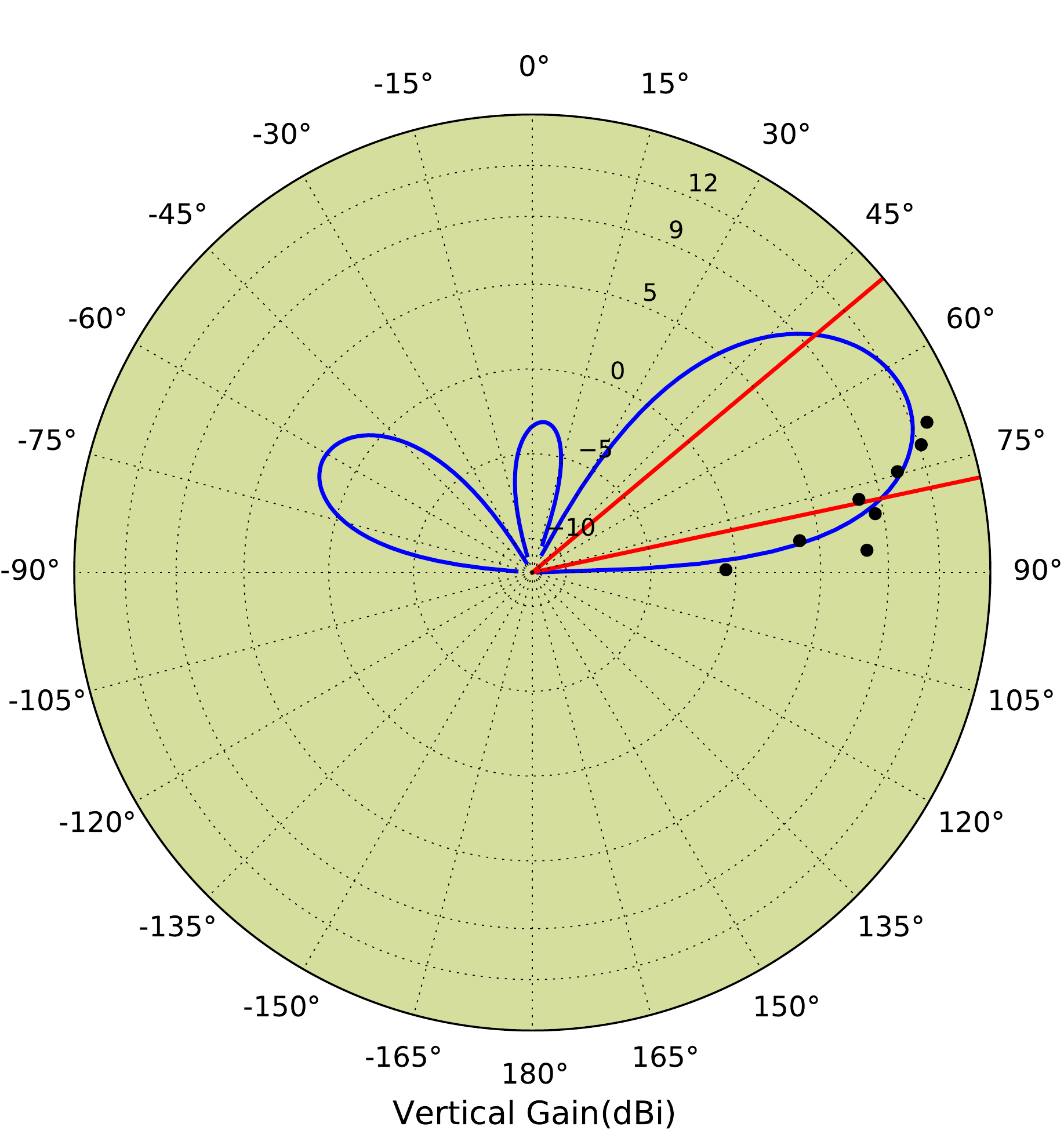}
}
\caption{
Simulated horizontal (top) and vertical (bottom) radiation pattern of a horizontally polarized TARA LPDA at the transmitter sounding frequency of 54.1~MHz. Beamwidths ($-3$~dB below peak gain) are shown with red lines. Peak gain is 12.6~dBi.
}
\label{fig:LPDA_rad_pattern}
\end{figure}

\begin{equation}
\label{eq:hkraus}
	h(\nu) = 2*\sqrt{\frac{Gc^2|Z_{in}|}{4\pi\nu^2Z_0}}\,.
\end{equation}

In the effective height expression, $G$ is the measured gain of 12.6~dBi (see Figure~\ref{fig:LPDA_rad_pattern}), $c$ is the speed of light, $Z_{in}$ is the complex antenna impedance, $\nu$ is the frequency, and $Z_0 = 120 \pi$ is the impedance of free space.  In terms of the measured complex reflection coefficient $S_{11}$, the impedance is given by $|Z_{in}| = \left| \frac{1+S_{11}}{1-S_{11}} \right| 50\, \Omega$.  The frequency-dependent magnitude of the effective height is plotted in Figure~\ref{fig:heff}.

\begin{figure}[h!]
\centerline{\mbox{\includegraphics  [width=0.45\textwidth]{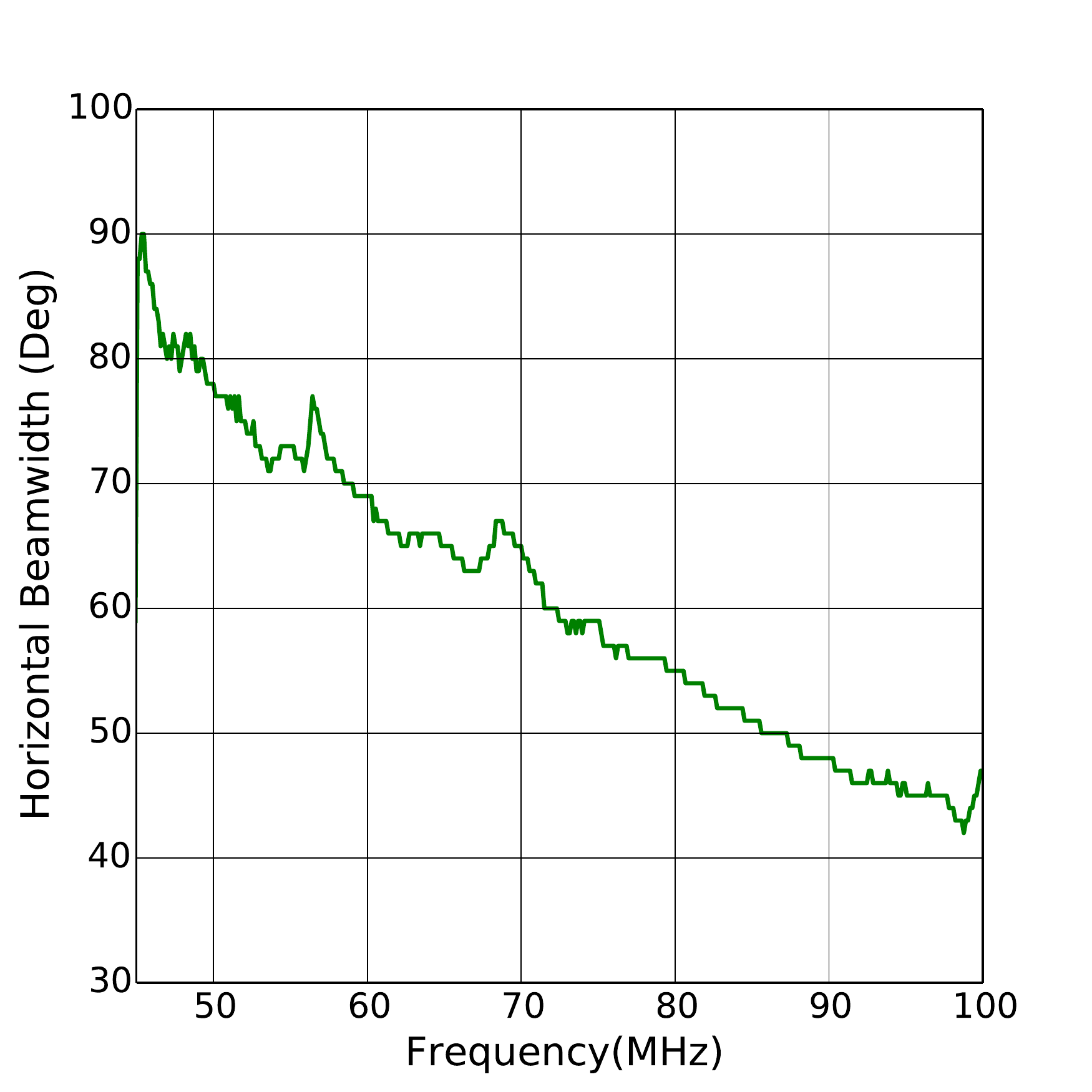}}}
\caption{Beamwidth of a single channel LPDA as measured in an anechoic chamber at the University of Kansas. 
\label{fig:LPDA_beamwidth}}
\end{figure}

Receiver antenna gain is a factor in the bi-static radar equation that affects detection threshold.  NEC was used in simulating the radiation pattern of the antenna to confirm directionality (see Figure~\ref{fig:LPDA_rad_pattern}). Simulated forward gain is 12.6~dBi and the vertical beamwidth is $23^\circ$ at the carrier frequency, 54.1~MHz. Figure~\ref{fig:LPDA_beamwidth} displays measured beamwidth in the band of interest.

%\begin{figure}[h!]
%\centering
%\begin{subfigure}{0.5\textwidth}
%\includegraphics[trim=9.5cm 0.0cm 9.5cm 0.0cm,clip=true,width=0.95\linewidth]{LPDA_rad_pattern_h.pdf}
%\end{subfigure}

%\begin{subfigure}{0.5\textwidth}
%\includegraphics[trim=9.5cm 0.0cm 9.5cm 0.0cm,clip=true,width=0.95\linewidth]{LPDA_rad_pattern_v.pdf}
%\end{subfigure}
%\caption{Simulated horizontal (top) and vertical (bottom) radiation pattern of a horizontally polarized TARA LPDA at the transmitter sounding frequency of 54.1~MHz. Beam widths are shown with red lines.
%\label{fig:LPDA_rad_pattern}}
%\end{figure}

%%The front-to-back ratio quantifies the peak %forward radiated power relative to the reverse lobe power.  %%Figures~\ref{fig:LPDA_beamwidth}) displays beam width over the %%frequency band %of interest.
%%and %\ref{fig:LPDA_frontback} the display beam width and front-to-back ratio results over the %%frequency band %of interest.

%%\begin{figure}[h!]
%%\centerline{\mbox{\includegraphics[width=0.45\textwidth]{LPDA_frontback.pdf}}}
%%\caption{Front-to-back ratio of a single channel LPDA measured in the CReSIS anechoic chamber.
%%\label{fig:LPDA_frontback}}
%%\end{figure}
 % Sam, Jordan
%%

%%
\section{Receiver Front-end}
\label{sec:rxfrontend}
%There are three receiver antennas at Longridge. Antenna 2 horizontal polarization and vertical polarization are connected to channel 0 and channel 1 of the FlexRIO respectively. Antenna 3 horizontal polarization and vertical polarization are connected to channel 2 and channel 3 of the FlexRIO respectively. Both the horizontal and vertical polarization signals from antenna 2 pass through a lightning arrestor, an RF limiter, a 40 dB amplifier, a bias tee, a low pass filter and a FM Band stop filter in that order. All the components except the low pass and the band stop filter, are placed in two weather proof boxes(one for each polarization) outside the building at longridge. The output from this RF enclosure is fed into a enclosure housing the low pass and the band stop filters, before they are fed into channel 0 and channel 1 of the FlexRIO. The horizontal and vertical polarization signals from antenna 3 pass through a lightning arrestor, an RF limiter, a 30 dB amplifier, a low pass filter and a FM Band stop filter in that order. All the components are placed in an enclosure inside the building at longridge and hence bias tees are not needed. They are then fed into channel 2 and channel 3 of the FlexRIO.

There are three dual-polarization antennas at the receiver site, two of which are currently connected to the DAQ (Section~\ref{sec:rxdaq}). RF signal from the antennas pass through a bank of filters and amplifiers. The components include an RF limiter (VLM-33-S+; Mini-Circuits), broad band amplifier, low pass filter (NLP - 100+; Mini-Circuits), high pass filter and an FM band stop filter (NSBP-108+; Mini-Circuits). Both polarizations from one antenna are filtered (37~MHz cutoff frequency high pass filter, SHP-50+; Mini-Circuits) and amplified (40~dB, ZKL-1R5+; Mini-Circuits) at the antenna, where a bias tee (ZFBT-4R2G+; Mini-Circuits) is used to bring DC power from the control room. The second antenna's channels are filtered (25~MHz high pass filter, NHP-25+; Mini-Circuits) and amplified (30~dB, ZKL-2R5+; Mini-Circuits) inside the control room. The lightning arrester (LSS0001; Inscape Data) minimizes damage to sensitive amplifiers by electric potentials that accrue during thunderstorms. The RF limiter prevents damage by transient high amplitude pulses (see Section~\ref{sec:challenges}).

\begin{figure}[h!]
\centerline{\mbox{\includegraphics[scale=0.50]{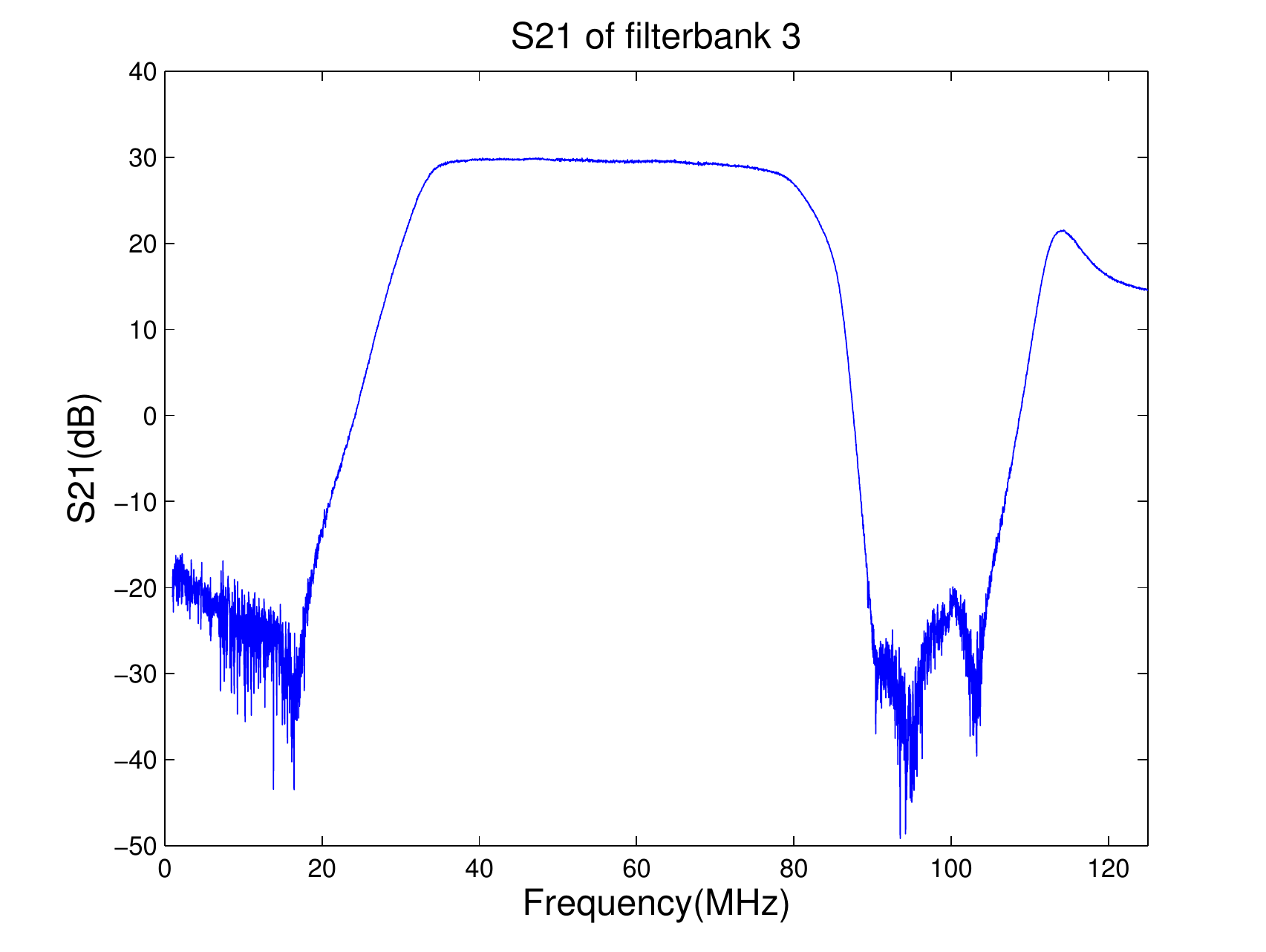}}}
\caption{$S_{21}$ (transmission coefficient) of the filter and amplifier bank connected to the triggering channel of the DAQ.
\label{fig:s21_of_filterbank3}}
\end{figure}

Signal conditioning in the amplifier/filter banks is characterized by the transmission coefficient (Figure~\ref{fig:s21_of_filterbank3}) $S_{21}$. It is a measure of the ratio of the voltage at the end of a transmission line relative to the input. Impedance mismatch relative to a 50~$\Omega$ transmission line, insertion loss for the various devices and gain from the amplifiers are combined in $S_{21}$ data. Of note in Figure~\ref{fig:s21_of_filterbank3} is the flat, high-gain (30~dB), broadband ($\simeq 40$~MHz) passband necessary for Doppler-shifted radar echoes.
%The lightning arrestor (LSS0001, Inscape Data) minimizes transient voltage damage to sensitive RF equipment. It has a VSWR of 1:1.5 from 0 to 5GHz and a VSWR of 1:2 from 5GHz to 6GHz. It has an impedance of 50 ohm and an insertion loss of 0.5dB. The RF limiter is used to protect other devices in the receiver frontend form damage caused by high input power. It has wide frequency of operation from 30 Hz to 3000 MHz. It has a VSWR of 1.05:1. It has an impedance of 50 ohm and an insertion loss of 0.23dB. The amplifier used is a wide band (10 Hz to 2500 MHz) amplifier. The bias tee is used to supply the amplifier with the required voltage to make the amplifier operational. The following figure shows the typical S21 values of the filterbanks.

%\clearpage
\begin{figure}[h!]
\centerline{
\includegraphics[trim=1.0cm 1.5cm 8.0cm 4.0cm,clip=true,width=0.48\textwidth]{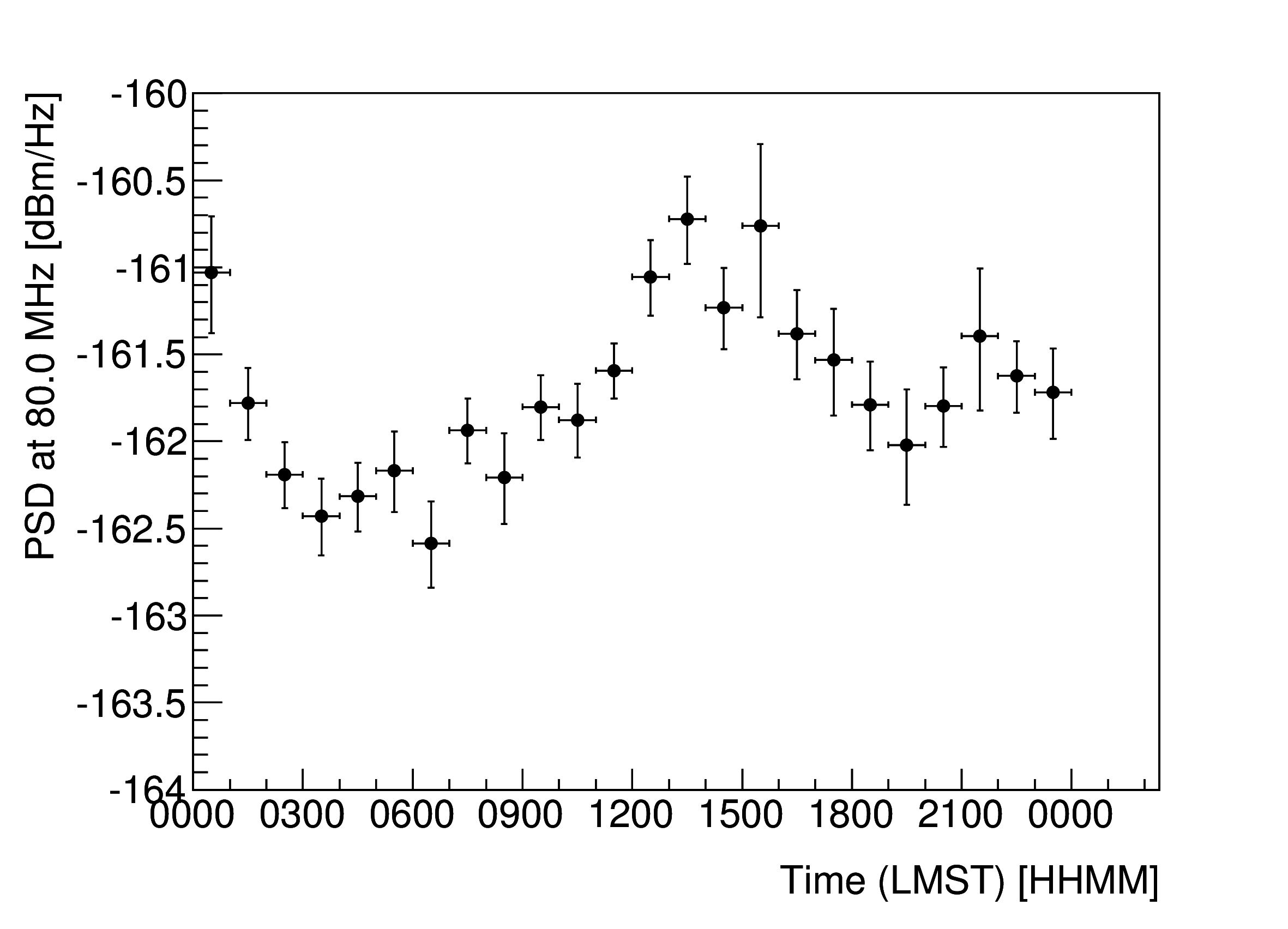}
}
\centerline{
\includegraphics[trim=1.0cm 1.5cm 8.0cm 4.0cm,clip=true,width=0.48\textwidth]{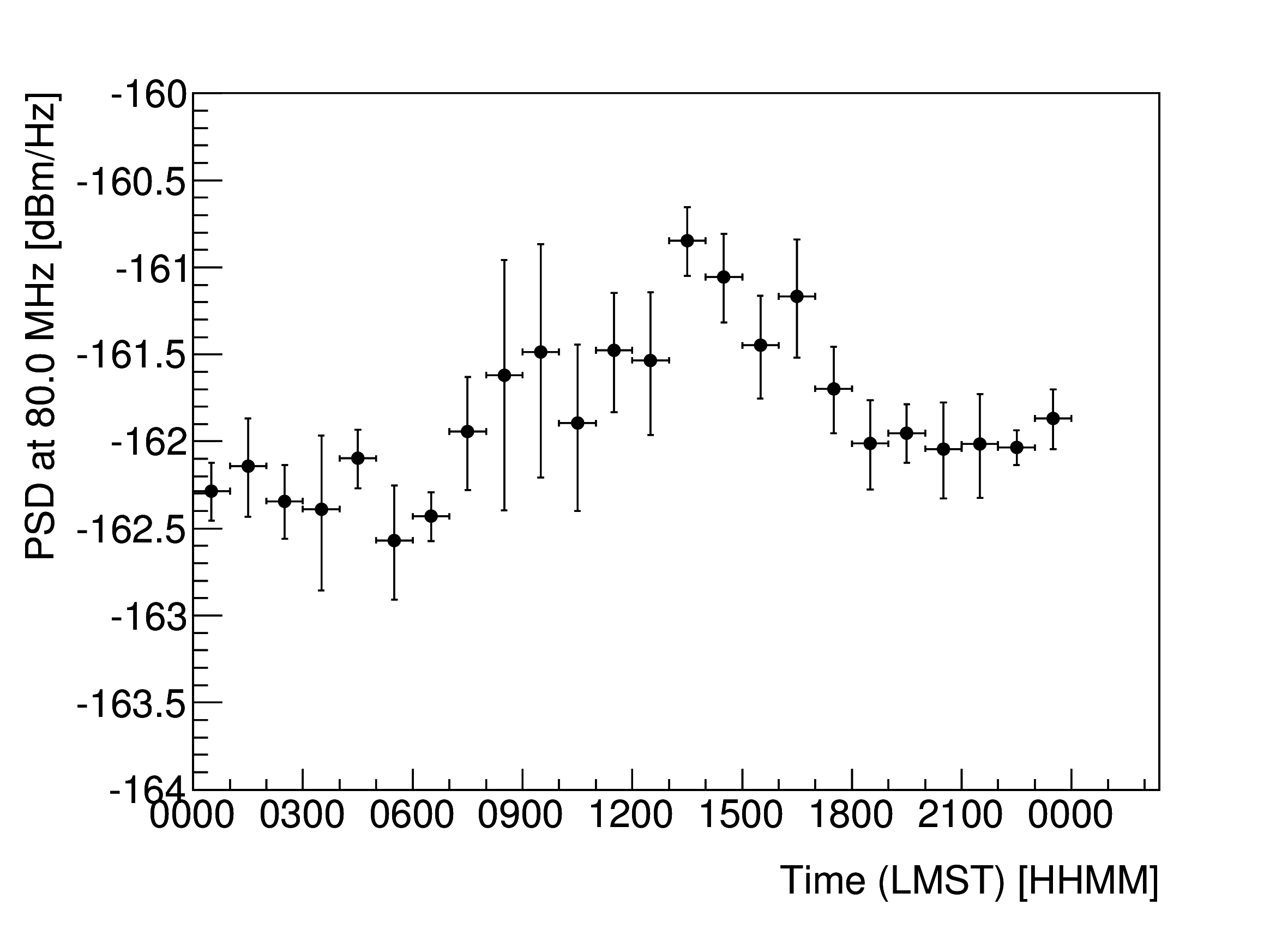}
}
\caption{Snapshot (forced trigger) Power Spectral Density (PSD) at 80.0~MHz averaged over eight days versus Local Mean Sidereal Time (LMST). Top: Data taken in December, 2013. Bottom: Data taken in May, 2014. There is strong correlation in peak PSD and sidereal time which indicates the signal is galactic in origin. Horizontal error bars show bin width. Vertical error bars are std. dev. in the mean.}
\label{fig:diurnal_variation}
\end{figure}

In any RF receiver system, sensitivity is limited by the combination of external noise entering through the antenna and internal noise from various sources like low noise amplifiers and other resistive losses from filters, cables and couplers. Noise entering the antenna is generated by the sky, earth and antenna resistive loss.  Diffuse radio noise from the galactic plane is non-polarized and is the dominant noise source in the TARA frequency band. Figure~\ref{fig:diurnal_variation} shows diurnal variation in the snapshot (forced trigger, 1~min$^{-1}$) spectrum that remains consistent in data taken six months apart. Each plot shows the Power Spectral Density (PSD, units of dBm/Hz) averaged over eight days versus Local Mean Sidereal Time (LMST). Horizontal and vertical error bars are bin width and std. dev. of the mean, respectively. The effect of amplifiers and cable losses have been removed such that absolute received power is shown. Data taken in December, 2013 are shown in the top plot, with those recorded in May, 2014 shown in the bottom plot. We observe that the peak occurs at the same time and power in each plot. Our conclusion is diurnal fluctuations are caused by changing perspective on the galactic center. 
% IJM: Not sure about this line: Correlating the peak with specific galactic features is difficult because the galactic disk is an extended object which interacts with rear and side antenna lobes.

\begin{figure}[h!]
\centerline{\mbox{\includegraphics[trim=0.0cm 0.0cm 1.1cm 0.8cm,clip=true,width=0.48\textwidth]{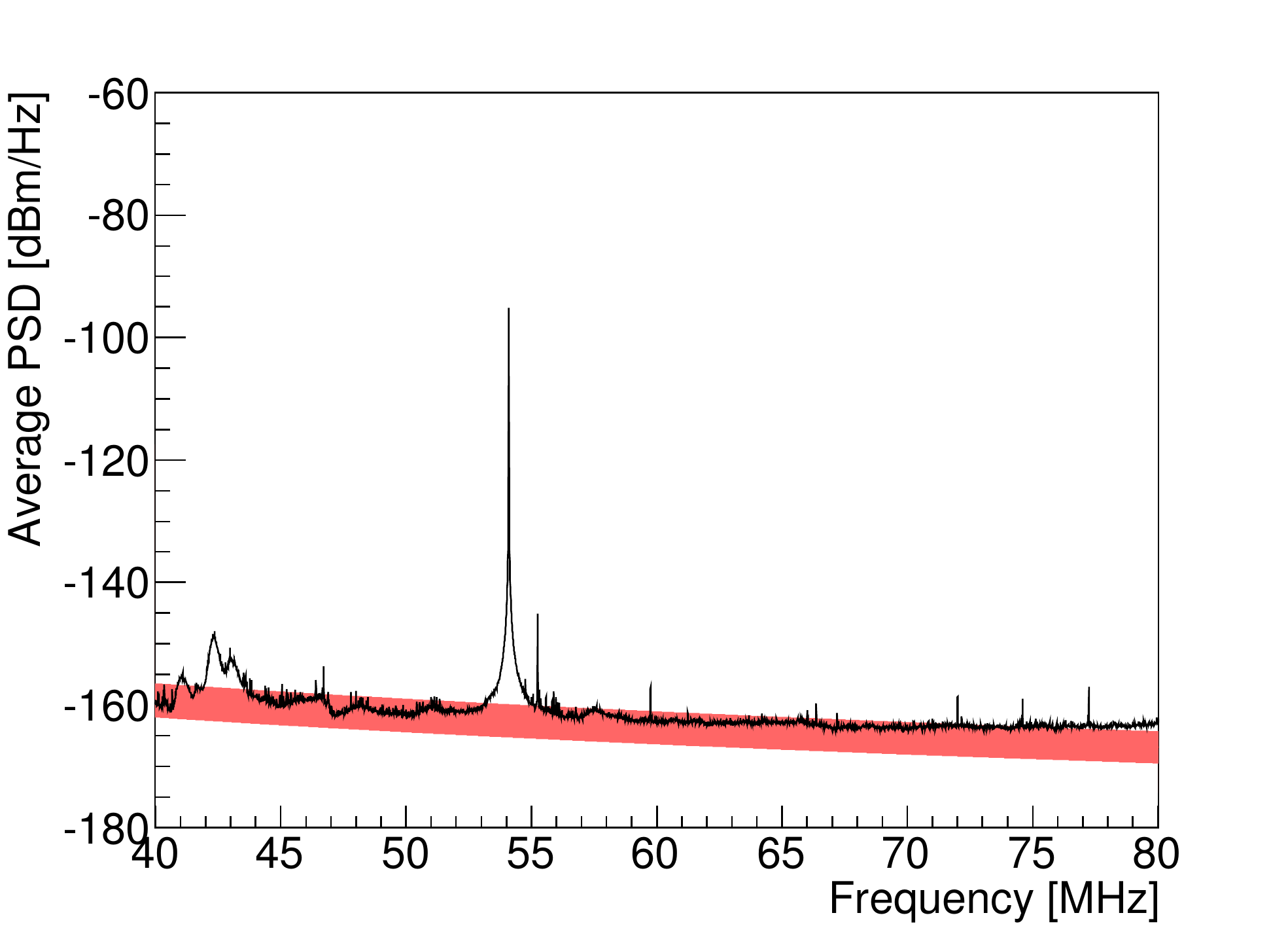}}}
\caption{Average receiver system noise floor (black) Power Spectral Density (PSD) in dBm/Hz superimposed with a fit to measured galactic background noise and its associated error~\cite{cane} (red band). System attenuation, filters and amplifiers were accounted for to determine absolute received power. No other calibration or scaling was applied to the receiver data.
\label{fig:LPDA_noise_floor}}
\end{figure}

By accounting for amplifier and instrumental gains and losses, the observed noise background can be compared with the irreducible galactic noise background~\cite{Dulk_galactic_noise} across the passband. Our average measured system noise is calibrated by removing the effects of individual components in the receiver RF chain from average snapshot spectra to determine the absolute received power. Without any other scaling, our corrected received power compares nicely with the galactic noise standard~\cite{cane} (Figure~\ref{fig:LPDA_noise_floor}). Important components for which adjustments were made include filters and amplifiers via the measured transmission coefficient $S_{21}$ and LMR-400 transmission line with attenuation data. Anthropogenic noise sources are transient and stationary noise is absent within our measurement band due to the receiver site's remote location. In this frequency region, galactic noise dominates thermal and other noise sources.  % Chetan
\section{Receiver DAQ}
\label{sec:rxdaq}

\subsection{DAQ Structure}
\begin{figure*}[t]
	\centering
         \includegraphics[width=0.8\textwidth]{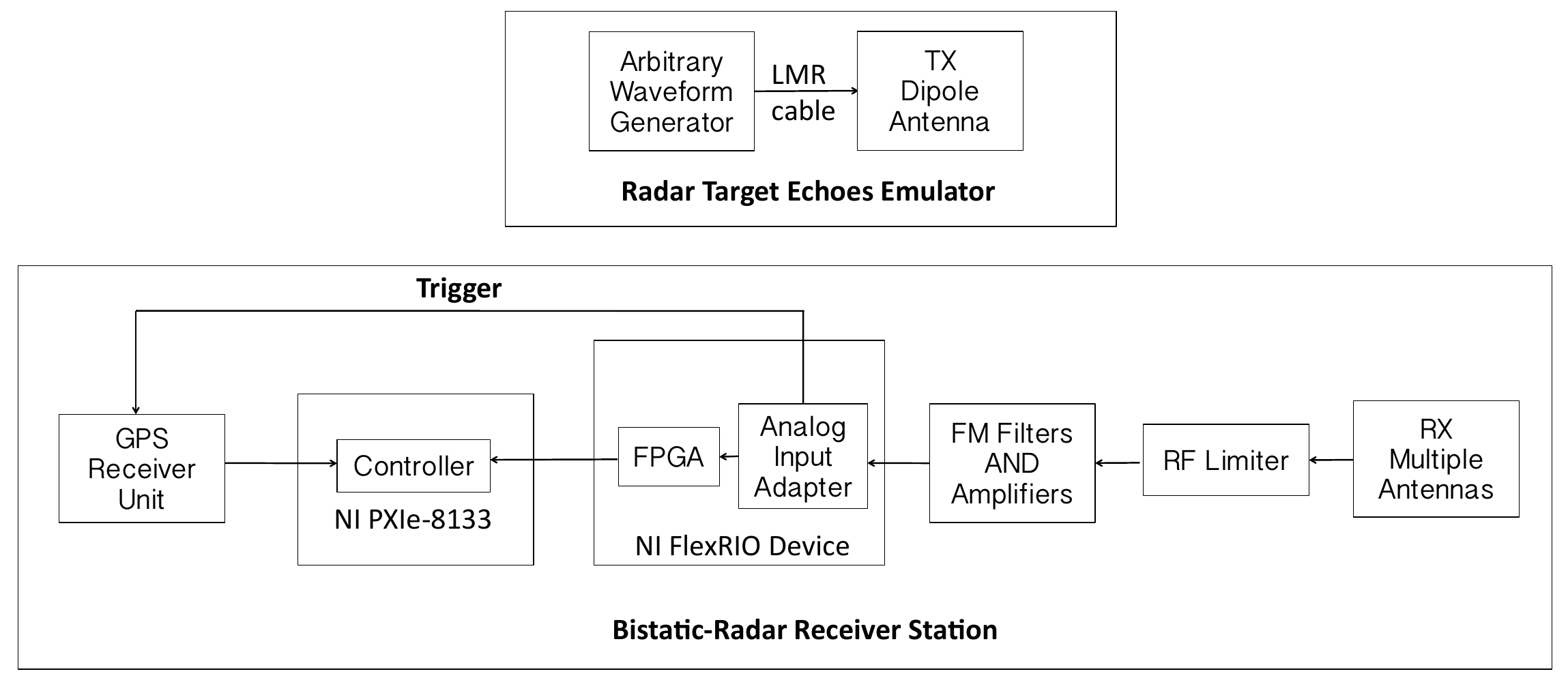}
	\caption{Elements of the radar receiver station.}
	\label{fig:sysdiagram}
\end{figure*}
%\clearpage % too many unprocessed floats
% IJM
% Suggest we eliminate all of the subsections and simply have several paragraphs.
The National Instruments FlexRIO system provides an integrated hardware and software solution for a custom software defined radio DAQ. It is composed of three basic parts: adapter module, FPGA module and host controller (as shown in the lower box of Figure~\ref{fig:sysdiagram}). A description of each of these subsystems follows.

The NI-5761 RF adapter module is a high-performance digitizer that defines the physical inputs and outputs of the DAQ system. It digitizes four analog input channels at a rate of 250~MS/s with 14-bit resolution. Eight TTL I/O lines are available for additional control, some of which are used in custom DAQ triggering schemes.

The NI-7965R FPGA module is based on the PXI express platform which uses a Xilinx Virtex-5 FPGA with 128MB on board DRAM. FPGA design provides accurate timing and intelligent triggering. The PXI-express platform has a high-speed data link to the host controller, which is connected to the development machine, a Windows based computer, which uses the LabVIEW environment to design and compile FPGA code. A host controller application, also designed in LabVIEW, runs on the development machine.

\subsection{Design Challenges}
\label{sec:challenges}

Based on the high velocity of the radar target, echoes are excepted to be characterized by a rapid phase modulation-induced frequency shift, covering tens of MHz in 10~$\mu$s. As the magnitude of the Doppler blue shift decreases as the shower develops in the atmosphere, these signals sweep (approximately) linearly from high to low frequency and are categorized as linear-downward chirp signals. Echo parameters are dependent on the physical parameters of the air showers. Thus, unlike existing chirp applications, we are interested in the detection of chirp echoes of variable amplitude, center frequency and frequency rates within a relatively wide band. In addition, the detection threshold must be minimized in order to increase the probability of detecting radar echoes with SNR less than one.

Furthermore, UHECR events are rare and random in time. TA receives only several $>10^{19}$~eV events per week, so background noise and spurious RF activity dominate.

\begin{figure}[h] 
	\begin{center}
		\includegraphics[trim=0.7cm 0.3cm 0.5cm 0.1cm,clip=true,width=0.5\textwidth]{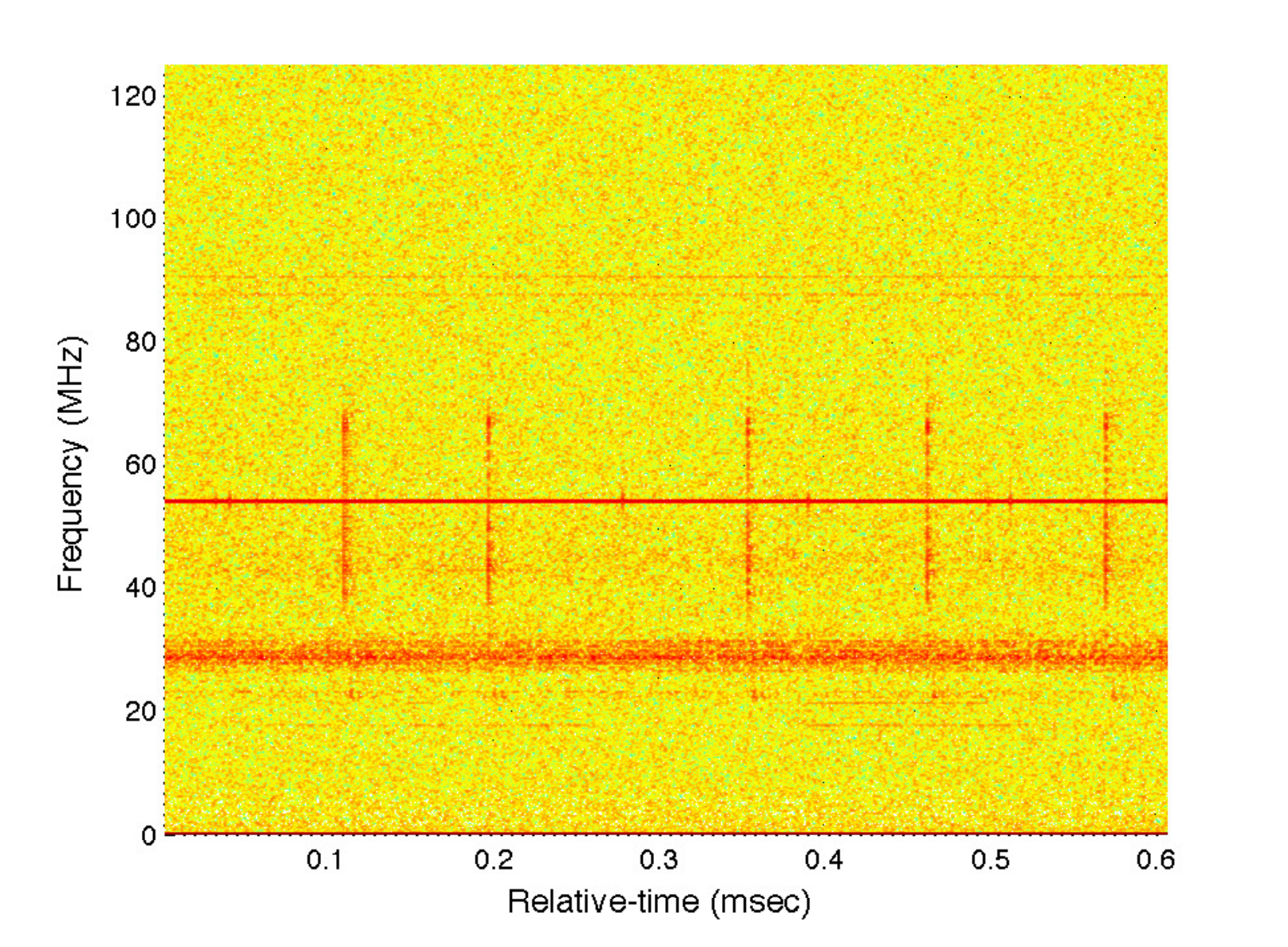}
	\end{center}
	\vspace{-10pt}
	\caption{Spectrogram of background noise at the receiver site. Frequency and time are on the vertical and horizontal axes, respectively, with color representing the power in a particular frequency component. The carrier signal is represented by the horizontal line at 54.1~MHz. Broadband transients are the vertical lines and stationary noise sources are the horizontal band near 30~MHz.}	
	\label{fig:sample_field}
\end{figure}

Figure~\ref{fig:sample_field} shows a spectrogram of data acquired in the field using the complete receiver and test system (Figure~\ref{fig:sysdiagram}), where FM radio and noise below $\sim30$~MHz are filtered out. The time-frequency representation shows that the background noise of our radar environment is rich with multiple undesirable components including stationary tones outside the 40-80~MHz effective band located at 28.5~MHz and, inside the band, the carrier at 54.1~MHz as well as broadband transients. Sudden amplitude modulation of stationary sources and powerful, short-duration broadband noise can cause false-alarms. % IJM - I don't think we should mention any sort of detection threshold here because it might be confused with a simple voltage threshold. Let's just say that such signals can cause false alarms and then after MF have been introduced explain why. [agreed]
A robust signal processing technique is needed to confront these challenges~\cite{6585981}.

\subsection{DAQ Implementation}
\label{sub:daqimp}
The DAQ is designed to detect chirp echoes and confront the problem of a variable noise environment. Two antennas feed the DAQ's four input channels. Each antenna is a dual-polarized LPDA (Section~\ref{sec:rxantenna}) with one output channel each for horizontal and vertical polarization. Data are collected simultaneously from each of the four analog channels with one horizontal channel considered the triggering channel, then sampled using a 250~MS/s ADC (Texas Instruments; ADS62P49). Analog to digital conversion is followed by fast digital memory storage on the FPGA chip, which stores the incoming samples from each channel sequentially, in a 131~$\mu$s (32744 sample) continuous circular buffer such that data in each buffer are continually overwritten. Three distinct trigger modes are implemented: ``snapshot'', ``Fluorescence Detector (FD) external", and ``matched-filter bank". 

\begin{figure}[!h] 
	\begin{center}
		\includegraphics[trim=0.0cm 0.3cm 0.0cm 0.0cm,clip=true,width=0.47\textwidth]{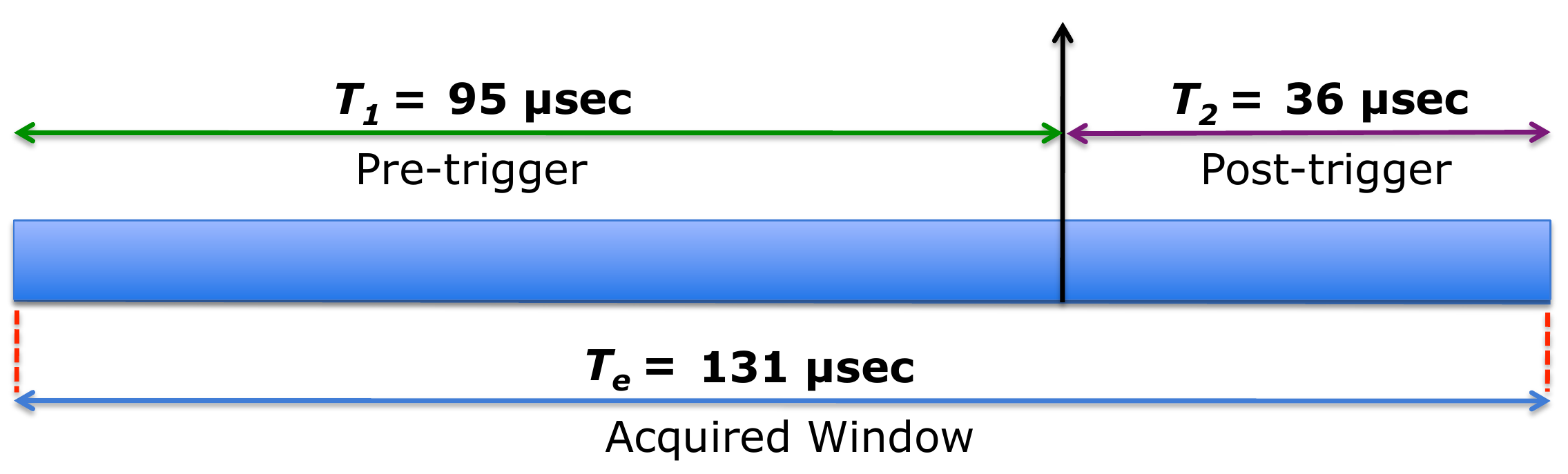}
	\end{center}
	\caption{Position of the triggering pulse within the data window that is written to disk.}	
	\label{fig:trig_timing_diag}
\end{figure}

When a trigger occurs, the circular buffer information is sent to the host controller to be permanently stored on the computer's disk. A 320~$\mu$s dead-time is required to transfer data from a buffer to FPGA memory, during which the DAQ cannot accept triggers. Sustained maximum trigger rate is 50~Hz due to FPGA-to-host data transfer limitations. As depicted in Figure~\ref{fig:trig_timing_diag}, \textit{pre/post} trigger acquisition is set to 95~$\mu$s and 36~$\mu$s, respectively, to allow for delay and jitter in the FD trigger timing (33~$\mu$s delay, 1~$\mu$s jitter) and sufficient post-trigger data to see an entire echo wave form. A GPS time stamp is retrieved from a programmable hardware module~\cite{gpsy2} and recorded for each trigger with an absolute error of $\pm 20$~ns. 

The snapshot trigger is an unbiased trigger scheme initiated once every minute that writes out an event to disk. These events will (likely) contain background noise only. Unbiased triggers are crucial for background noise estimation and analysis.

During active FD data acquisition periods, the Long Ridge FD (the location of the TARA receiver site) emits a NIM (Nuclear Instrumentation Module) pulse for each low level trigger with a typical rate of $\sim$~ 3--5~Hz or much higher  during FD calibration periods. The low level trigger is an OR of individual FD telescope mirror triggers. Dead time due to high FD-trigger rates are as high as several milliseconds during calibration periods. This does not reduce data acquisition time significantly because these periods occur only for several minutes and less than half a dozen times per FD data acquisition period. Further, FD operation only amounts to 10\% duty cycle on average. The FlexRIO is forced to trigger by each pulse received from the FD. Each FD run will result in many thousands of triggers which can be narrowed to several events that coincide with real events found in reconstructed TA data.

\begin{figure}[h]
	\centering
         \includegraphics[width=0.30\textwidth]{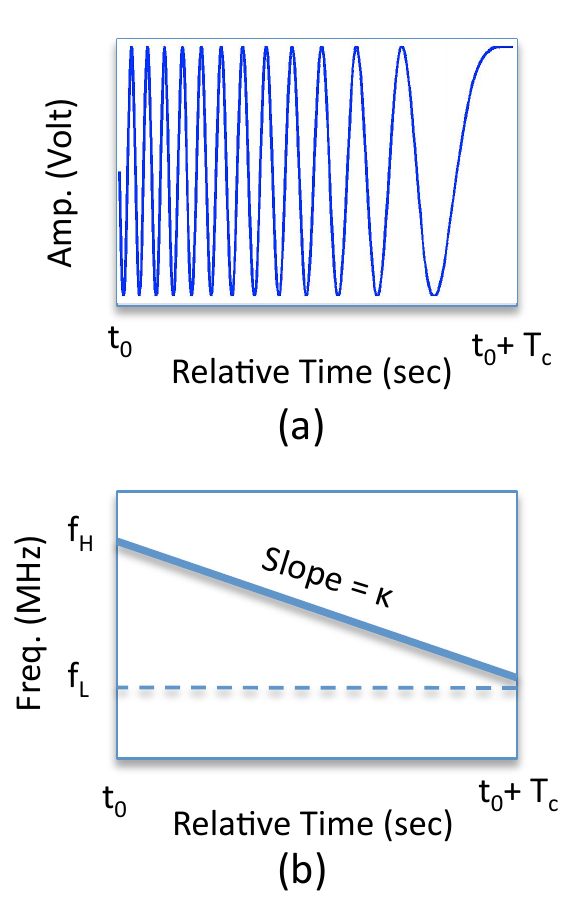}
	\caption{Linear down-chirp signal. (a) Signal in time-domain. (b) Signal in time-frequency domain.}
	\label{fig:chirpexample}
\end{figure}

The matched filter (MF) bank is a solution for the problem of detecting radar chirp echoes in a challenging noise background using signal processing techniques. The signal of interest is assumed to be a down-chirp signal that has duration $T_{c}$ seconds with a constant amplitude, start (high) frequency $f_{\rm H}$, center frequency $f_{\rm C}$, end (low) frequency $f_{\rm L}$ and chirp rate $\kappa$~Hz/sec. An example of the signal of interest is shown in Figure~\ref{fig:chirpexample}. Assuming that it is centered around time $t=0$, such a chirp signal is written as
\begin{equation}
s(t)=\text{rect}{\left( \frac{t}{T_{c}} \right)} \cos(2\pi{f_{\rm C}}t-\pi{\kappa}{t^{2}}).
\end{equation}
where $\text{rect}(x)$ is the rectangle function and $t$ is the time in seconds.

We limit our interest to detecting the presence of $s(t)$ within a certain bandwidth, without prior knowledge of the chirp rate $\kappa$. Based on simulation of the physical target, reflected echoes are expected to have a peak amplitude within or near the range [60-65]~MHz. Thus, we consider $f_{\rm H}$ to be 65~MHz and $f_{\rm L}$ to be 60~MHz.

\begin{figure}[h]
	\begin{center}	
		\includegraphics[width=0.48\textwidth]{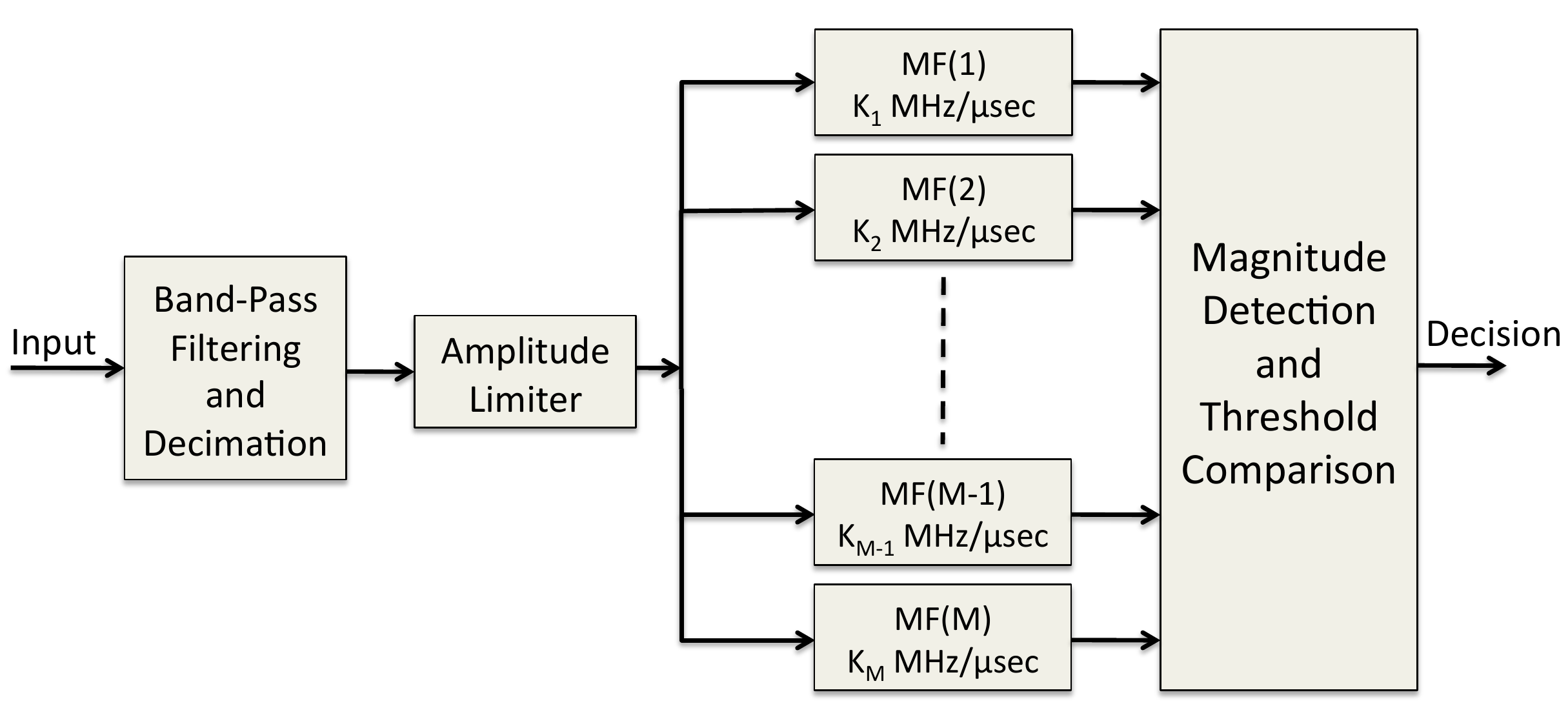}
	\end{center}
	\caption{Block diagram of the matched-filter-type detector.}
	\label{fig:MF_blockdiagram}
\end{figure}

Since the chirp rate varies, we use a bank of filters matched to a number of quantized chirp rates, $\kappa_1$, $\kappa_2$, $\cdots$, $\kappa_M$. A functional block diagram of the detection process is illustrated in Figure~\ref{fig:MF_blockdiagram}.

Let $\textbf{y}_{m}$ denote the output samples of the $m$th matched filter and $\gamma_{m}$ the threshold at the filter output. As depicted in Figure~\ref{fig:MF_blockdiagram}, a trigger decision is made at the output of the matched-filter bank by comparing magnitudes of the elements of $\bf y_1, \bf y_2, \cdots, \bf y_M$, each, against the corresponding threshold levels $\gamma_1,\gamma_2,\cdots,\gamma_M$, respectively. 

Threshold levels are defined as $n_\gamma$ units of the signal level (equivalently, noise standard deviation) at the output of each filter, denoted by $\sigma_m$ for the $m$th matched filter. Every time a trigger condition (the presence of a chirp) is met, an event  is declared. Since the background noise level varies with time, $\sigma_m$ is measured every five seconds to maintain a constant data acquisition rate. 

%The false alarm rate depends on the varying background. Clearly, probability of false alarm depends on the noise variance. Therefore, to maintain a constant level of false-alarm rate, an adaptive procedure is implemented on the FPGA to calculate the noise variance every short amount of time and then update the threshold level.

The most probable chirp-rate interval for a distribution of simulated radar echoes is ${\cal K}=[-3,-1]$~MHz/$\mu$s. We choose $M=5$ and the chirp rates (in MHz/$\mu$s) as
$$\kappa_1=-1.1161, \kappa_2=-1.3904, \kappa_3=-1.7321,$$ 
$$\kappa_4=-2.1577, \kappa_5=-2.6879\,.$$

\subsubsection{Amplitude Limiter} \mbox{}\\
Radio background at the remote receiver site is clear of stationary interference signals in the frequency band of interest, 40--80~MHz. Therefore, the broadband transients mentioned in Section~\ref{sec:challenges} are the primary source of false alarms. Consequently, the threshold of the MF detector must be raised in order to maintain the desired false alarm rate. The result is high data rate in return for low trigger thresholds. A digital amplitude limiter applied immediately before the input to the MF detector helps to minimize false alarms while keeping the detection threshold as low as possible and without significantly degrading detection efficiency. 

%\begin{figure}[h]
%	\centering
%         \includegraphics[width=0.45\textwidth, height=0.5\textwidth]{rxdaq_amplim.jpg}
%	\caption{Sample wave form replete with high amplitude transients clipped at an arbitrary level to demonstrate the function of the amplitude limiter.}
%	\label{fig:ampliminp}
%\end{figure}

The amplitude limiter clips the amplitude of the received signal to a fraction $k$ of its RMS value before clipping. Its mathematical expression is 
\begin{eqnarray}
\left\{\begin{array}{l}\displaystyle
y=x,\ \ |x|<k\sigma_s \vspace{1mm}\\
\displaystyle y=k\sigma_s,\ \ x>k\sigma_s \vspace{1mm}\\
\displaystyle y=-k\sigma_s,\ \ x<-k\sigma_s.\end{array}\right.
\end{eqnarray}
where $x$ is the raw input, $y$ is the amplitude limited output, and $\sigma_s$ is the RMS value of the signal before clipping. The result is a reduced relative power ratio of the spurious impulses to the non-perturbed background. Clipping also lowers the waveform RMS in proportion to the clipping level. 

\subsubsection{Band-Pass Filtering} \mbox{}\\
\label{subsub:bandpass}
We observe considerable CW noise within the 40--80~MHz band, including the carrier signal. The carrier and other persistent tones can have large amplitudes and lead to high matched filter RMS output which can, as shown in the next section, prevent detection of low SNR chirp signals. Such tones, including the carrier, can be easily filtered out. Before the amplitude limiter, a narrow band-pass filter eliminates all frequencies outside a 60-65~MHz band with -80~dB stop band attenuation. Data stored in the ring buffer are not filtered this way.
% The frequency response of the digital band-pass filter is shown in Fig.~\ref{fig:bpf_response}.
%\begin{figure}[h]
%	\centering
%         \includegraphics[width=0.7\textwidth, height=0.5\textwidth]{rxdaq_bpf_response.pdf}
%	\caption{Digital band-pass filter frequency response.}
%	\label{fig:bpf_response}
%\end{figure}

\subsection{Performance Evaluation}
% IJM: I don't see how the following paragraph is relevant to this section. It is restating what has been said in previous sections and doesn't lend any information that helps complete the current section.  [agreed]
%Figure~\ref{fig:sysdiagram} shows the block diagram of the receiver system. We utilize the NI-5761 adapter module with a sample rate ($F_s$) 250 MSPS. As mentioned earlier, our system-on-chip design is implemented over the high performance Virtex-5 FPGA which is integrated with PXIe interface for host connectivity.

Detection performance of the MF detector has been evaluated under two test signal conditions: noise only or signal plus noise. For each test, the Boolean result of the threshold comparison with the MF outputs is recorded. The probability of signal plus noise exceeding MF thresholds is the efficiency and the average rate of erroneous detection decisions caused by filtered noise is false alarm rate. 

The ability to detect a received chirp signal in background noise depends on the ratio of the signal power to the background noise power. Radar carrier power dominates the background so two quantities are used to describe the background noise. First, we define the ratio of the test chirp signal power to the radar carrier power as the signal-to-carrier ratio (SCR). Second, we use either the SNR (Equation~\ref{eq:snr}) or ASNR (Equation~\ref{eq:asnr}), depending on the type of test chirp signal input to the matched-filter bank, after filtering out the powerful carrier signal.

% IJM: Is the following paragraph necessary? It seems to me like this has been said before. What do you think?
%The false alarm rate depends on the varying background. Clearly, probability of false alarm depends on the noise variance. Therefore, to maintain a constant level of false-alarm rate, an adaptive procedure is implemented on the FPGA to calculate the noise variance every short amount of time and then update the threshold level.

Consider the following observations about performance analysis. First, it is clear that system performance depends on the chosen threshold level $n_{\gamma}$ (user defined, a multiple of $\sigma_{m}$ as defined previously) for each SNR value. False alarm rate is expected to decrease as the threshold level increases, at the expense of detection efficiency of low SNR chirp signals.
% A very large threshold level will decrease the false alarm rate, but also limit detection of low amplitude chirp signals. 
Conversely, detection efficiency increases as the threshold decreases. Second, the false alarm rate is expected to decrease as the amplitude limiter level decreases because high amplitude transients are effectively removed. To this date, radar echoes from CR air showers have not been detected, so it's unlikely that the EAS cross section is large enough to produce such large amplitude impulses. Therefore such signals are dismissed \emph{a priori}. Our strategy is to choose the threshold and amplitude limiter level that gives high detection efficiency for a given SNR and low false alarm rate.

\begin{figure}[!h]
	\centering
         \includegraphics[width=0.48\textwidth]{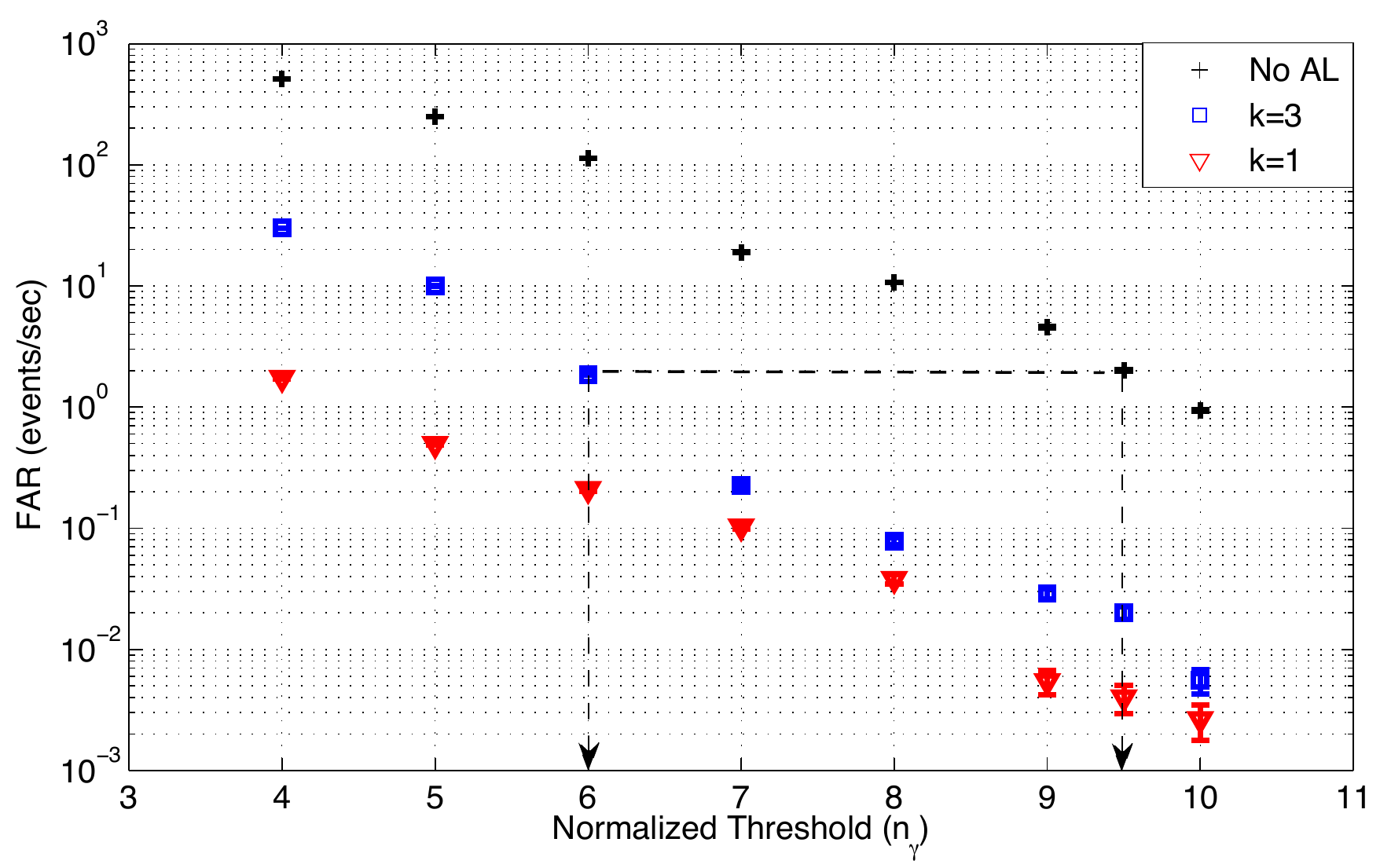}
	\caption{False-Alarm Rate versus relative threshold ($n_\gamma$ units of the standard deviation at each filter output) for different amplitude limiter levels.}
	\label{fig:farVSthr}
\end{figure}

Two tests are conducted to determine the ideal amplitude limiter level and the efficiency as a function of MF threshold. The goal of the first test is to measure the average false alarm rate of the non-Gaussian noise environment and evaluate the improvement that could be achieved by adding the amplitude limiter.  Results are shown in Figure~\ref{fig:farVSthr} for three different amplitude limiter levels, which clearly show that the limiter level has a significant effect on the false alarm rate. Efficiency curves for different amplitude limiter levels (described in the next paragraphs) show that the amplitude limiter does not \emph{decrease} detection performance of chirp signals, although they are also clipped. 

Consider the following interpretation of Figure~\ref{fig:farVSthr}. In order to achieve a 2~Hz false alarm rate, $n_\gamma$ has a value of six for $k=3$ and 9.5 for $k=10$ (black dashed line). Thus, detection thresholds can be decreased which enhances positive detection of low SNR signals.

%To tune the param the performance of the LRT detector under the existing non-Gaussian environment. The goal from this test is to measure the average false-alarm rate (FAR) acquired by the detector and evaluate the expected improvement that could be achieved by adding the amplitude limiter. Second, we evaluate the detection performance of the proposed detector for a typical chirp signal versus SNR under a specified threshold level that corresponds to a proper level of false-alarm rate. We present results for the most probabilistic interval of chirp slopes [-3,-1]~MHz/$\mu$s, different SNR values in the range of [-25,20]~dB, and a range of threshold levels 4 to 12 $\sigma_s$.

The second test applies a theoretical chirp signal with various chirp rates and SNR values that correspond to a reasonable false alarm rate. Based on data storage and post-processing computational requirements, we have decided that a false alarm rate of $\sim1$~Hz is reasonable. Artificially generated chirp signals are transmitted \emph{in situ} to the receiving antennas by an arbitrary waveform generator (AFG 3101; Tektronix, Inc.) and a dipole antenna. Both linear chirp signals and a simulated radar echo (see Section~\ref{sec:eas_echoes}) are used in measuring detection performance.

\subsubsection{Linear chirp signal}
\begin{figure}[!h]
	\centering
         \includegraphics[width=0.48\textwidth]{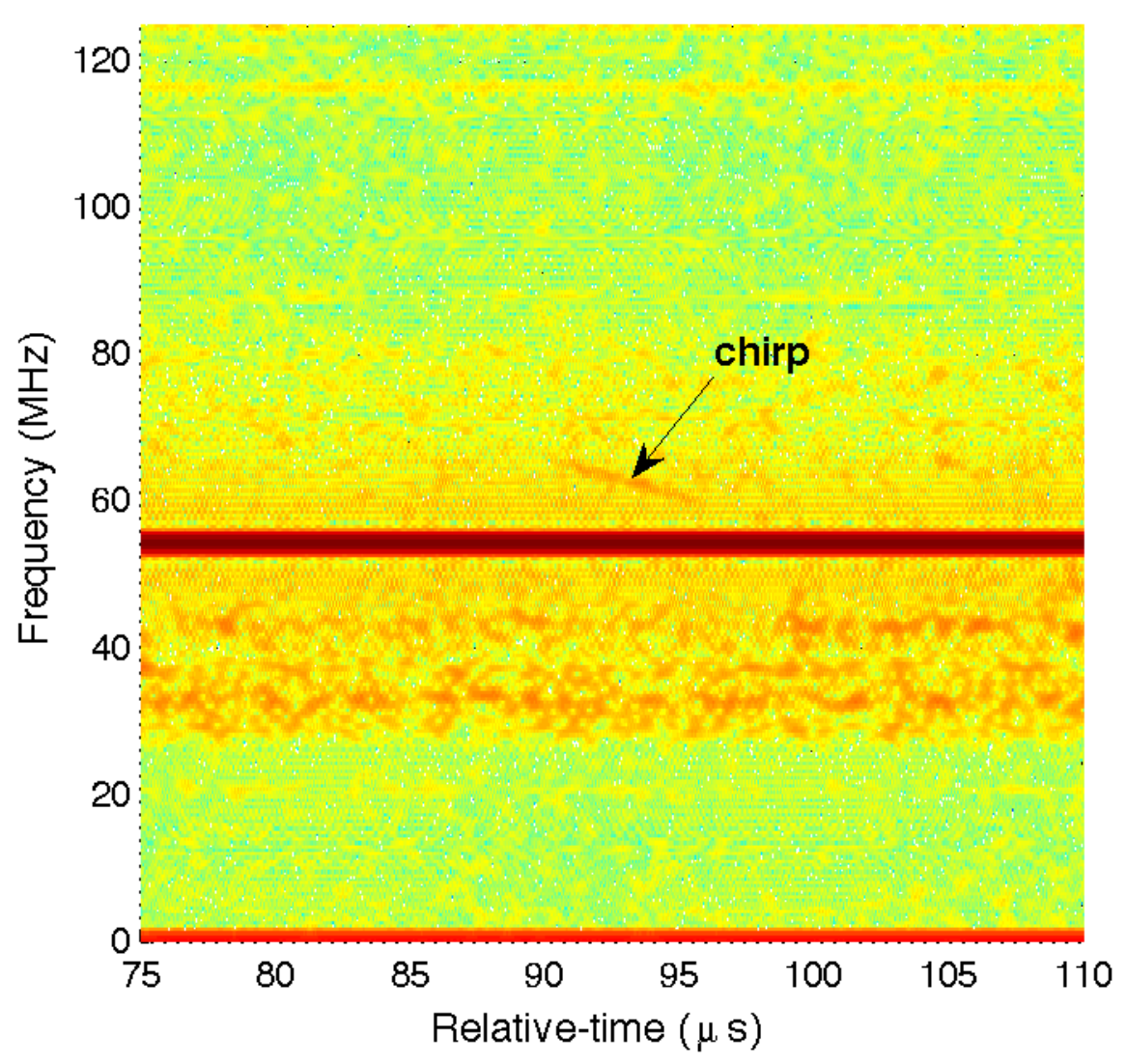}
	\caption{Time-frequency (spectrogram) representation of a linear, $-1$~MHz/$\mu$s, -10~dB SNR received chirp signal as recorded by the DAQ system.}
	% Linear, -1~MHz/$\mu$s, -10~dB SNR chirp signal superimposed on background noise recorded at the receiver site. Left: time domain represenation. Right: time-frequency (spectrogram) representation.
	\label{fig:neg10dB_chirp}
\end{figure}
A periodic, linear chirp with -1~MHz/$\mu$s rate is embedded in a real receiver site background wave form. Figure~\ref{fig:neg10dB_chirp} shows the spectrogram of a chirp embedded with -10~dB SNR and -40~dB SCR value.

\begin{figure}[h!]
	\centering
         \includegraphics[width=0.48\textwidth]{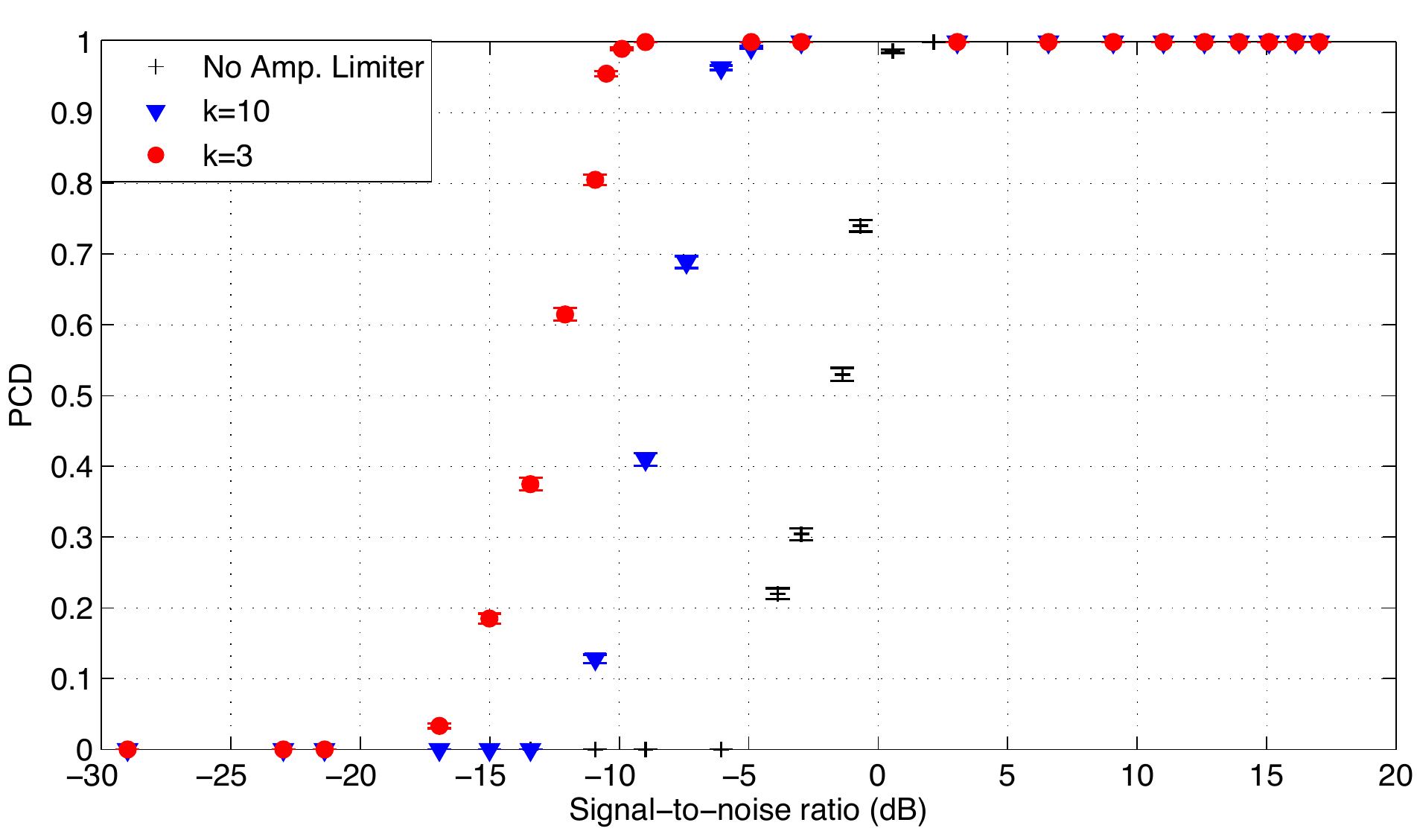}
	\caption{Probability of detection for the matched-filter-type detector with $n_{\gamma}=6$.}
	\label{fig:PCD}
\end{figure}

Figure~\ref{fig:PCD} shows detection performance for a 2~Hz false alarm rate. Efficiency is shown for cases where the amplitude limiter is removed and at two different levels that result in the same false alarm rate, each with different threshold levels. The minimum SNR for which complete detection is achieved is 5~dB when no amplitude limiter is applied, 0~dB for $k=10$ (\textit{soft clipping}), -6~dB for $k=3$ (\textit{hard clipping}). These results imply that by using the amplitude limiter, high detection performance can be achieved with low complexity. To maximize detection ability, the amplitude limiter is currently fixed at $k=3$.

\subsubsection{Simulated Air Shower}
\label{subsub:simshower}
\begin{figure}[h!]
	\centering
         \includegraphics[width=0.48\textwidth]{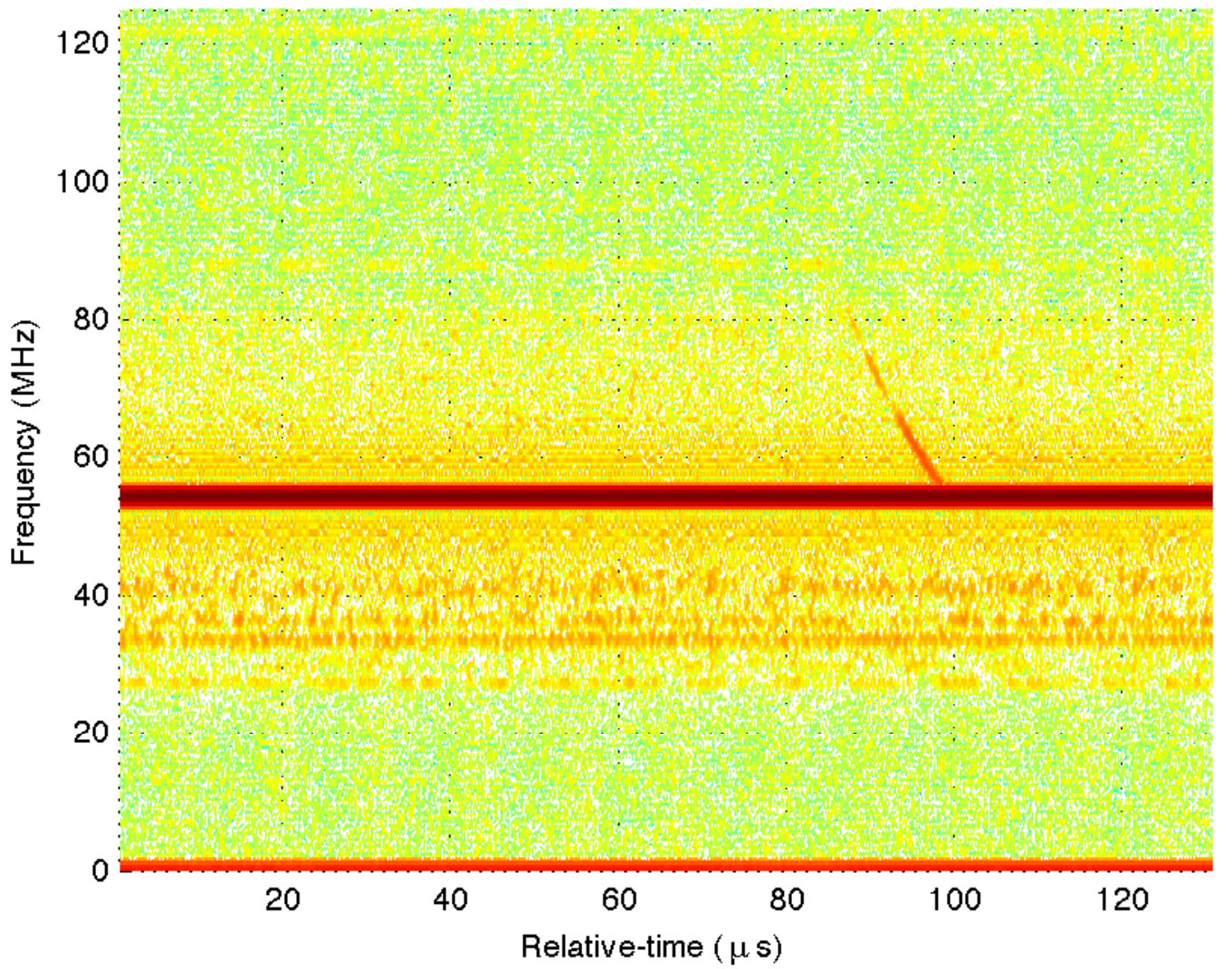}
	\caption{Spectrogram of simulated air shower radar echo with 5~dB ASNR. The radar echo is from a simulated shower inclined $30^{\circ}$ out of the $TX\rightarrow RX$ plane and located midway between the transmitter and receiver.}
	\label{fig:5dB_chirpecho_airshower}
\end{figure}

\begin{figure}[h!]
	\centering
         \includegraphics[width=0.48\textwidth]{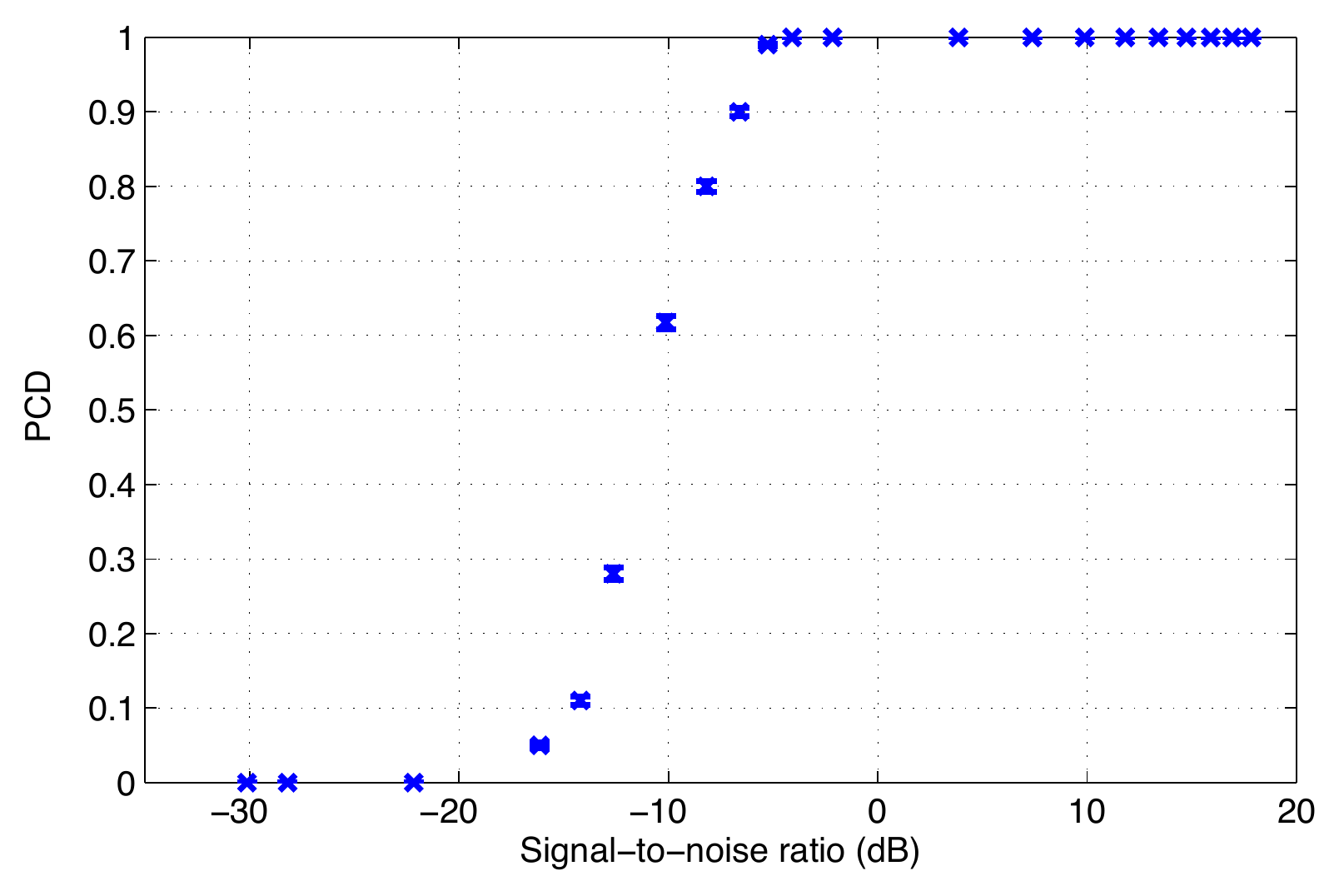}
	\caption{Probability of correct detection for the matched-filter detector using $n_{\gamma}=6$ for a simulated air-shower echo that is scaled and emulated with a function generator.}
	\label{fig:rxdaq_pcd_simAS}
\end{figure}
In a more realistic test, a simulated radar echo from a 10~EeV air shower inclined $30^{\circ}$ out of the $TX\rightarrow RX$ plane and located midway between the transmitter and receiver is scaled and transmitted to the receiving antennas using a function generator. Figure~\ref{fig:5dB_chirpecho_airshower} shows a spectrogram of the received waveform with 5~dB ASNR and -25~dB SCR. The echo is broadband (about 25~MHz) and short in duration (10~$\mu$s). Detection efficiency of the emulated chirp is shown in Figure~\ref{fig:rxdaq_pcd_simAS}. The minimum ASNR for which complete detection is achieved is -7~dB.

%\clearpage
 % Mohamed
%%

%%
\section{Remote Receiver Station}
\label{sec:remoterx}

\begin{figure}[h!]
\begin{minipage}{0.48\textwidth}
\centerline{\mbox{\includegraphics[trim=0.5cm 0.3cm 0.5cm 1.2cm,clip=true,width=0.95\textwidth]{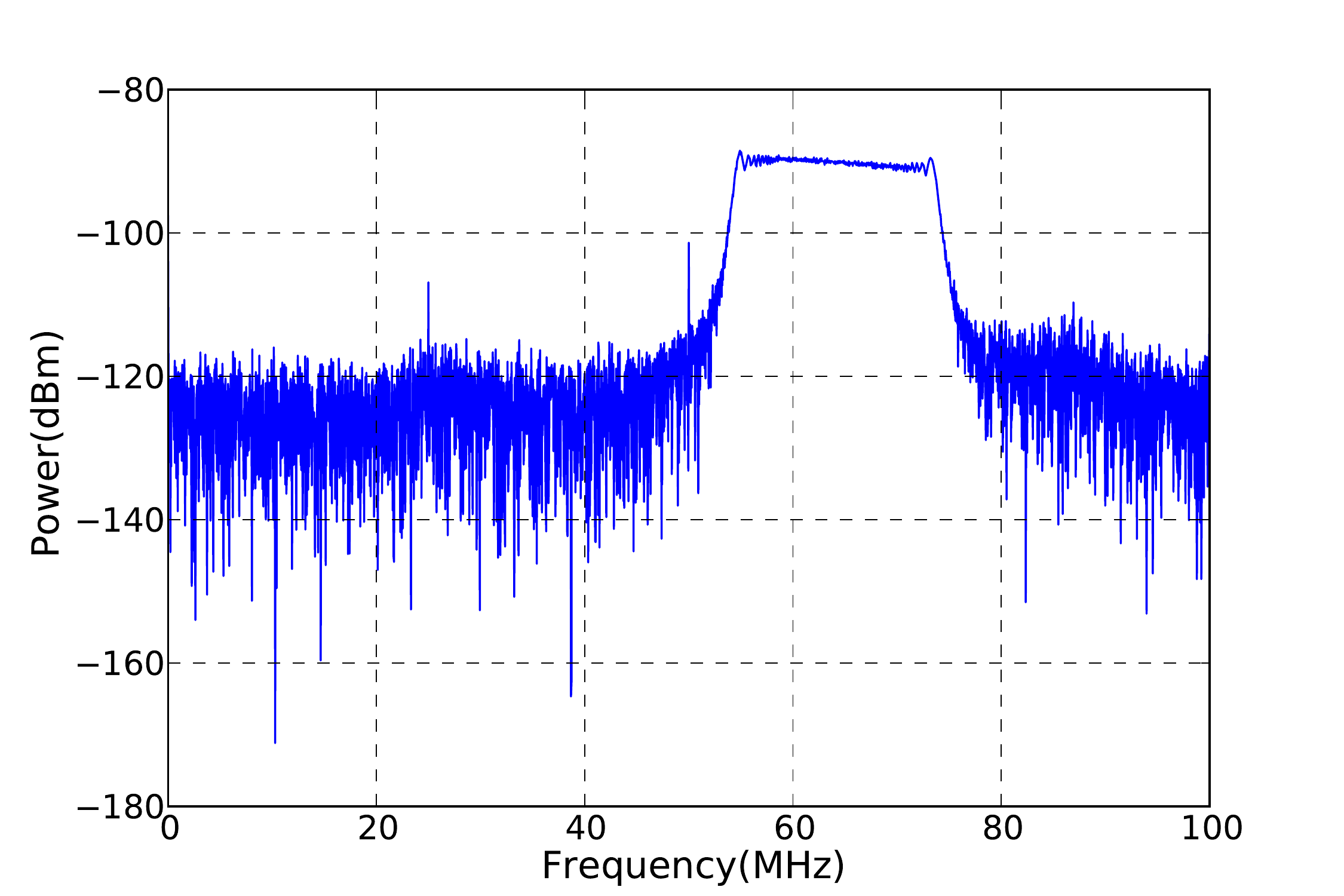}}}
\vspace{0.5cm}
\centerline{\mbox{\includegraphics[trim=0.5cm 0.3cm 0.5cm 1.2cm,clip=true,width=0.95\textwidth]{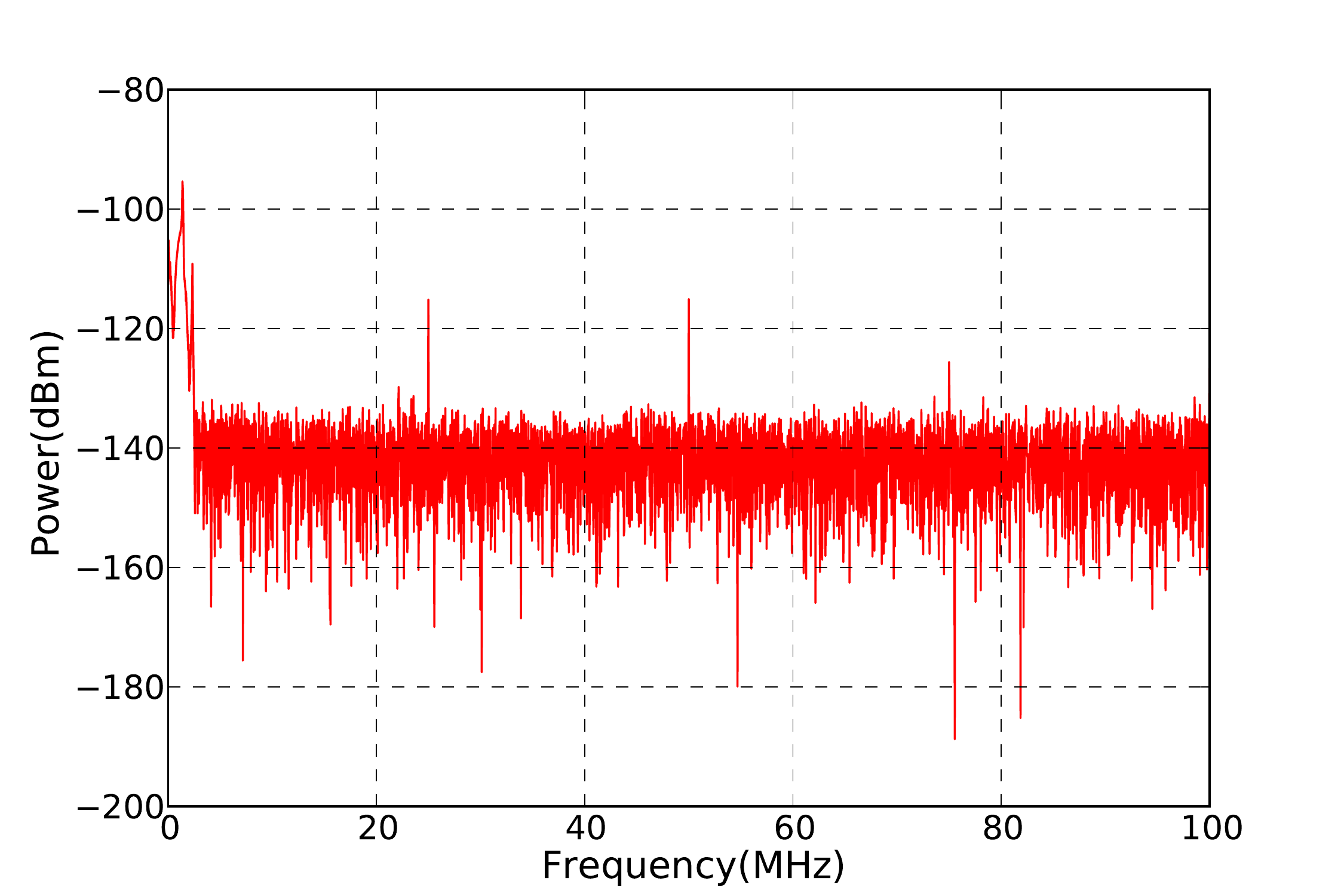}}}
\caption{Top: the power spectrum of a -10~MHz/$\mu$s chirp created by a signal generator, prior to mixing. Bottom: the power spectrum of a 1~MHz monotone signal after signal mixing and passing through a low pass filter. The chirp is evident as the left-most peak in this distribution.
\label{fig:rs_chirp_fft}}
\end{minipage}
\end{figure}

In addition to signal detection using matched filtering in the FlexRIO, an alternative technique has been developed that accomplishes chirp detection using a primarily analog signal chain.  Remote stations, by definition, are generally subject to less radio interference, and add stereoscopic measurement capabilities which theoretically allow unique determination of CR geometry and core location.  In contrast to the FlexRIO system, a mostly analog data acquisition system has lower power consumption at a cost which is also comparatively inexpensive. Triggering logic for our remote receiver station and some specific details of hardware components are discussed in the next couple subsections. 

\subsection{Remote Triggering}
\label{sub:triggering}

\begin{figure}[!ht]
\begin{center}
\begin{minipage}{0.47\textwidth}
\centerline{\mbox{\includegraphics[trim=1.5cm 0.0cm 3.3cm 1.5cm,clip=true,width=0.95\textwidth]{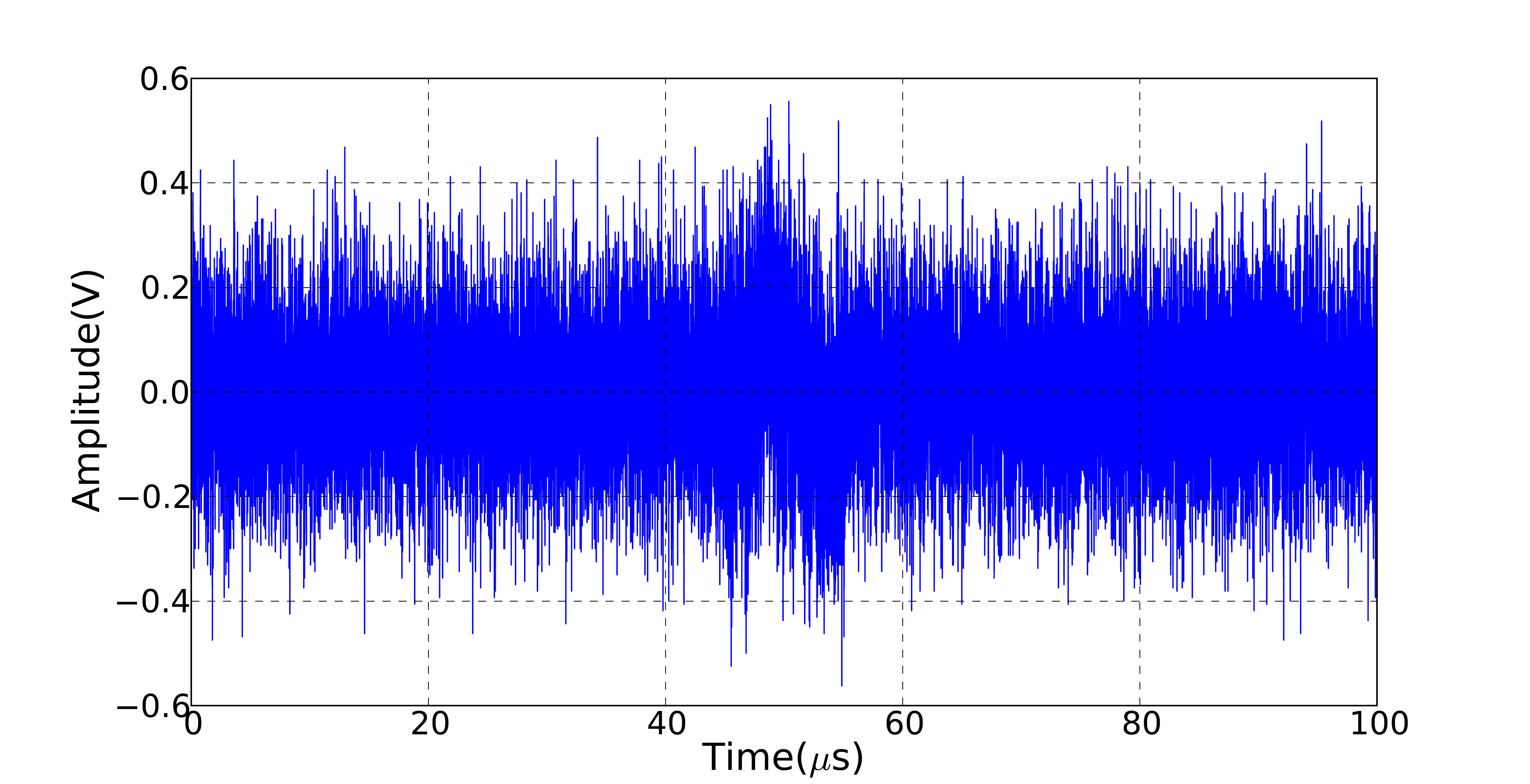}}}
\label{fig:rs_orig_td}
\centerline{\mbox{\includegraphics[trim=1.3cm 0.0cm 3.3cm 1.5cm,clip=true,width=0.95\textwidth]{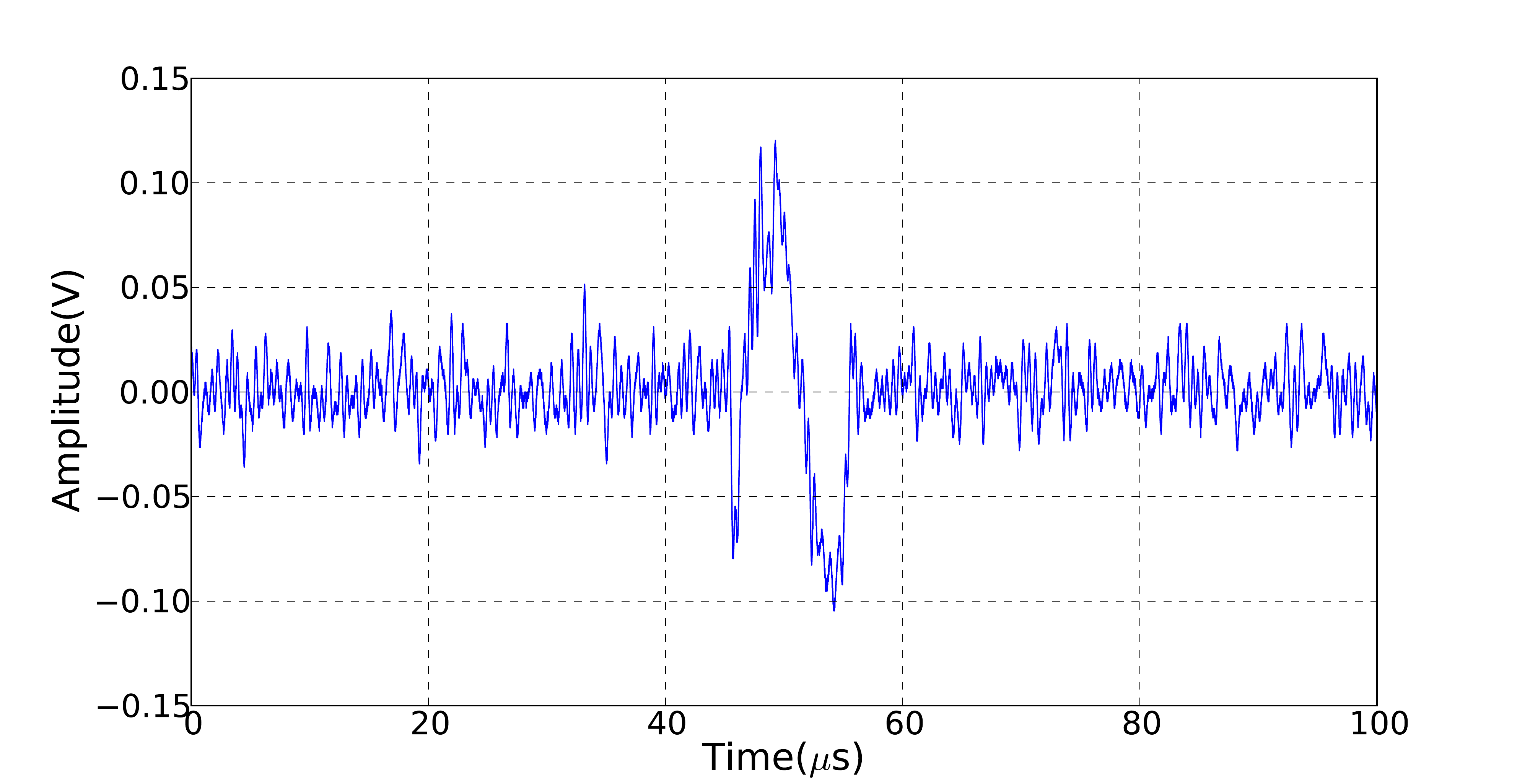}}}
\label{fig:rs_dc_td}
\centerline{\mbox{\includegraphics[trim=0.8cm 0.0cm 3.3cm 1.5cm,clip=true,width=0.95\textwidth]{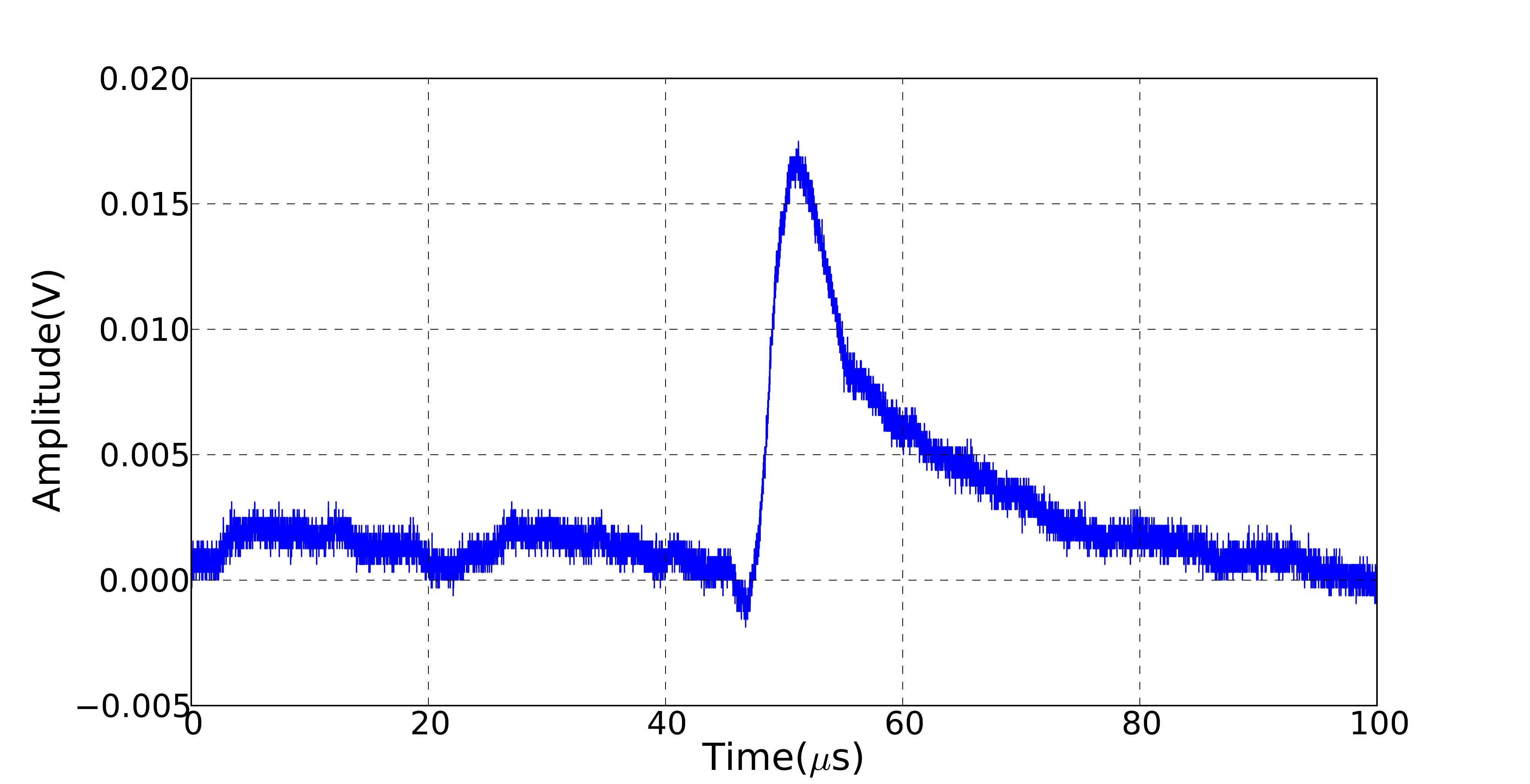}}}
\label{fig:rs_envdet_td}
\caption{Top: A 0~dB SNR and 1~MHz/$\mu$s chirp embedded in noise prior to mixing.  Second from top: The monotone signal after input chirp is mixed with delayed copy of itself and passed through a low-pass filter.  Bottom: Monotone passed through the Agilent 8471D power detector.
\label{fig:rs_blockdia}}
\end{minipage}
\end{center}
\end{figure}

\begin{figure*}[t]
\begin{center}
\begin{minipage}{0.9\textwidth}
\centerline{\mbox{\includegraphics[trim=0.2cm 2.8cm 0.4cm 3.0cm,clip=true,width=0.85\textwidth]{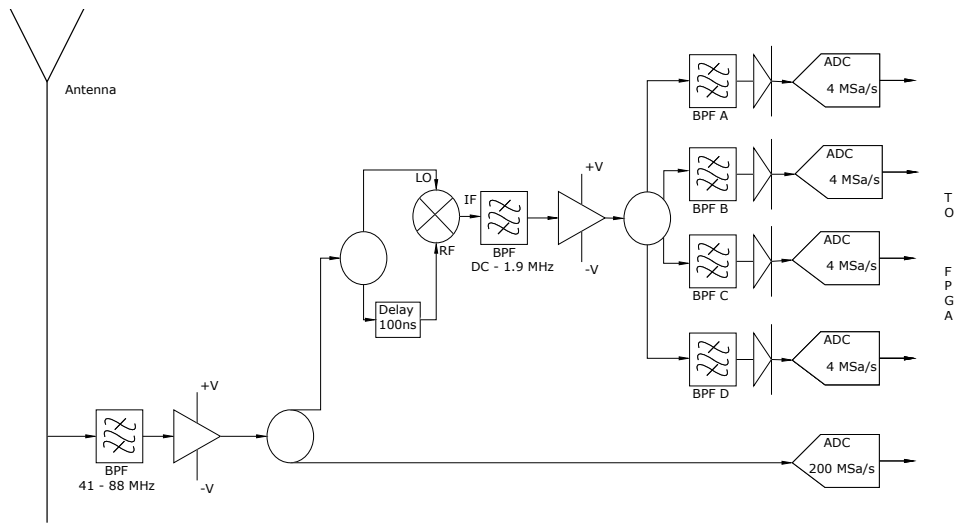}}}
\caption{Block diagram of the event triggering to be employed in the remote station.
\label{fig:rs_blockdia2}}
\end{minipage}
\end{center}
\end{figure*}

The alternative approach is based on an analog frequency mixer.  The input signal is mixed with a delayed copy of itself, i.e, $s(t)\otimes s(t - \tau)$.  For an incident chirp signal, the non-linear components in the mixer result in a product term that yields a monotone at a beat frequency $f_{\text{beat}} = \kappa \tau$;  dependent only on the delay time $\tau$ and the chirp rate $\kappa$.  The delay is created with 100~ft of LMR-600  cable, which produces negligible losses and removes the need for power consuming active components.  With appropriate filtering, the problem of chirp detection is ultimately reduced to that of detecting a down-converted monotone.  This is illustrated in Figure~\ref{fig:rs_chirp_fft}. Portrayed here with an oscilloscope is a -10~MHz/$\mu$s chirp which has been converted to a 1~MHz monotone by  mixing with a delayed copy of itself.

After mixing, the signal is passed through an envelope detector (8471D; Agilent, Inc.). The  entire time domain signal chain is illustrated in Figure~\ref{fig:rs_blockdia}. In this oscilloscope based example, a chirp with 0~dB SNR at a rate of -1~MHz/$\mu$s is first band-pass filtered (41-100~MHz) and then amplified by 20~dB. The signal is then mixed and low-pass filtered (DC-1.9~MHz) and passed through the Agilent power detector.

%\begin{figure}[h]
%\centerline{\mbox{\includegraphics[width=0.5\textwidth]{rs_blockdia.png}}}
%\caption{Block diagram of the event triggering to be employed in the remote station.
%\label{fig:rs_blockdia2}}
%\end{figure}

The expected value of chirp rates from EAS echos is typically between -1 to -10~MHz/$\mu$s (see Section \ref{sec:eas_echoes}).  Consequently, with 100~ns delay, the down-converted signal has a frequency between 100~kHz and 1~MHz.  To trigger on such signals, the mixed signal is split into multiple copies.  Each copy is then passed through custom band-pass filters and an envelope detector.  Different frequency bands are then compared by majority logic in an FPGA, requiring no more than one band to form a trigger in order to suppress impulsive noise. Each of the frequency banded outputs corresponds to a separate range of chirp rates. The block diagram in Figure~\ref{fig:rs_blockdia2} outlines this triggering procedure.

\subsection{Remote Station Electronics}
\label{sub:rstation}

\begin{figure*}[!ht]
\begin{center}
\begin{minipage}{1.0\textwidth}
\centerline{\mbox{\includegraphics[trim=0.0cm 0.0cm 0cm 0cm,clip=true,width=0.8\textwidth]{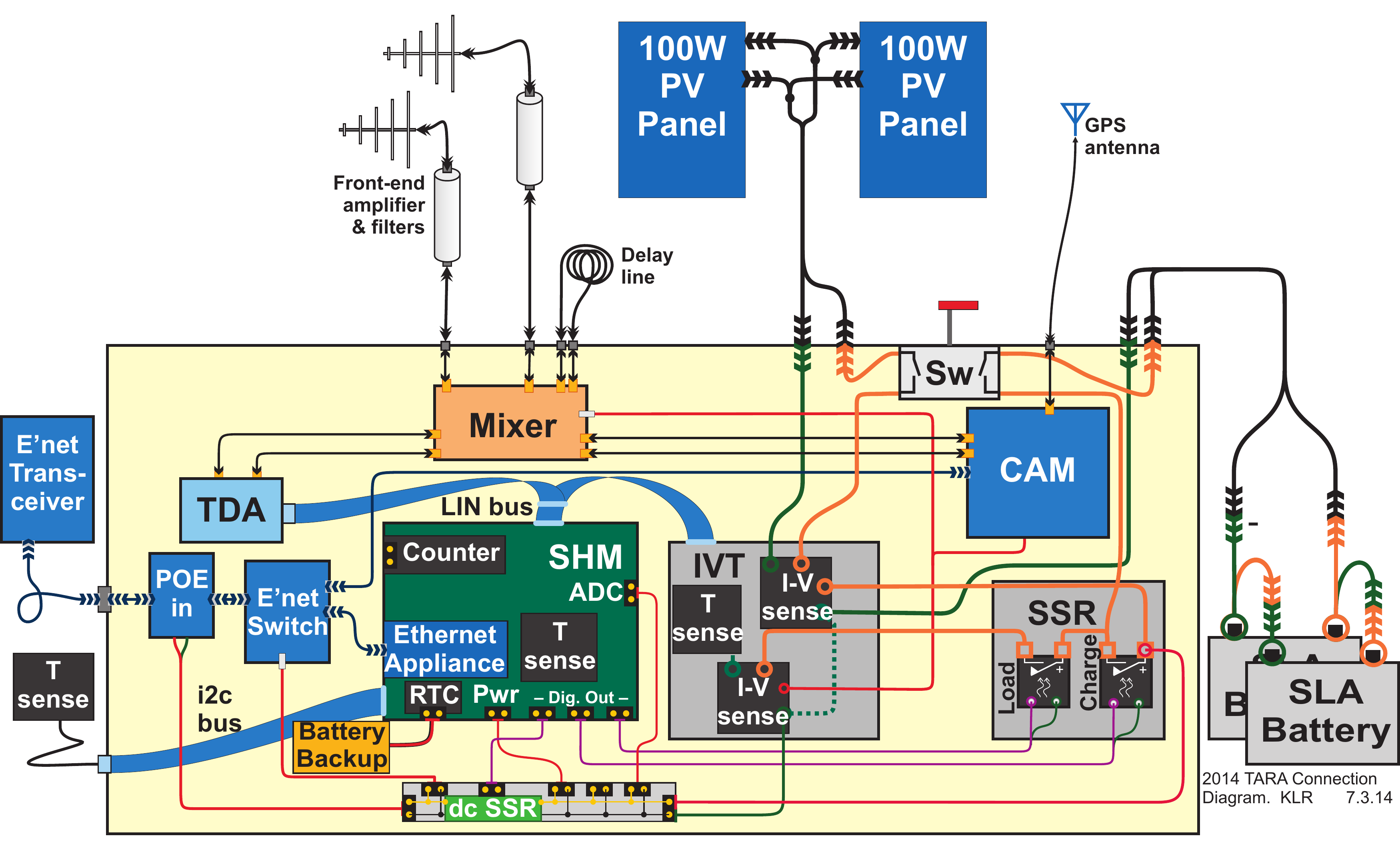}}}
\caption{Schematic block diagram of the remote detector electronics.  Chirp acquisition module (CAM), power systems, acquisition electronics, and communications blocks are all shown in this figure.
\label{fig:rs_proto_2014}}
\end{minipage}
\end{center}
\end{figure*}

The layout of the full remote station, currently in the field outside of Delta, UT, including the Chirp Acquisition Module (CAM), power systems, acquisition electronics, and communications blocks is shown in Figure~\ref{fig:rs_proto_2014}. 

The Chirp Acquisition Module has a modular design enabling quick debugging of the constituent components. This unit is comprised of a custom triggering board encompassing four band pass filters and envelope detectors, and a four channel ADC (AD80066; Analog Devices) with 16~bit resolution sampling at 4~MS/s per channel to sample the signal out of the envelope detectors. A high speed ADC (AD9634 evaluation board; Analog Devices) sampling at 200~MS/s directly samples the raw data from the Antenna and an FPGA (Spartan - 6~LX16; Xilinx) performs the majority comparison logic to trigger and capture triggered data before transferring to a single board computer. A Raspberry~Pi (Rev.~2) single board computer stores triggered data in an SD card along with GPS time stamps (M12M; i-Lotus). 

\begin{table}[h]
\centering
\begin{tabular}{|l|r|r|}\hline
Component              & Power Consumption(W)  \\ \hline\hline
Single Board Computer  &  5.0                  \\ \hline
Low Speed 4 Ch. ADC    &  0.5                  \\ \hline
High Speed 1 Ch. ADC   &  0.4                  \\ \hline
FPGA                   &  3.0                  \\ \hline
RMS Counter            &  2.0                  \\ \hline
System Health Monitor  &  1.0                  \\ \hline
60 dB Amplifier (x2)   &  4.0                  \\ \hline
25 dB Amplifier (x2)   &  0.4                  \\ \hline
GPS                    &  0.2                  \\ \hline
GPS and GPS Antenna    &  0.4                  \\ \hline
Communication Antenna  &  3.0                  \\ \hline\hline
\textbf{Total}         & 19.9                  \\ \hline
\end{tabular}
\caption{Estimated power budget for the remote station.\label{tab:remoterx_power_budget}}
\end{table}

Another major component is the System Health Monitor~\cite{arashm} (SHM), which both monitors performance and controls, via Solid State Relays (SSRs), the system solar (two 100~Watt photo-voltaic panels) and battery (sealed lead acid) power. The SHM also records antenna data digitized by the TDA receiver on local SD flash memory. The TDA (Transient Detector Apparatus) receiver has two channels with front-end amplifiers, followed by filters and a logarithmic amplifier. Finally, the SHM and CAM are connected to a 5~GHz Ethernet transceiver via a switch for remote system control and data access.

\subsection{Remote Station Prototype Studies}
\label{sub:power source}
To understand the required power budget (Table~\ref{tab:remoterx_power_budget}) from the perspective of solar resources in Western Utah, a prototype with system requirements nearly identical to those of the current, full scale remote detector station was deployed at the Telescope Array Fluorescence Detector site at Long Ridge, Utah in the spring of 2013.  

The deployed hardware included a system health monitor (SHM) to monitor performance and power provision, four data acquisition channels, a 12~W dummy-load and Ethernet communications. The first prototype remote site was deployed several hundred meters from the LR site.  Four detector channels include horizontal and vertical polarizations of the standard TARA receiver LPDA, a spiral (frequency-independent) antenna, and 50~$\Omega$ terminator for comparison to system noise.  Antennas feed four bulkhead connectors through LMR-400 coaxial cable, where the signals are amplified and fed into TDA detectors.

\begin{figure}[h]
\begin{center}
\centerline{\mbox{\includegraphics[trim=0.2cm 0.1cm 0.1cm 0.0cm,clip=true,width=0.48\textwidth]{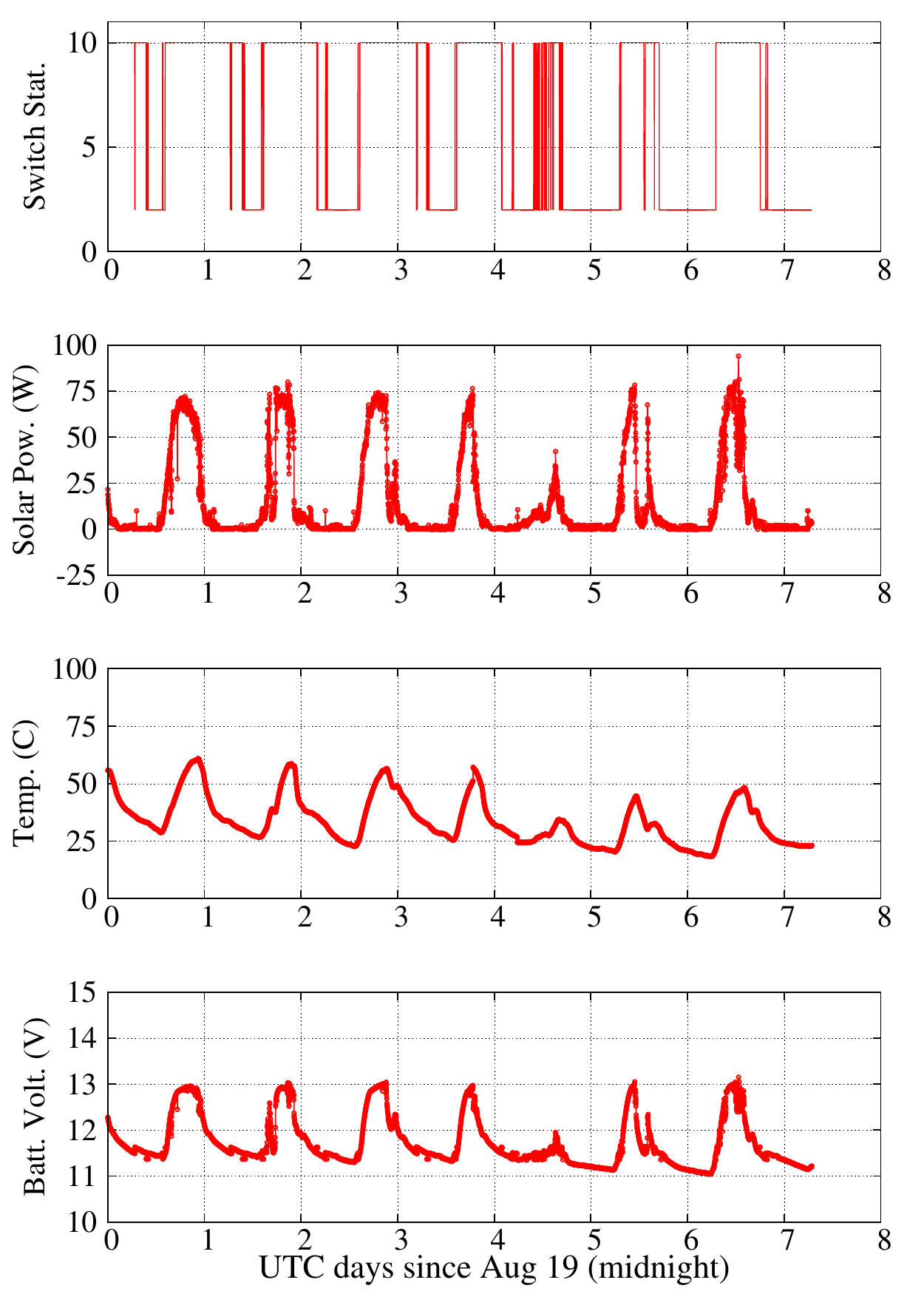}}}
\caption{Remote station solar power data for August.
\label{fig:solarPower1}}
\end{center}
\end{figure}

TDA detectors record a hit when voltage rises above a tunable threshold (set to 100~mV). Independent of the presence of ``hits'', the trigger rate is reported in software over regular 10~s intervals.  The software controlling the detectors and power management of the station is located in micro-controllers on the System Health Monitor (SHM).  

The SHM also supports remote communications, however the prototype station was connected to the FD facilities through 200~m of CAT6 Ethernet cable, with power-over-Ethernet (PoE) offering ample capability to include system monitoring. Several environmental variables in addition to antenna TDA voltage rates are recorded: solar panel current, voltage, battery voltage, the status of the SSRs (the dummy load), temperature measurements, and support for an anemometer.  The prototype remote station recorded solar panel power throughout the summer of 2013, quantifying the amount of solar energy available over time.  Figure~\ref{fig:solarPower1} displays the results.  Each day, the station consumed approximately 13~W, or $\approx 26$~AHr.  Data were transmitted via Ethernet and stored locally.

The data show a clear diurnal variation.  With the 100~W rated solar panel oriented toward the sun at 12~p.m. on June 1st, approximately 75~W peak power delivery was observed.  Solar power curves (second from top in Figure~\ref{fig:solarPower1}) have a 3.6~hr full-width half-maximum, meaning the station collected 21.5~AHr per day.  After 40~days the station began to switch off the dummy load at night via the SSRs as the battery became depleted.  The y-axis of the upper graph is a binary number representing the switch status; 10 (1010 in binary) indicates that both the solar power and dummy load are connected and 2 (0010 in binary) indicates that the dummy load switch has been disconnected by the SHM.  

Two improvements to the remote station have been implemented as a result of these prototype data.  First, solar photo-voltaic power has been doubled (relative to the prototype) to 200~W using two panels.  The power requirement is 20~W for the current remote station (see Table~\ref{tab:remoterx_power_budget}).  After accounting for other prototype inefficiencies, two 100~W panels result in a positive power budget.  Second, fine-tuning of SSR shutdown and start-up voltages has been implemented in the new remote station to protect the batteries.

The first autonomous remote station was deployed in June 2014, approximately 5~km NNE of the Long Ridge receiver station. This station and its performance will be described in greater detail in a forthcoming submission to this journal.

\section{Conclusion}
\label{sec:conclusion}
The TARA detector is designed to search for unique cosmic ray radar echoes with very small radar cross sections (RCS). Specifically, the following key characteristics strongly reduce the minimum detectable RCS: high transmitter power (40~kW, Section~\ref{sec:transmitter}), high-gain transmitter antenna (22.6~dBi, 182~linear, Figure~\ref{fig:txantenna_radiation_patterns}, Section~\ref{sub:txantenna_theoretical_performance}), low noise Radio Frequency (RF) environment consistent with galactic backgrounds (Figure~\ref{fig:LPDA_noise_floor}, Section~\ref{sec:rxantenna}), innovative triggering scheme that permits detection of signals 7~dB below the noise (Section~\ref{subsub:simshower}), and broadband reciever antenna (12.6~dBi gain, 18.2~linear, Figure~\ref{fig:LPDA_rad_pattern}, Section~\ref{sec:rxantenna}). 

\begin{figure}[h]
\centerline{\mbox{\includegraphics[trim=0.2cm 0.0cm 1.2cm 0.0cm,clip=true,width=0.48\textwidth]{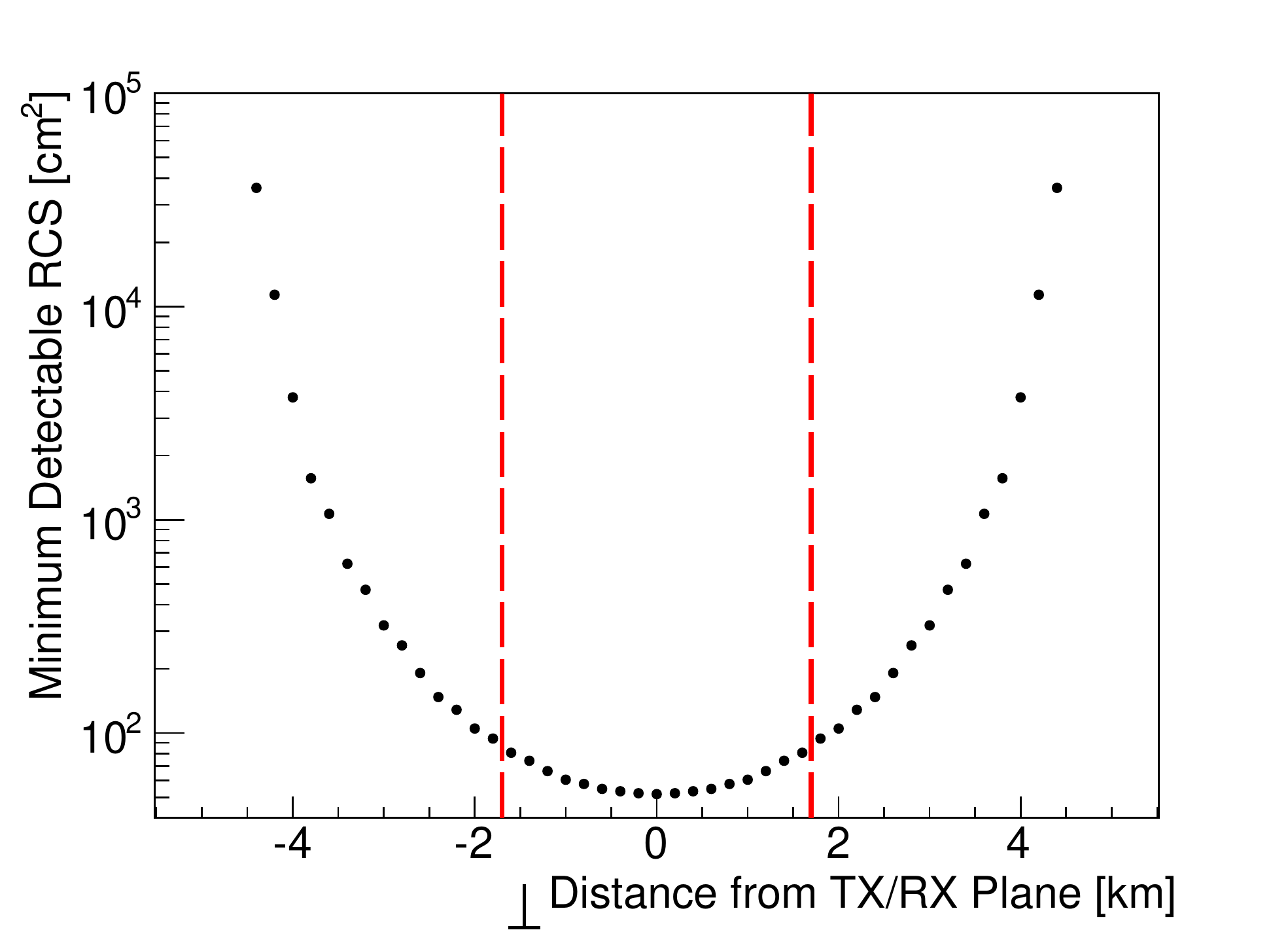}}}
\caption{Minimum detectable radar cross section (RCS) as a function of distance perpendicular to the plane connecting the transmitter and receiver. The transmitter antenna main lobe points along this plane. For simplicity, the minimum RCS is calculated from the bi-static radar equation (Equation~\ref{eq:bistatic_radar_equation_1}, Section~\ref{sec:eas_echoes}) for a cosmic ray air shower midway between transmitter and receiver with maximum transmitter and receiver gains. The 5~MHz FlexRIO passband trigger scheme (Section~\ref{subsub:bandpass}) was assumed to detect a constant amplitude radar echo with chirp rate in [-3,-1]~MHz/$\mu$s (Section~\ref{sub:daqimp}) and signal-to-noise (SNR) ratio 7~dB (Section~\ref{subsub:simshower}) below background noise (Figure~\ref{fig:LPDA_noise_floor}, Section~\ref{sec:rxantenna}), the empirical detection performance for the 5~MHz DAQ passband. Futher assumptions are ground-level detection and constant wavelength $\lambda$. Vertical dashed red lines show the -3~dB beamwidth of the transmitter antenna.
\label{fig:minrcs}}
\end{figure}

Figure~\ref{fig:minrcs} shows a calculation of the minimum detectable TARA RCS for a cosmic ray Extensive Air Shower (EAS) located in several positions along a line perpendicular to the transmitter/receiver plane, midway between the transmitter and receiver. The bi-static radar equation (Equation~\ref{eq:bistatic_radar_equation_1}, Section~\ref{sec:eas_echoes}) permits this simple calculation that assumes a constant power radar echo self-triggered in the DAQ 5~MHz band (Section~\ref{subsub:bandpass}) with chirp rate in [-3,-1]~MHz/$\mu$s (Section~\ref{sub:daqimp}). Maximum transmitter/receiver gains are used for each point, given the azimuthal position of the shower core location. Further, the signal is assumed to have constant wavelength and is Doppler-shifted into the DAQ [60,65]~MHz band, for which the -7~dB noise floor correction is appropriate, and scattered near the ground (to simplify distance calculation). 

The TARA project represents the most ambitious effort to date to detect the radar signature of cosmic ray induced atmospheric ionization. These signals will be characterized by their low power, large Doppler shift (several tens of MHz), and short duration ($\sim10$~$\mu$s). TARA combines a high-power transmitter with a state-of-the-art high sampling rate receiver in a low-noise environment in order to maximize the likelihood of cosmic ray echo detection. Importantly, TARA is co-located with the Telescope Array astroparticle observatory, which will allow for definitive confirmation that any echoes observed are the result of cosmic ray interactions in the atmosphere. 

\section{Acknowledgments}
\label{sec:ack}
This work is supported by the U.S. National Science Foundation grants NSF/PHY-0969865 and NSF/MRI-1126353, by the Vice President for Research of the University of Utah, and by the W.M.~Keck Foundation. L.~B. acknowledges the support of NSF/REU-1263394. We would also like to acknowledge the generous donation of analog television transmitter equipment by Salt Lake City KUTV Channel~2 and ABC Channel~4, and the cooperation of Telescope Array collaboration. 

We would like to specifically thank D.~Barr and G.~McDonough from Telescope Array for their services.  


%% References


%\clearpage

\bibliography{refs}


\end{document}